\PassOptionsToPackage{pdfpagelabels=false}{hyperref}
\documentclass[useAMS,usenatbib]{mnras}
\pdfoutput=1
\usepackage[pdftex]{graphicx}
\usepackage{amsmath}
\usepackage{amssymb}
\usepackage{color}
\usepackage{times}

\definecolor{portlandorange}{rgb}{1.0, 0.35, 0.21}
\definecolor{ForestGreen}{RGB}{20, 180, 20}

\newcommand{\bea}{\begin{eqnarray}}
\newcommand{\eea}{\end{eqnarray}}
\newcommand{\Ms}{{\rm M}_\odot}
\newcommand{\MS}{M_{\rm stars}}

\volume{{\rm submitted}}

\title[TNG Quenched Fractions vs. Observations]{Quenched fractions in the IllustrisTNG simulations: \\comparison with observations and other theoretical models}

\author[Donnari et al.]
{Martina Donnari$^{1}$\thanks{E-mail: donnari@mpia-hd.mpg.de}, 
Annalisa Pillepich$^{1}$, 
Dylan Nelson$^{2,3}$, 
Federico Marinacci$^{4}$,  \newauthor 
Mark Vogelsberger$^{5}$, and
Lars Hernquist$^{6}$
\\\\
$^{1}$Max-Planck-Institut f{\"u}r Astronomie, K{\"o}nigstuhl 17, D-69117 Heidelberg, Germany\\
$^{2}$Max-Planck-Institut f{\"u}r Astrophysik, Karl-Schwarzschild-Str. 1, D-85748, Garching, Germany\\
$^{3}$Universit\"{a}t Heidelberg, Zentrum f\"{u}r Astronomie, Institut f\"{u}r theoretische Astrophysik, Albert-Ueberle-Str. 2, 69120 Heidelberg, Germany\\
$^{4}$ Department of Physics and Astronomy "Augusto Righi", University of Bologna, Via Gobetti 93/2, I-40129, Bologna, Italy \\
$^{5}$ Kavli Institute for Astrophysics and Space Research, Massachusetts Institute of Technology, Cambridge, MA 02139, USA\\
$^{6}$ Harvard-Smithsonian Center for Astrophysics, 60 Garden Street, Cambridge, MA 02138, USA 
}

\begin{document}
\maketitle


\begin{abstract}
We make an in-depth comparison of the IllustrisTNG cosmological simulations with observed quenched fractions of central and satellite galaxies, for $\MS=10^{9-12}~\Ms$ at $0\leq z\leq3$. We show how measurement choices (aperture, quenched definition, star formation rate (SFR) indicator timescale), as well as sample selection issues (projection effects, satellite/central misclassification, and host mass distribution sampling), impact this comparison. The quenched definition produces differences of up to 70 (30) percentage points for centrals (satellites) above $\sim 10^{10.5} \Ms$. At $z \gtrsim 2$, a larger aperture within which SFR is measured suppresses the quenched fractions by up to $\sim$\,50 percentage points. Proper consideration of the stellar and host mass distributions is crucial: naive comparisons to volume-limited samples from simulations lead to misinterpretation of the quenched fractions as a function of redshift by up to 20 percentage points. Including observational uncertainties to theoretical values of $\MS$ and SFR changes the quenched fraction values and their trend and/or slope with mass. Taking projected rather than three-dimensional distances for satellites decreases the quenched fractions by up to 10 percent. TNG produces quenched fractions for both centrals and satellites broadly consistent with observations and predicts up to $\sim$\, 80 (90) percent of quenched centrals at $z=0$ ($z=2$), in line with recent observations, and higher than other theoretical models. The quantitative agreement of TNG and SDSS for satellite quenched fractions in groups and clusters depends strongly on the galaxy and host mass range. Our mock comparison highlights the importance of properly accounting for observational effects and biases.
\end{abstract}



\section{Introduction}
\label{intro}

At low redshift, galaxies in the Universe can be classified in two main populations according to their star formation activity: ``star-forming'' galaxies with on-going star formation, young stellar populations, and blue colors on the one hand and ``quenched'' galaxies which have ceased to form young stars, have older stellar populations, and redder colors on the other.

The balance between these two populations -- or the quenched fraction -- is a strong function of galaxy mass, environment, and cosmic time. Overall, the fraction of quenched galaxies has been shown to increase with galaxy stellar mass at all redshifts \citep[$z\lesssim3$,][]{2013Muzzin,2017Darvish,2018Fang}. In dense environments such as groups and clusters low-mass satellites with $\MS \lesssim 10^{10}~\Ms$ are frequently quenched, due to environmentally-driven processes such as the inability to accrete new gas \citep{1980Larson} combined with the tidal \citep{1983Merritt} and ram-pressure \citep{1972Gunn,2017Poggianti} stripping of pre-existing gas reservoirs. At the same time, more massive galaxies with \mbox{$\MS \gtrsim 3 \times 10^{10}~ \Ms$} can quench regardless of environment \citep{2012Wetzel,2013Wetzel,2015Peng,2019Davies}. In this case the physical mechanism most often invoked is feedback from a supermassive black hole, or Active Galactic Nuclei \citep{2005DiMatteo,2005Hopkins,2006Croton}.

Although there exists a broad qualitative consensus on the quenched fractions of galaxies versus mass, environment, and redshift, several quantitative discrepancies remain across observational compilations \citep{2008Haines,2012Geha,2014Lin,2018Jian,2017Fossati,2017Wagner}. This is due in part to the different approaches that are taken by various studies, particularly to separate quenched and star-forming galaxies. Depending on redshift and wavelength coverage, different tracers are available to derive star-formation rates of galaxies \citep{2014Speagle}, and these inherit systematic biases \citep{2019Leja}. Different group finding algorithms lead to different environmental measures and classifications of central versus satellite status \citep{2007Yang,2015Campbell,2020Gao}. Different surveys contain various selection effects, integrate galaxy star-formation out to different radii and apply different dust corrections \citep{2007Salim}, and sample a diversity of galaxy and host halo mass distributions versus redshift \citep{2019PintosCastro}.

Comparison of observational data with theoretical models can shed light on the impact of the systematic uncertainties, assumptions, and biases \citep{2019Donnari, 2020Katsianis}. Modern large-volume cosmological hydrodynamical simulations such as Illustris, EAGLE, Horizon-AGN, IllustrisTNG, SIMBA, Romulus, and FABLE fundamentally aim to reproduce the observed dichotomy of the galaxy population, of both star-forming and quenched galaxies, through a physical quenching mechanism. As a result, they broadly recover the trend of increasing quenched fraction versus mass \citep{2015Furlong,2019Donnari,2021Donnari}. At the same time, they also self-consistently capture gas dynamical processes such as stripping, enabling a view into the quenching of satellites  \citep{2016Bluck,2017Bahe,2019Tremmel,2020Appleby,2021Donnari}. 

Recent semi-analytical models continue to refine the processes which generate the quenched galaxy population \citep{2015Brennan,2016Croton,2019Baugh,2020Fontanot,2020Henriques}, although -- despite explicit calibration against osbervationally-inferred quenched fractions -- discrepancies remain \citep{2018Asquith}. This highlights the difficulty of producing a full galaxy population with consistent stellar mass functions and SFR distributions across cosmic epochs. \cite{2018Wang} and \cite{2020Ayromlou} compare the quenched fractions between the Munich SAM L-Galaxies \citep{2015Henriques} and the EAGLE and IllustrisTNG simulations, respectively, quantifying differences that arise due to details of the quenching processes.

We focus herein on the IllustrisTNG simulations and the associated TNG physical model for galaxy formation \citep{2017Weinberger,2018Pillepich_model}, making comparisons to observations as well as other theoretical models. The outcomes of the TNG simulations have been validated against a broad set of observations: the color distribution of red versus blue galaxies at $z=0$ \citep{2018Nelson}, the distributions of metals in the intra-cluster medium \citep{2018Vogelsberger}, the galaxy mass-metallicity relation \citep{2018Torrey}, the galaxy size-mass relation across redshifts and for star-forming versus quenched galaxies \citep{2018Genel}, the OVI content of the circumgalactic medium \citep{2018NelsonB}, the dark matter fractions within massive galaxies \citep{2018Lovell}, the environmental dependencies of the cold gas contents of satellites \citep{2019Stevens}, and the star formation main sequence and quenched fractions of central galaxies \citep{2019Donnari}.

In our recent companion paper \citep{2021Donnari} we present the diverse pathways that TNG galaxies take towards quenching within the hierarchical growth of structure. Having postponed observational comparisons there, in this paper we now aim to properly and robustly compare the outcome of the TNG simulations to empirical inferences of central and satellite quenched fractions as a function of stellar masses, halocentric distance, host halo mass, and redshift. We focus mainly on galaxies with stellar mass between $10^{9-12} \Ms$ residing in groups and clusters, and we primarily compare to SDSS data by creating mock catalogs following the procedures adopted in  \cite{2012Wetzel,2013Wetzel}.

This paper is organized as follows. In Section \ref{method} we introduce the simulations and analysis methods. In Section \ref{choices} we show how different observational issues impact the inferred quenched fractions. In particular, we contrast several definitions of star-forming versus quenched galaxies, physical apertures, and star-formation rate tracers, also considering observational uncertainties, host mass estimates, satellites selections, and central/satellite misclassification. A comparison to other theoretical models and observations is presented in Section \ref{observations}, with a particular focus on SDSS data. We conclude and summarize in Section \ref{summary}.


\section{Methods and Definitions}
\label{method}

\subsection{The IllustrisTNG simulations}

In this work, we chiefly use the TNG100 and TNG300 cosmological magneto-hydrodynamical simulations of the IllustrisTNG\footnote{\url{http://www.tng-project.org}}  project \citep{2018Naiman,2018Marinacci,2018Springel,2018Pillepich,2018Nelson}. These two volumes have side lengths of $\sim 100$\,Mpc and $\sim 300$\,Mpc, and mass resolutions (average gas cell mass or star particle mass) of $m_{\rm baryon}\sim 1.4\times 10^6 \Ms$ and $\sim 1.1\times 10^7 \Ms$, respectively. The third and highest resolution volume is TNG50 \citep{2019Pillepich_50,2019Nelson_50}, which simulates a smaller volume of $\sim 50$\,cMpc across, with roughly fifteen times higher mass resolution, $m_{\rm baryon}\sim 8.5\times 10^4 \Ms$.
In this paper we use TNG50 in comparison to other models and observations.

The TNG simulations are performed with the \textsc{Arepo} code \citep{2010Springel}, which solves for gravity and ideal magnetohydrodynamics \citep{2011Pakmor}. They include a well-explored and well-validated physical model for the key galaxy formation physics. All details of this underlying galaxy formation model are described in the two TNG method papers \citep{2017Weinberger,2018Pillepich_model}. In short, the simulations evolve dark matter, gas, stars, and supermassive black holes, accounting for the physical processes of gas radiation (heating, cooling, and the background radiation field), star formation \citep[above a density threshold following the two-phase interstellar medium model of][]{2003Springel}, stellar population evolution and chemical enrichment (tracking supernovae types Ia, II, and AGB stars), galactic-scale winds from stellar feedback \citep[see][]{2018Pillepich_model}, supermassive black holes (formation, mergers, and gas accretion), and black hole feedback \citep[combining a high accreiton rate thermal mode with a low accretion rate kinetic mode; see][]{2017Weinberger}.

\subsection{Central and satellite populations}
\label{sec:sample}

Substructures corresponding to haloes and subhaloes are identified via the Friends-of-Friends \citep[FoF;][]{1985Davis} and \textsc{Subfind} algorithms \citep{2001springel,2009Dolag}. In this work we differentiate between galaxies residing in groups and clusters, from the general or `field' population. Unless otherwise stated we define the virial radius as $R_{200c}$, the radius within which the mean enclosed mass density is 200 times the critical density of the Universe. The total mass enclosed within this radius is $M_{200c}$, and in general we consider all haloes with $M_{200c} \geq 10^{12} \, \Ms$ to be hosts. 

We refer to hosts with mass in the range $10^{12-14} ~\Ms$ as groups, and to those above this mass as clusters. All galaxies residing within one virial radius of the center of their parent host FoF halo are satellites, except for the single galaxy centered at the minimum of the gravitational potential, which is typically the most massive and labeled as the central. We also call ``central'' the galaxies that are associated with the field population, i.e. those that are not associated with any groups or clusters.
Throughout this analysis, the central sample may include backsplash galaxies, namely galaxies that might have been satellites in the past but are not at the time of inspection.

\begin{figure*}
\centering
\includegraphics[width=\textwidth]{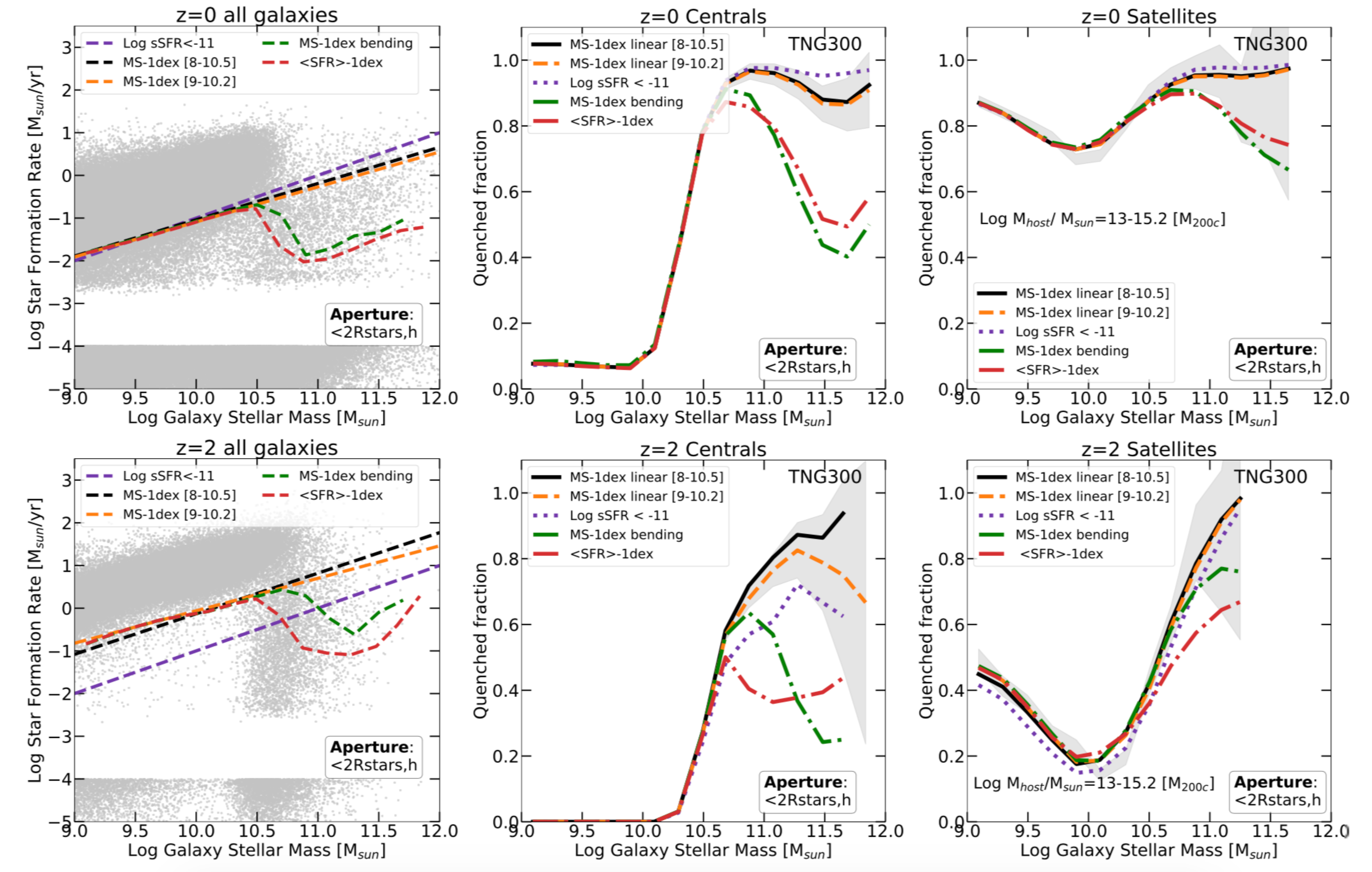}
\caption{\label{fig:definitions} {\bf Quenched definitions}. The SFR-$\MS$ plane for the TNG300 simulation (left panels), as well as the quenched fraction versus stellar mass for centrals (middle panels) and satellites (right panels), at $z=0$ (top) and $z=2$ (bottom). In the left panels, colored dashed lines define star-forming galaxies and thus trace the lower end of the MS, as labeled in the legend. These same colors show the corresponding quenched fractions in the middle and right panels. The grey shaded area represents the Poissonian errors (shown for the black line only for clarity). Although the definition of `quenched' has little impact for low-mass galaxies below $10^{10.5} \Ms$, above this mass the inferred quenched fractions are highly sensitive to this choice. This holds for both centrals and satellites, is true at both $z=0$ and $z=2$, and can qualitatively change the interpretation of high-mass quiescence. These considerations must hence be taken into account whenever results from different observational or simulated samples are compared.}
\end{figure*}

At $z=0$, TNG300, TNG100, and TNG50 have 3733, 182, 24 groups and clusters above $M_{200c} \geq 10^{13} \, \Ms$, respectively, and more than 39,000, 2,800, and 1,600 satellites with stellar mass $\MS\geq 10^9\Ms$. The most massive clusters in TNG300, TNG100, and TNG50 has a halo mass of $10^{15.2} ~ \Ms$, $10^{14.6}$, and $10^{14.3} ~ \Ms$ -- see also Table 1 of \citet{2021Donnari}. The minimum stellar mass we consider in this paper is $\MS \geq 10^9 ~ \Ms$, corresponding to at least about 100, 1000, 1500 star particles in TNG300, TNG100, and TNG50, respectively.

\subsection{Stellar masses and star formation rates}
\label{SFRinTNG}

We adopt throughout a common definition of galaxy stellar mass as the sum of all gravitationally bound stellar particles within twice the stellar half-mass radius.
We note that this is not comparable to any observational choice and differences can be large in general, especially at the high-mass end \citep[see][for a discussion on this]{2018Pillepich}. However, for the purposes of the quenched fractions, the stellar mass choice implies a somewhat negligible shift of the functions along the x-axis: see Appendix~\ref{appendix_A}.

We characterize galaxies by their `instantaneous' SFR by taking, directly from the simulation output, the sum of the SFRs of all gas cells that are gravitationally bound and within $2 R_{\rm star,h}$, twice the stellar half mass radius. 

We also consider the SFR measured over different timescales, measured indirectly from the mass of recently formed stellar particles within the same aperture. We consider time intervals of both 200 Myr and 1000 Myr (see Section \ref{sec:SFRs}). We refer to other upcoming works the task of extracting SFRs from the simulated galaxies by mocking what is typically done in observations.  

Finite numerical resolution implies that there is a minimum resolvable star-formation rate. To properly handle this issue, we randomly assign a SFR value in the range $(10^{-5} - 10^{-4}) \, \Ms \rm yr^{-1}$ to those galaxies with log(SFR) below the resolution limit of the simulation, which otherwise would have SFR $\sim 0$ (see \citealt{2019Donnari} for more details). The existence of such galaxies is of the essence for those authors who simply average SFRs instead of distinguishing between star-forming and quiescent galaxies in some manner. Even if their actual distribution is unknown, those galaxies would distribute well below the main sequence, thus populating the quenched region.
Additionally, we note that we do not identify starburst or green-valley galaxies.

\begin{figure*}
\centering
\includegraphics[width=0.49\textwidth]{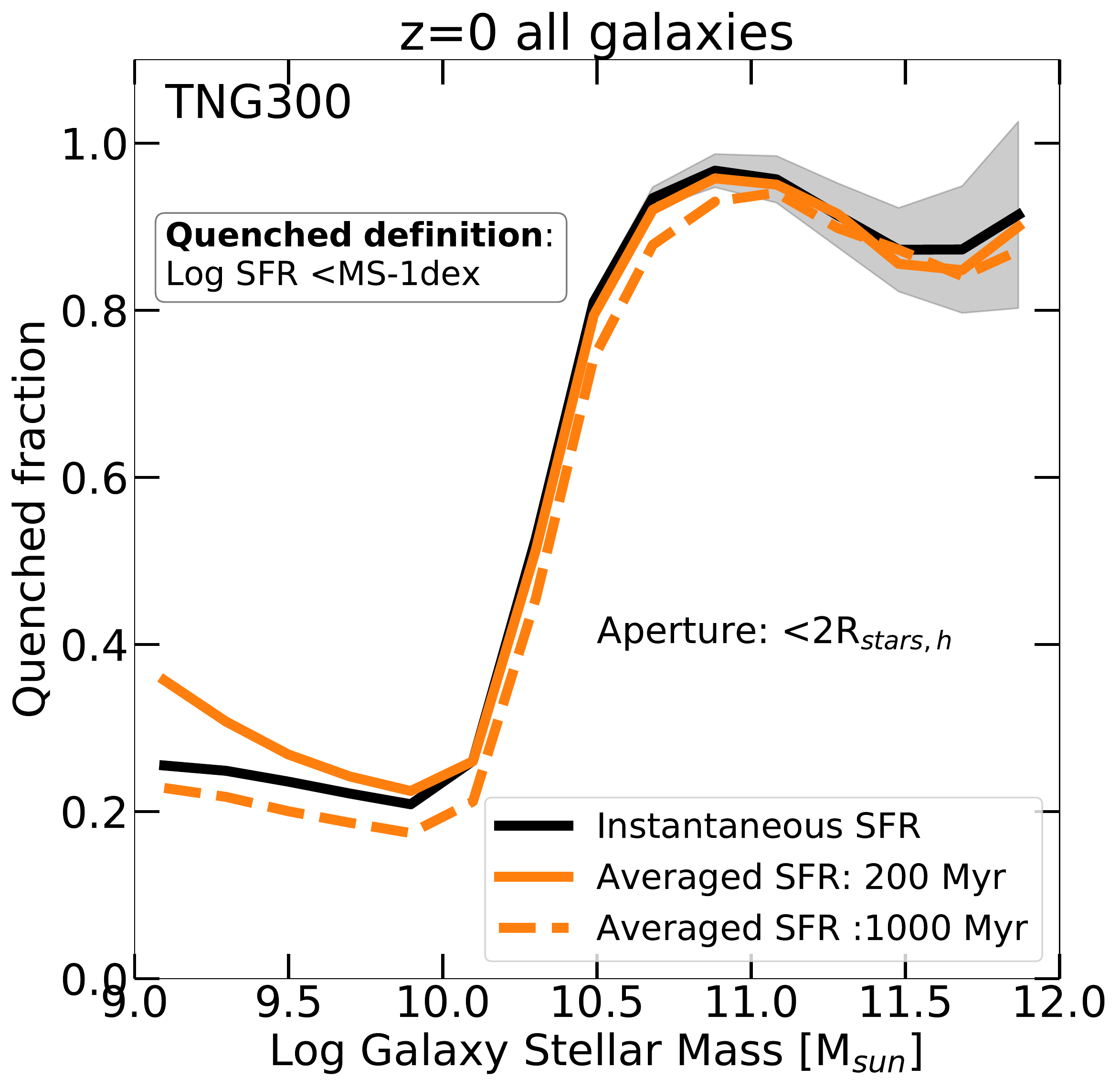}
\includegraphics[width=0.49\textwidth]{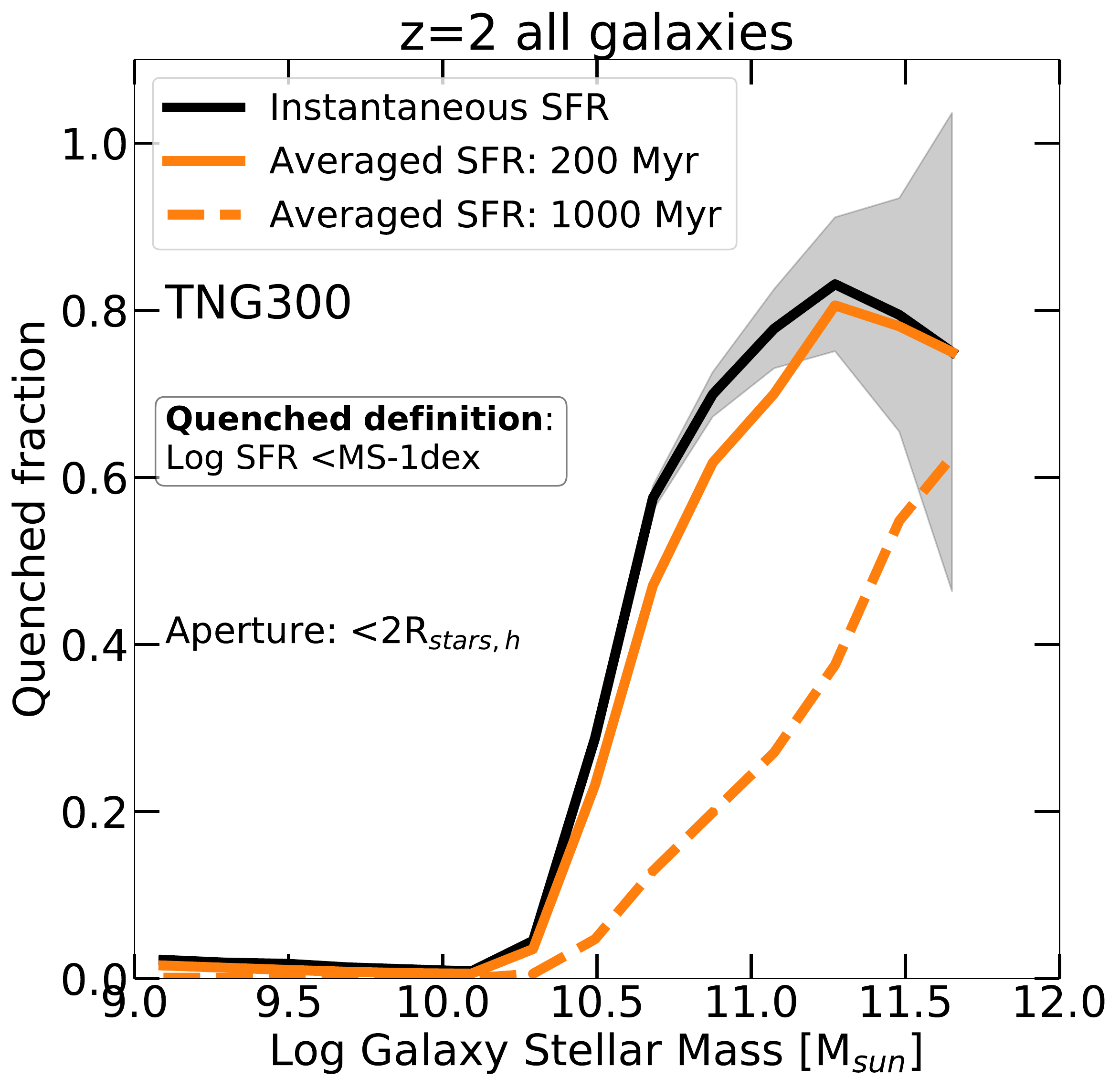}
\caption{\label{fig:timescales} {\bf Averaging timescales for SFR} Quenched fractions of TNG300 galaxies at $z=0$ (left) and $z=2$ (right), for the instantaneous SFR measured from the gas (black) and from the amount of stellar mass formed over the last 200 Myr (orange solid) and 1000 Myr (orange dashed). In all cases, quenched galaxies are selected via the linearly extrapolated MS definition (see text). While at redshift zero the impact of averaging timescales is negligible, for $z>2$ where SFRs are higher and have higher variability, it can be significant.}
\end{figure*}
\section{Results: The Systematic Uncertainties of Galaxy Quenched Fractions}
\label{choices}

Different measurement decisions and sample selection effects can have a critical impact on the comparison of observational data with theoretical models. In fact, they can have critical effects in whatever comparison, whether across observational samples, among simulated results and for the simulation vs. observations comparisons. Here we undertake an analysis of how key choices shape the inferred quenched fractions of TNG300 galaxies. Under the assumption that the TNG galaxies are sufficiently realistic, the systematic uncertainties quantified in the following apply to any observational and simulated sample.

\subsection{The effects of different measurement choices}
\label{sec:choices}

\subsubsection{Star-forming vs. quenched definitions}

We explore several criteria to separate quenched versus star-forming galaxies. Throughout the paper we do not make distinctions between {\it quiescent} and {\it quenched} galaxies, which will all be called quenched. 

Figure \ref{fig:definitions} shows how the star-formation main sequence (MS) and quenched fractions, both as a function of galaxy stellar mass, depend on definition. The grey shaded area represents the Poissonian errors\footnote{Strictly speaking, the errorbars for the quenched fractions should be estimated via a binomial proportion confidence interval. We have compared the Poissonian errors that we adopt throughout to those from a Wilson score interval, finding them consistent in the majority of the regimes.}, shown for the black curve only for clarity. All galaxies (left panels) are contrasted against centrals (middle panels) and satellites only (right panels), at $z=0$ (top) and $z=2$ (bottom). The gray point clouds show the full underlying galaxy population, while colored curves indicate the boundary between the MS and the quenched galaxies or the quenched fraction medians, respectively, according to the definitions in the legends. The `quenched' definitions we consider are all are based on SFR, and defined as follows \citep[see][for UVJ-based criteria]{2019Donnari}.

{\it Specific star formation rate $< 10^{-11}$\,yr$^{-1}$:} according to a commonly adopted definition in the literature \citep{2008Franx,2009Fontanot,2020Sherman}, `quenched' galaxies are defined as those having sSFR $\leq 10^{-11} \, \rm yr^{-1}$ (purple  lines). We note from the onset that, at $z>0$ this fixed threshold fails to capture the evolution of the median MS, and a redshift-dependent threshold should be preferred to capture galaxy quenching across cosmic epochs. 

{\it One dex below the extrapolated star-forming main sequence:} following \cite{2019Donnari, 2019Pillepich_50}, we iteratively locate the locus of the MS and label galaxies as quenched if they fall 1 dex or more below it. Specifically, we stack all galaxies in 0.2 dex bins of stellar mass, measure the median SFR in the bin, and iteratively remove quenched systems, re-calculate a new median, and repeat until the median SFR in the mass bin converges. Beyond a certain stellar mass the locus of the MS in TNG, as well as in observations, bends downwards towards lower sSFR values. We therefore extrapolate the MS, by adopting a linear fit within two different stellar mass ranges -- $\Ms= 10^{8-10.5}\MS$ (black lines) and $\Ms= 10^{9-10.2}\MS$ (orange lines). We also use the fit MS directly without extrapolation (green lines). 

{\it One dex below the median SFR:} we define the MS as the median value of the SFR values in a given  stellar mass bin, using the same 0.2 dex bins as above, and label quenched galaxies as those 1 dex or more below the median (red lines).

Besides the methods listed above, other criteria have been used in the literature, such as the Gaussian Mixture Model \citep{2018Bisigello,2019Hahn} and, more traditionally, colour-colour and colour-magnitude diagrams \citep[][and reference therein, and see also \citealt{2019Donnari} for the simulated UVJ diagram of TNG galaxies]{2009Williams,2010Whitaker,2011Wuyts,2012Quadri}.

Figure \ref{fig:definitions} shows that all the definitions are largely in agreement for galaxies below $\sim 10^{10.5} \Ms$ at $z=0$. As a result, they all give the same quenched fractions of $\sim$\,10 percent for centrals and $\sim$\,80 percent for satellites, on average. However, for galaxies above $\sim 10^{10.5} \Ms$ the behavior becomes more complex. At $z=0$ the linear extrapolations of the MS (black and orange lines) as well as the fixed threshold \mbox{log(sSFR/yr)$ < -11$} (purple lines) return similar quenched fractions of $\sim$\,90-100 percent for both centrals and satellites\footnote{It is important to note that this agreement depends on sampling, especially at the highest mass end. In TNG100, for example, where there are fewer high-mass galaxies than in TNG300, fitting the MS over different stellar mass ranges implies quenched fractions at  $\MS \simeq 10^{11.0-11.5}\,\Ms$ that differ by up to 20 percentage points.}. On the other hand, by following the bending main sequence or the median SFR values (green and red lines), the inferred quenched fractions are lower by up to $\sim\,40-50$ percentage points for centrals, and up to $\sim\,20-30$ percentage points for satellites.

The results are similar at $z=2$, but with even more dramatic possible differences. It should be noted that at higher redshift the SFR-$\MS$ plane is shifted toward higher SFR values:  as a result, the fixed criterion \mbox{log(sSFR/yr) $< -11$} fails to follow the evolution of the MS and is a poor separator of the quenching boundary. 
Whereas we discourage the usage of such cut at $z>0$, here we show it for reference and to raise awareness among the readers by contrasting it against a more appropriate choice. Throughout the paper, we also we use the fixed criterion whenever the TNG outcome is compared to other works that adopt this method, in order to make a significant comparison.

As at redshift zero, also at higher redshifts the low-mass end quenched fractions are insensitive to definition. However, the separation between quenched vs. star-forming galaxies becomes critical for galaxies more massive than $\gtrsim 10^{10.7}~\Ms$. For these massive systems the $z=2$ quenched fraction of centrals (satellites) can vary from $30-100$ percent for centrals, and from $60-100$ percent for satellites taken collectively. The simulation predicts that $z=2$ high-mass galaxies with $\MS \simeq 10^{11.5}\,\Ms$ have a low quenched fraction of only $30$ percent if one does not extrapolate from the low-mass MS (green lines), or a high quenched fraction of $80$ percent if doing so (orange and black lines, lower middle panel).

Overall the shape and the actual values of the star-formation main sequence, as well as the quenched fraction of galaxies, sensitively depend on definition at the high-mass end. This is particularly severe for central galaxies, and to a lesser degree, also for satellites. We hence conclude that the only way to make meaningful comparisons across samples is by matching the adopted star-forming vs. quenched definitions.

\begin{figure*}
\centering
\includegraphics[width=0.33\textwidth]{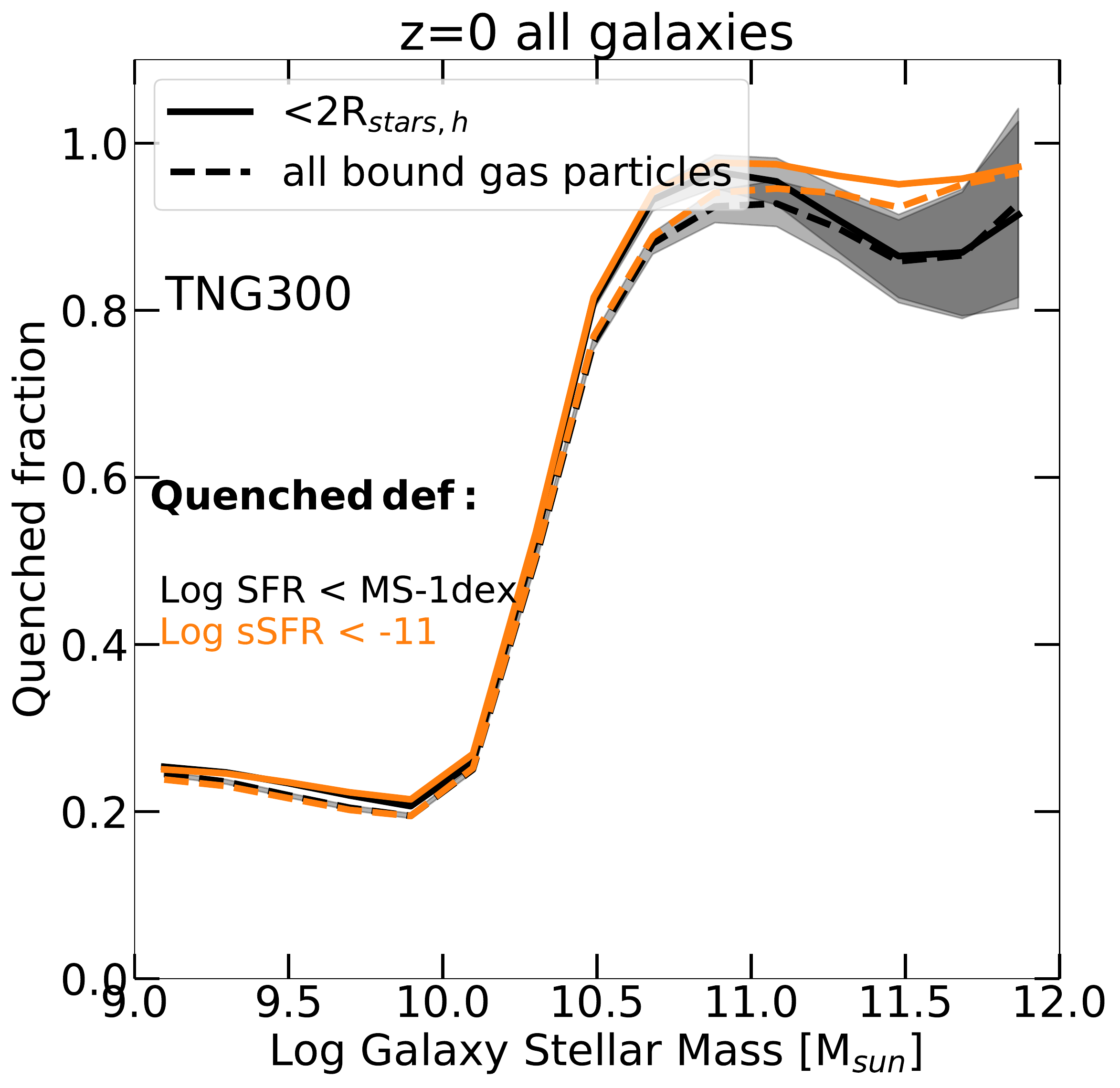}
\includegraphics[width=0.33\textwidth]{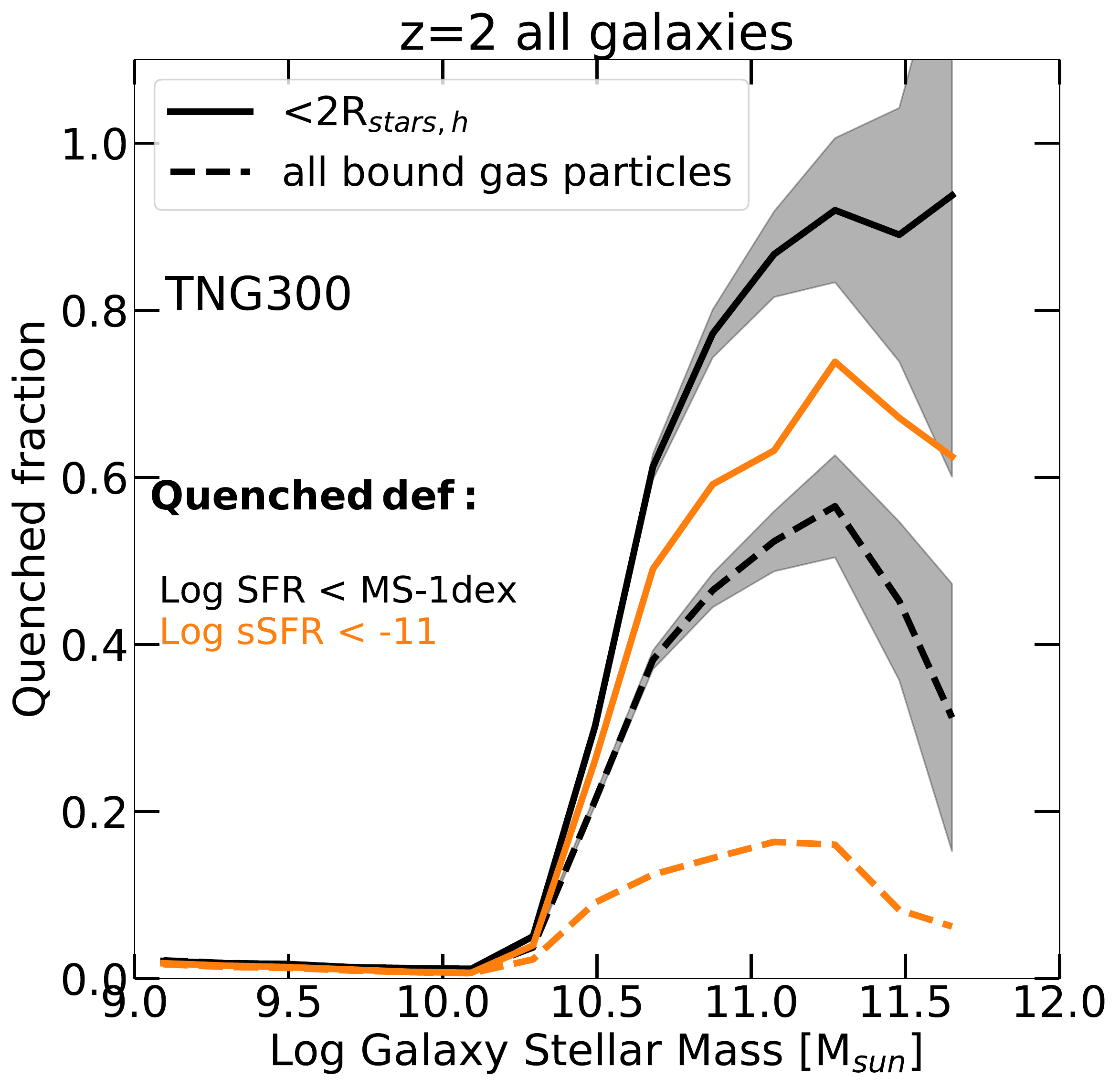}
\includegraphics[width=0.33\textwidth]{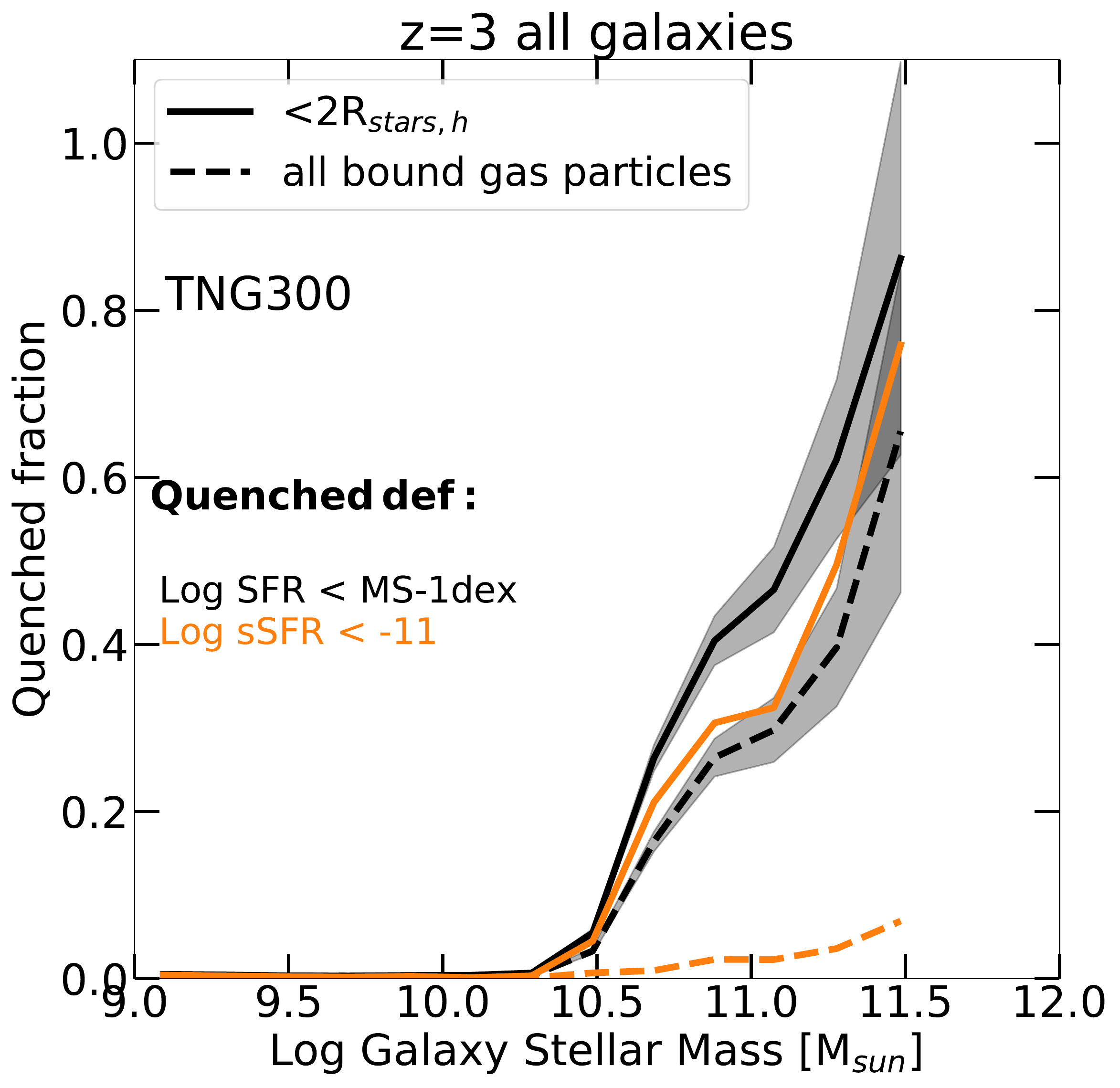}
\caption{\label{fig:apertures} {\bf Different apertures and quenched definitions.} We show the trend of quenched fraction versus stellar mass for all galaxies in TNG300 at $z=0$ (top left), $z=2$ (top middle), $z=3$ (top right). In each case, we contrast two different physical apertures within which star-formation is measured, either restricting to $2R_{\rm stars,h}$ (solid), or including all gravitationally bound gas (dashed). We also compare the impact for two different quenched definitions, MS-1dex (black) and sSFR$<$11 (orange). Measuring SFRs in larger apertures substantially lowers the inferred quenched fractions for high-mass $z \gtrsim 2$ galaxies.}
\end{figure*}
\subsubsection{Averaging timescales for SFRs} 
\label{sec:SFRs}

As discussed in Section \ref{SFRinTNG}, the most direct measurement of star-formation rates from a hydrodynamical simulation is the instantaneous SFR of gas, which is not a direct observable. In observed data the SFR of a galaxy can be derived from a variety of different techniques, typically based on available wavelength coverage of the SED. These different tracers are thought to be sensitive to recent star-formation rates over different timescales \citep[see also][]{2019Donnari}. 

In Figure \ref{fig:timescales} we compare the quenched fractions of galaxies inferred based on adopting different star-formation indicator timescales: instantaneous (black curves) versus time-averaged over 200 Myr (solid orange) and 1000 Myr (dashed orange). To label quenched galaxies we adopt the extrapolated MS definition, and again contrast results from $z=0$ (left panel) to $z=2$ (right panel), including all galaxies in TNG300.

At $z=0$ the impact of different averaging timescales compared to the instantaneous SFR is almost always negligible. The small differences at the low-mass end ($<10^{9.5}~\Ms$) are likely due to resolution effects \cite[][Appendix A]{2019Donnari}.

At higher redshifts, however, the star formation rates of galaxies are overall higher, and more time variable, making indicators more sensitive to different timescales. At $z=2$, while there is no appreciable difference between the instantaneous and 200 Myr SFRs (black versus solid orange curves), the longer Gyr timescale produces substantially lower quenched fractions.  For galaxies above $10^{10.5}~ \Ms$ this effect lowers the inferred quenched fraction by up to $\sim$\,40 percent. This results from how galaxies shift on the SFR-$\MS$ plane under different averaging timescales, particularly below the MS: while the locus of the MS is not strongly affected by the averaging timescales \cite[see Figure 2 of][]{2019Donnari_err}, the SFRs of galaxies in the process of being quenched (e.g. those that are falling off or are below the MS) can vary dramatically depending on the SFR tracer.

\subsubsection{Apertures}

\begin{figure}
\centering
\includegraphics[width=0.46\textwidth]{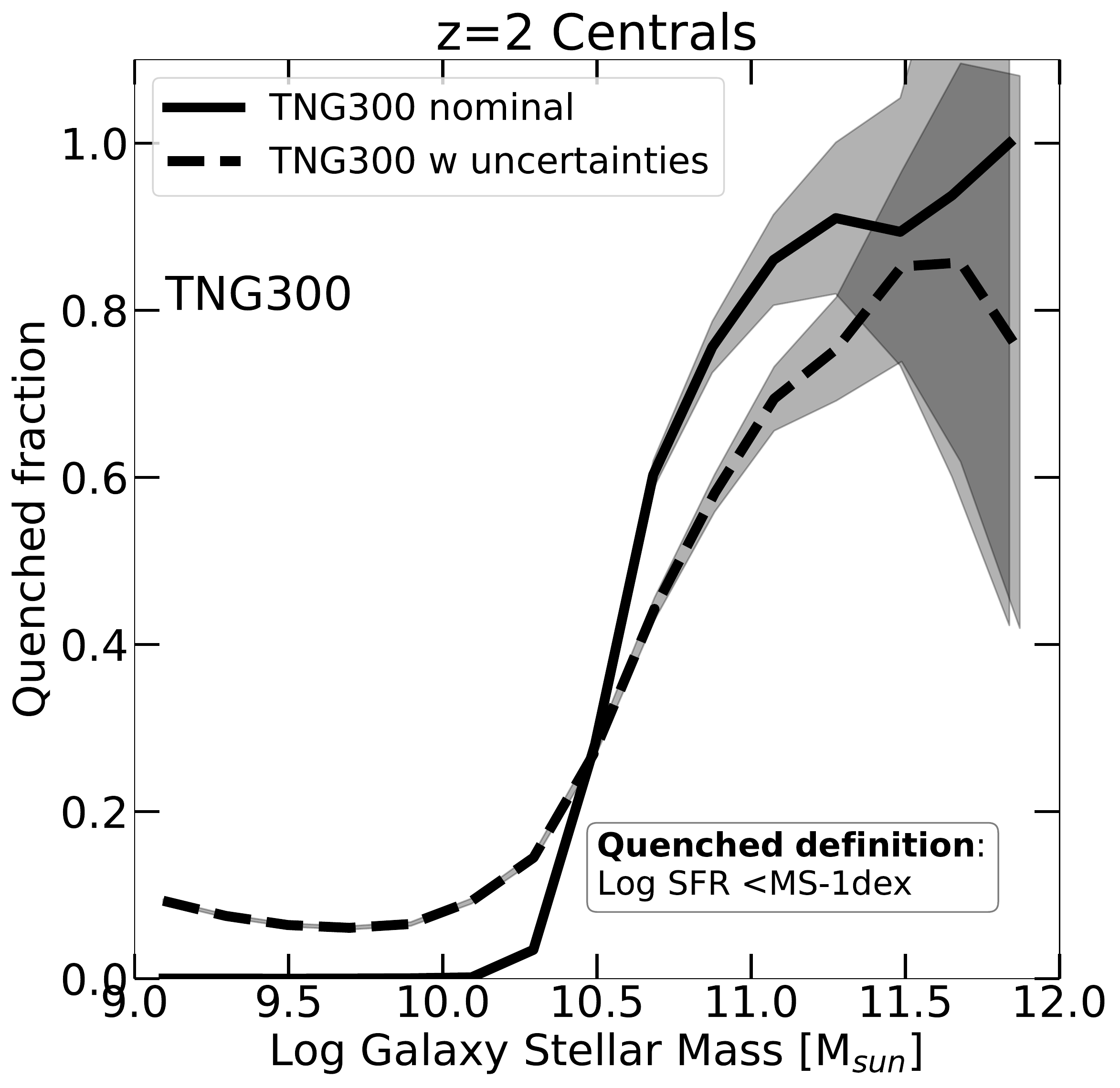}
\includegraphics[width=0.46\textwidth]{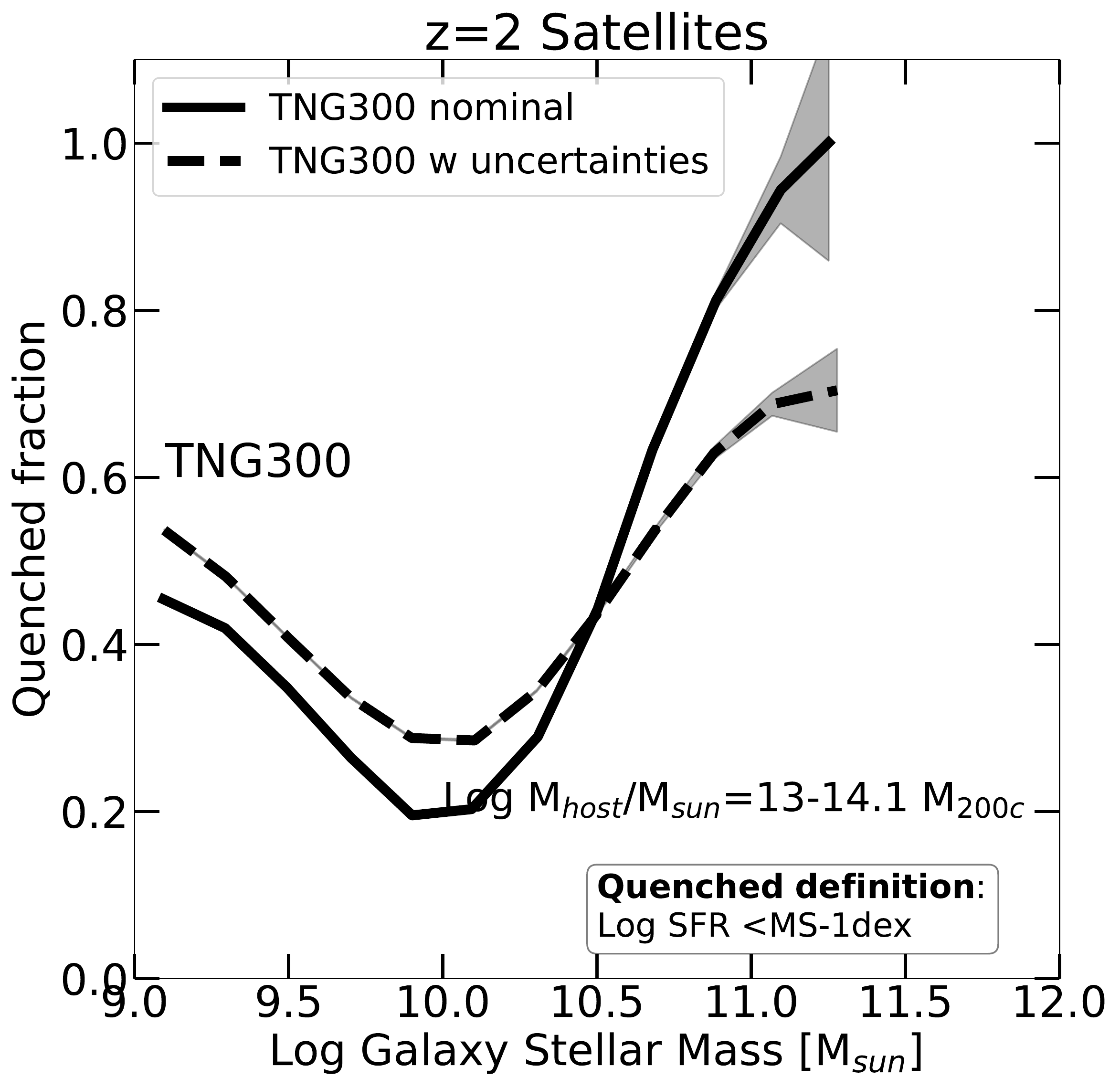}
\caption{\label{fig:errors} {\bf Observational uncertainties}. Quenched fractions of centrals (top panel) and satellites (bottom panel) at $z=2$. Solid curves show direct results from TNG300, assuming zero uncertainty, while dashed curves include observationally-motivated random errors on the galaxy stellar mass and SFR measurements. The most important effect of including these systematic uncertainties is to make the trend of quenched fraction with mass less steep, as galaxies are scattered between bins. The same findings apply almost unchanged at all redshifts studied here ($z\leq3$)}.
\end{figure}

The measured value of the SFR depends on the physical aperture within which it is measured. In the following, we show this using the outcome of TNG300: however, given the demonstrated realism of many properties of the TNG galaxies \citep[see][for a partial compendium]{2019Nelson_PublicRelease}, we expect this issue to apply a priori also to observed galaxies. 

To explore the impact of the aperture choice, Figure \ref{fig:apertures} shows the quenched fraction trends for an instantaneous SFR measured within twice the stellar half mass radius $2R_{\rm stars,h}$ (solid curves) versus for all gravitationally-bound resolution elements (dashed curves). The latter is an effective aperture that is typically straightforward to use with simulation data, given the functioning of most galaxy finders, but it is in practice never attained when dealing with observational data. We show results for $z \in \{0,2,3\}$ (different panels), including in each case two different quenched definitions, MS-1dex linearly extrapolated (black) and \mbox{log(sSFR) $ < -11\,\rm{yr}^{-1}$} (orange).

At $z=0$ the variations due to different physical apertures are minimal. Conversely, at higher redshift $z > 2$ and for galaxies more massive than $10^{10.5}\Ms$ we find that larger apertures lead to lower quenched fractions. That is, measuring SFRs within the galaxy body only (solid lines) results in a quenched fraction which is up to $\sim$\,60 percent higher, as compared to more global SFRs (dashed lines). It is clear that these differences substantially impact the interpretation of high redshift galaxy quenching. For instance, if one adopted the log(sSFR/yr) $ < -11 $ definition at $z=2$ and extracted an obvious value from the simulation -- the SFR of all gas cells -- the conclusion would be that only a small minority, $\lesssim 10$\, percent, of massive galaxies with $\MS \simeq 10^{11.5}\Ms$ are quenched. Changing the aperture to twice the stellar half mass radius leads to the opposite conclusion: that the majority of galaxies, $\gtrsim 60$\,percent, are quenched at the same mass and redshift.

In the TNG model, star formation occurs in extended regions beyond the main bright body of galaxies \citep{2019Donnari}. Quenching also occurs `inside-out' \citep[][their Figure 16]{2019Nelson_50}, with star-formation first suppressed in the centers of galaxies, and later in the outskirts. As a result, SFRs measured within for example radii of 5 pkpc versus 30 pkpc can be non-negligibly different. This is particularly relevant for transitional galaxies, and hence can especially impact high redshift results. As observations can only access a star-formation tracer within some aperture, defined by instrumental considerations \citep[e.g. a fiber-fed spectrograph such as SDSS;][]{2007Salim}, or by sensitivity limits \citep[e.g. resulting in loss of sufficient S/N in an emission line such as H$\alpha$;][]{2018ForsterSchreiber}, it is important to appropriately match this choice in any comparison: for theory vs. data, data vs. data, or theory vs. theory. Indeed, different choices in SFR measurements have been proposed as the underlying reason for the long-standing inconsistency between observed stellar mass functions from $0 < z < 2$ and observed star formation rates over these same epochs \citep{2020Leja}.

\subsubsection{Statistical errors for observationally-inferred values}

To explore the impact of statistical uncertainties on the two key observationally inferred quantities -- stellar masses and star formation rates -- Figure \ref{fig:errors} again shows the quenched fractions of centrals (top panel) and satellites (bottom panel) at $z=2$. The results at redshift zero are similar and omitted for brevity. The direct output of the simulation, with $\MS$ and SFR values assumed to be known exactly, are shown with solid black lines. We then add a random Gaussian component to the simulated stellar masses (with $\sigma = 0.2$ dex) and SFRs (with $\sigma = 0.6$ dex) in order to roughly match observational uncertainties in these quantities, as inferred from SED fitting and SFR calibrators, respectively \citep[see also][]{2018Genel,2019Leja,2021Anand}. The resulting quenched fractions are shown with dashed curves.

Regardless of the redshift and of whether a galaxy is a central or a satellite, the most important effect of including reasonable statistical uncertainties is a `shallower' trend of the quenched fraction with galaxy stellar mass. This is particularly clear for centrals, where the slope of the median line is smaller with uncertainties included. We think that, when measurement uncertainties are included, more low stellar mass galaxies scatter into higher stellar mass bins than the other way around.
The sharpness of the transition from predominately star-forming to predominantly quenched galaxies is also reduced, i.e. smoothed out, by the observational uncertainties: the latter is intricately linked with the physics of quenching in the simulations \citep[also producing the blue to red color transition][]{2018Nelson}, most notably through the high to low state black hole feedback transition \citep{2018Weinberger}. It has been suggested that this transition is somewhat too rapid in TNG \citep[e.g.][]{2020Terrazas}, but our result highlights that not only the normalization, but also the slope, of galaxy population relations can change due to systematic uncertainties \citep[see also][for a related conclusion with respect to the $M_{\rm HI}-M_\star$ relation for TNG]{2019Stevens}. 

Beyond this steepening effect, the inclusion of systematic errors also changes the actual values of galaxy quenched fractions, depending on mass. With respect to the error-free case (solid lines), galaxies below (above) $\sim 10^{10.5} ~\Ms$ show a higher (lower) quenched fraction, by $\sim$\, 10 percent. One consequence is the interpretation of whether or not low-mass quenched field centrals \citep{2012Geha} exist in the simulations.


\subsection{The effects of sample selection}

\subsubsection{Host halo mass distributions}
The most relevant quantity that determines environmental processes is the total halo mass of the host potential wells where satellite galaxies reside. Both observations and simulations agree in that higher host mass implies larger quenched fractions (at least for low-mass galaxies, see \citealt{2021Donnari} for an extended discussion). A multitude of observational techniques is adopted to infer host halo masses (via abundance matching or mass proxies like X-ray cluster properties or optical-based richness), each affected by its own systematic bias and/or random errors \citep{2014Old,2015Old,2018Old}. 
For example, \cite{2018Old} find a systematic bias for several methods used to infer halo masses, depending on the cluster mass. They are indeed significantly overestimated for lower mass clusters, by $\sim$ 10 percent at $10^{14} \Ms$ and larger than  20 percent for  $<10^{13.5} \Ms$ \citep[see also][for a related topic]{2014Old,2015Old}.

While a discussion of systematic bias and uncertainties of different halo mass measuring is beyond the scope of this paper, here we comment on the importance of matching the nominal host mass distributions when comparing the satellite quenched fractions across samples.

\begin{figure}
\centering
\includegraphics[width=0.48\textwidth]{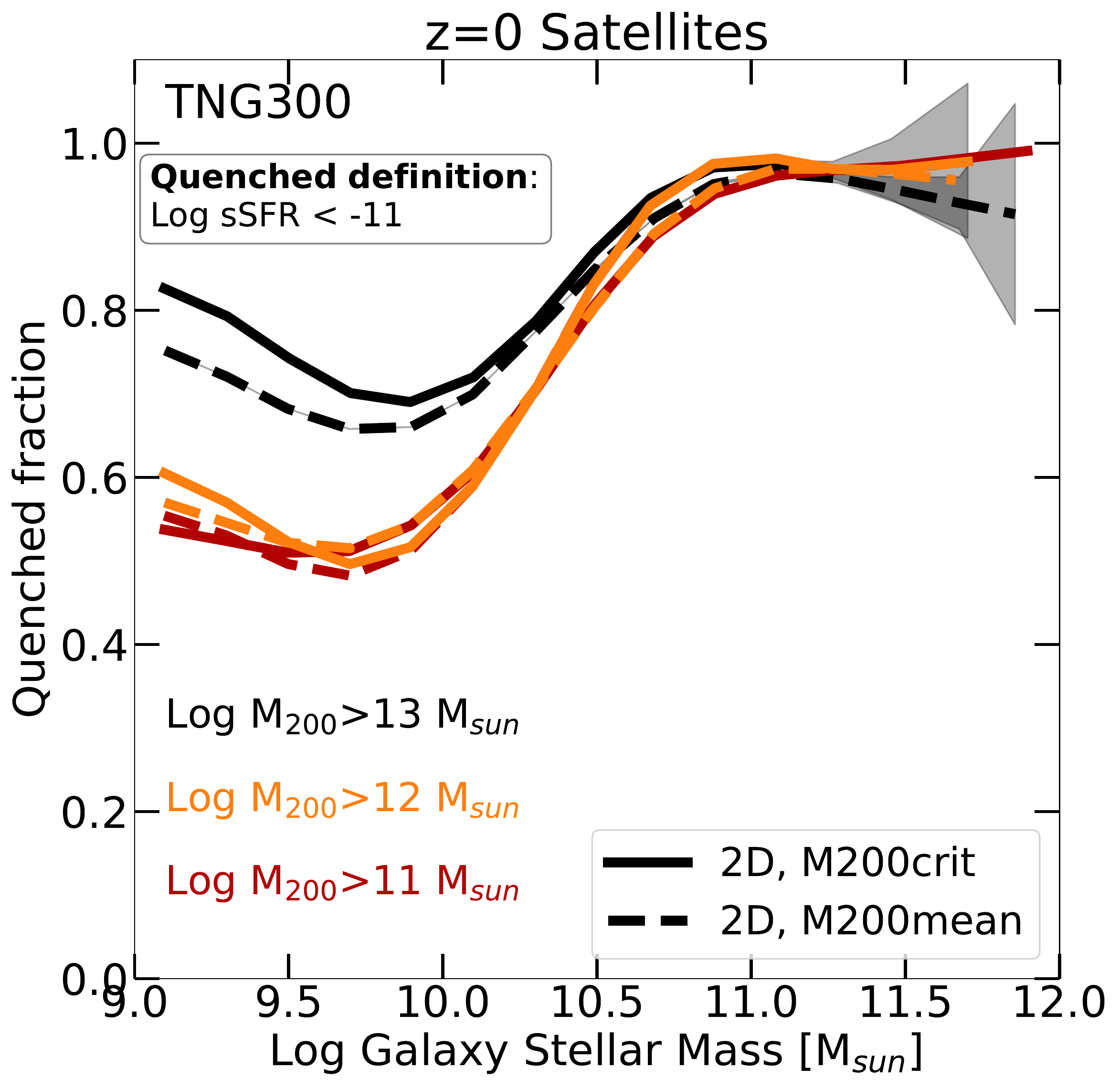}
\caption{\label{fig:hostmass} {\bf Host halo mass}. TNG300 quenched fraction of satellites at $z = 0$ for three host mass ranges where the minimum host mass reads $10^{13}, 10^{12},$ and $10^{11} ~\Ms$ for the black, orange, and red curves, respectively. In all cases, the upper limit of the host mass distributions is $10^{15.2}~\Ms$, the most massive host of the TNG300 volume. Solid and dashed curves denote M$_{200c}$ and M$_{200m}$ mass definitions, respectively. Not properly matching the lower host mass boundary implies systematic differences of up to 20-25 percentage points at the low-mass end.}
\end{figure}
Firstly, because the satellite quenched fractions are a strong function of host mass, it is essential to match at the very least the upper and lower boundaries of the sampled host mass distributions. Figure~\ref{fig:hostmass} shows the quenched fraction of satellites at $z=0$ for three host mass ranges with distributions inherited from the volume-limited nature of the TNG300 simulated box. The minimum host mass is $10^{13}, 10^{12},$ and $10^{11} ~\Ms$ for the black, orange, and red curves, respectively. In all cases the upper limit of the host mass distributions is $10^{15.2}~\Ms$, the most massive host of the volume. Solid and dashed curves denote M$_{200c}$ vs. M$_{200m}$ mass definitions, respectively, where the latter denote the mass enclosed in a sphere whose mean density is 200 times the \textit{mean} density of the Universe. It is evident that not properly matching the lower host mass boundary produces large systematic differences, up to 20 percentage points if the minimum host mass is e.g. $10^{13}$ vs. $10^{12}~\Ms$. 
\begin{figure*}
\centering
\includegraphics[width=0.48\textwidth]{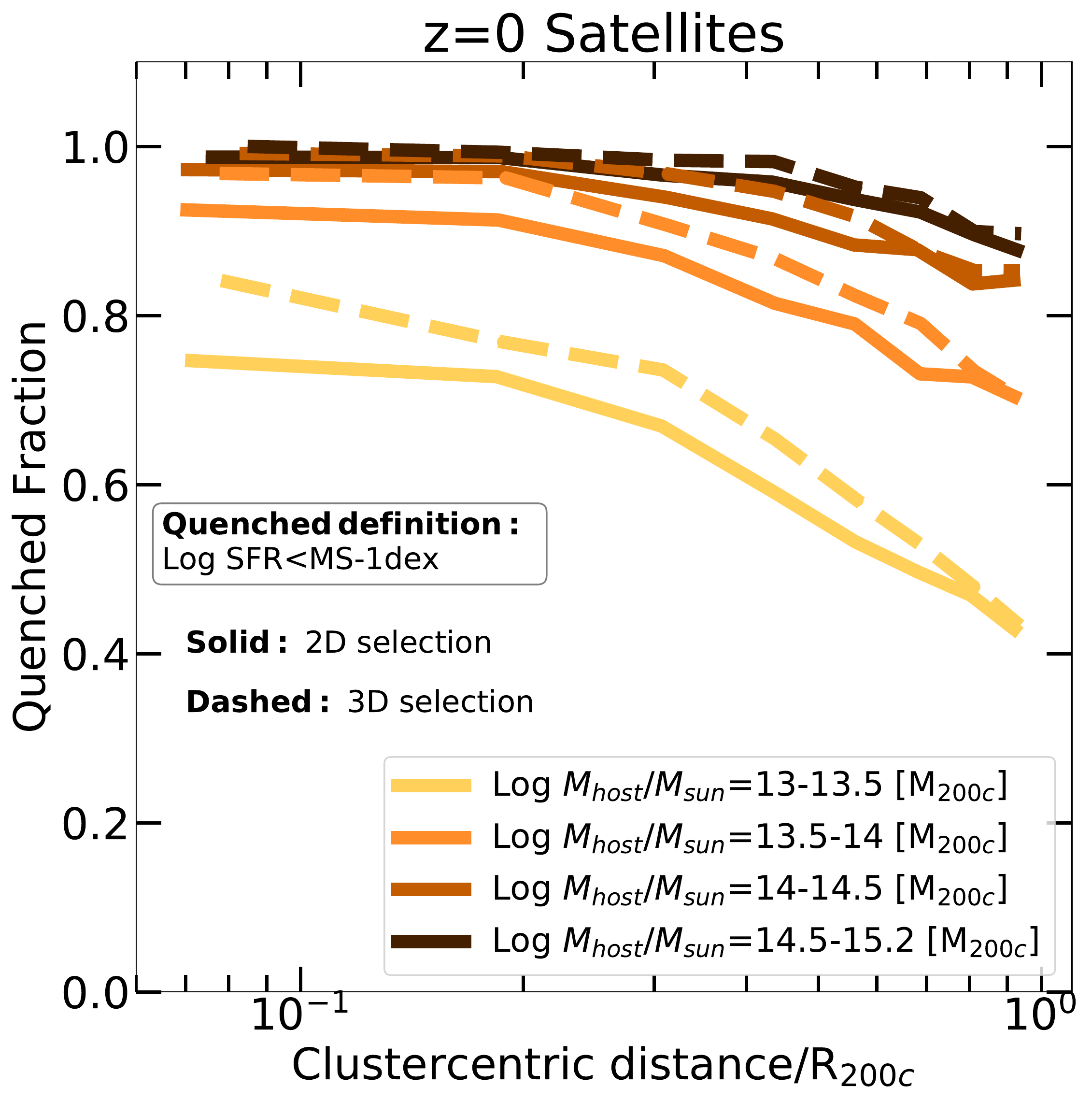}
\includegraphics[width=0.48\textwidth]{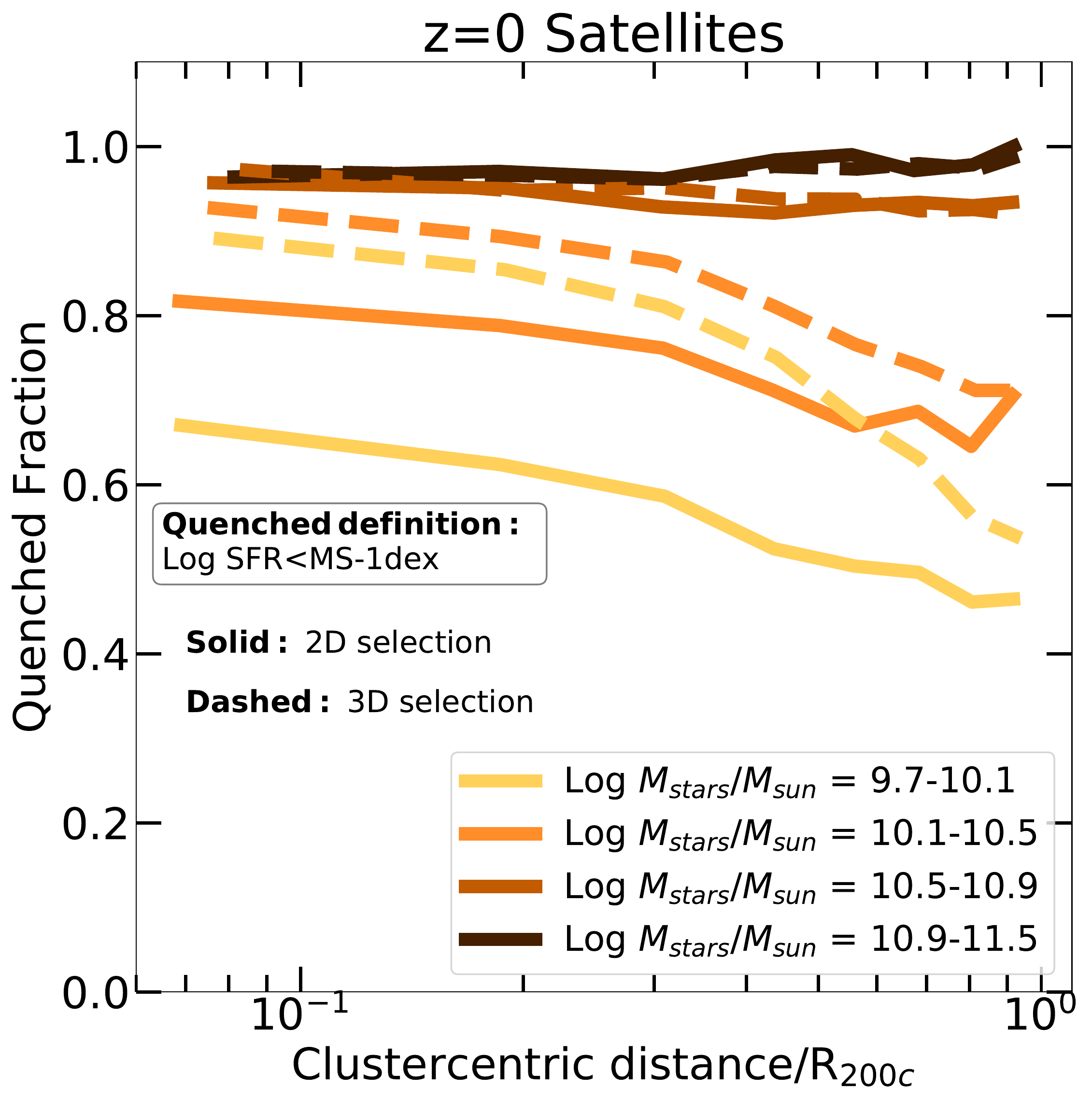}

\caption{\label{fig:2D3D} {\bf 3D versus 2D distances}. Radial profiles of satellite quenched fractions as a function of 3D membership identification and distances (dashed), versus 2D membership identification and projected distances (solid). In the former case, galaxies within a radius of $R_{200c}$ are classified as satellites, while in the latter case galaxies within a cylinder of this same radius and encompassing the entire FoF along the z-direction are labeled as satellites. We show two decompositions of the satellite+host population. First, including all satellites with $\MS = 10^{9-12}$, we split the sample into four host halo mass bins (left panel). Second, stacking all host halos with $M_{\rm host} = 10^{13-15.2}$ we split the satellites into four stellar mass bins (right panel). Quenched fractions are always lower for the 2D case because of contamination effects, by up to 20 percentage points, e.g. in the case of low-mass satellites (right panel, yellow curves).}
\end{figure*}
Secondly, whereas confusing the nominal definition of halo mass (e.g. M$_{200c}$ vs. M$_{200m}$) is a priori a problem, in comparison to what discussed so far it is a lesser issue, but still impacts the results by up to 10 percent. The situation is similar across redshifts, $z \le 2$. In the next sections, we further comment on the importance of matching the shape of the host mass function within the considered host mass range.

\subsubsection{3D vs. 2D projected distances for satellites}

After host halo mass, another important quantity for an environmental study of galaxy evolution is the halocentric distance: the distance between a satellite galaxy and the center of its host (or parent) dark matter halo. We can measure distances in two ways: in two-dimensional projection, on an imaginary sky plane perpendicular to the line of sight, and in three-dimensions. While the latter is the obvious and more common choice in simulations, it is fundamentally inaccessible in observations, which must either rely on pure sky-projected distance, or a combination of projected distance and line-of-sight velocity separation (i.e. an unknown mixture of line-of-sight peculiar velocity and line-of-sight physical distance). The same issues arise for group identification and membership assignment itself, i.e. finding host halos and associating satellites to them \citep{2007Yang,2017Lim}.

In Figure \ref{fig:2D3D} we show radial profiles of satellite quenched fractions, comparing these two choices: two-dimensional cylindrical satellite selection with radius $R_{200c}$ and encompassing the entire FoF along the z-direction and projected 2D distances (solid curves) versus three-dimensional spherical satellite selection within $R_{200c}$ and 3D distances (dashed curves). We slice the galaxy population in two different ways. In one case we include all satellites above a stellar mass threshold of $\MS > 10^{9}\Ms$, splitting the sample into four different host mass bins (left panel). Separately, we include all hosts above a halo mass threshold of $M_{\rm halo} > 10^{13}\Ms$, splitting the sample into four different satellite stellar mass bins (right panel).

In terms of global trends, for all satellite masses in all hosts, the fraction of quenched galaxies is higher closer to the host center (see also \citealt{2021Donnari}; Figure 4). This halocentric distance trend is stronger for lower mass hosts (yellow lines), because distant satellites have not yet saturated towards very high quenched fractions. The distance trends are also stronger for lower stellar mass satellites, and become negligible above $\MS \gtrsim 10^{10.5}~\Ms$ for the same saturation reason, i.e. because the quenched fraction of such massive galaxies in the field is already high due to supermassive black hole feedback based quenching.

The impact of two versus three dimensional sample selection and distance calculations is manifest in both panels. In fact, the differences in Figure \ref{fig:2D3D} probably underestimate the  effects, as in the 2D measurements along the z direction we here only account for galaxies within the FoF hosts, and not for any galaxy within a line-of-sight velocity range or photometric-redshift uncertainty that may correspond to projections along the line of sight of even many Mpc. Quenched fractions are always lower for the 2D case, which is due to two contamination effects. First, satellites at large physical distances (and thus lower quenched fractions) but small projected distances contaminate small radii bins and suppress the quenched fraction. Second, and similarly, field galaxies which are not physically associated with or affected by the host, due to large physical but small projected distances, likewise contaminate $\leq R_{\rm 200}$ bins with field systems, suppressing the inferred quenched fractions.

The quantitative change depends on host and satellite mass. For host halos less massive than $10^{14}~\Ms$ (left panel) the 2D based technique reduces the quenched fraction by $\sim\,$10 percent with respect to the 3D method, decreasing to zero towards higher host mass. For satellites with stellar masses less than $10^{10.5}~\Ms$, 
the quenched fractions with the 2D technique are lower by \mbox{$\sim\,20 - 30$}\,percentage points when compared to the 3D method. This difference increases towards zero distance -- near the center of the parent halo. We note (but do not show) that such differences are less clear if we consider quenched fractions versus satellite stellar mass, stacking all hosts together, likely because the statistics are dominated by large distance orbits where the effect is always smaller.

\begin{figure*}
\centering
\includegraphics[width=0.45\textwidth]{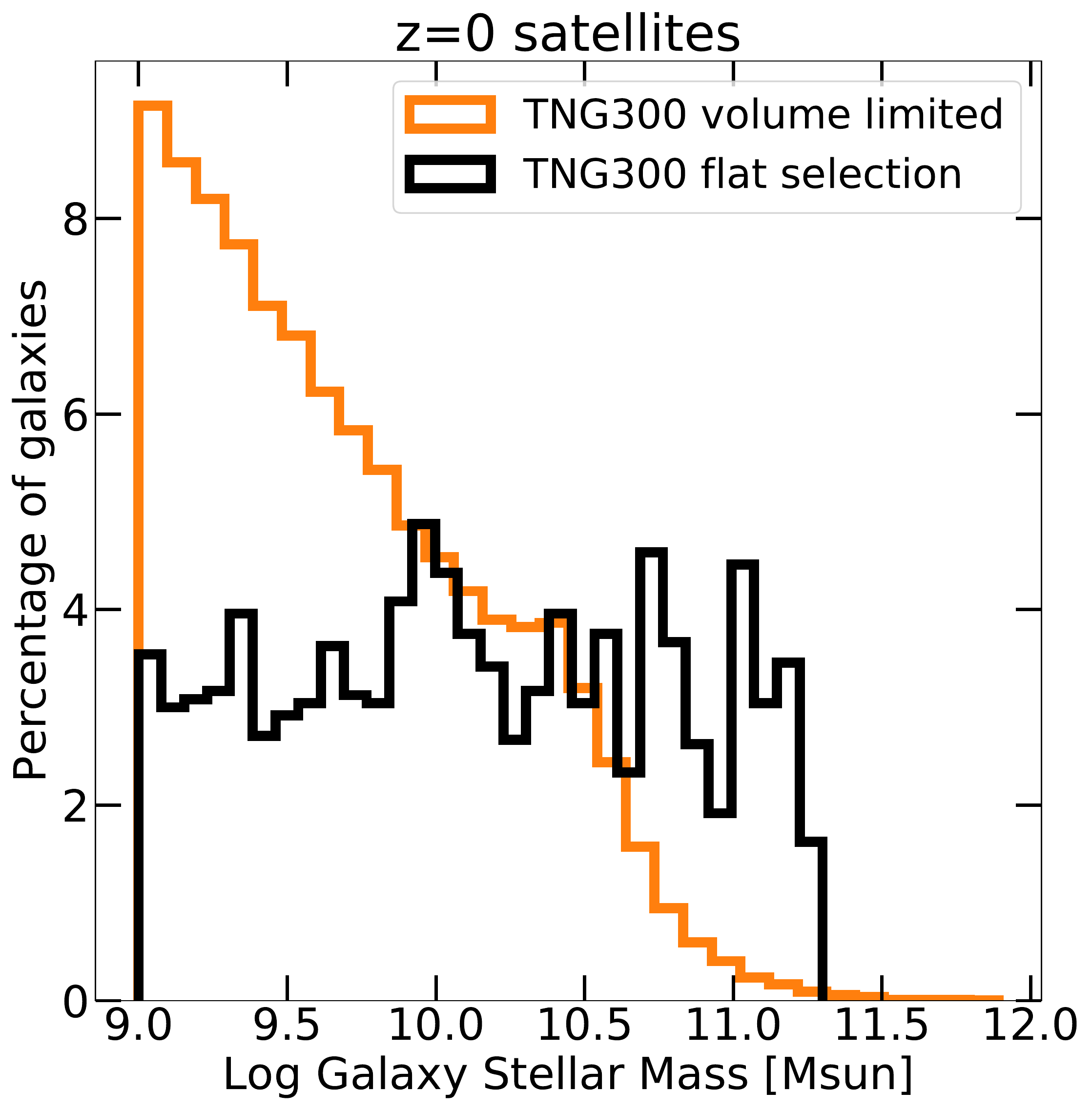}
\includegraphics[width=0.48\textwidth]{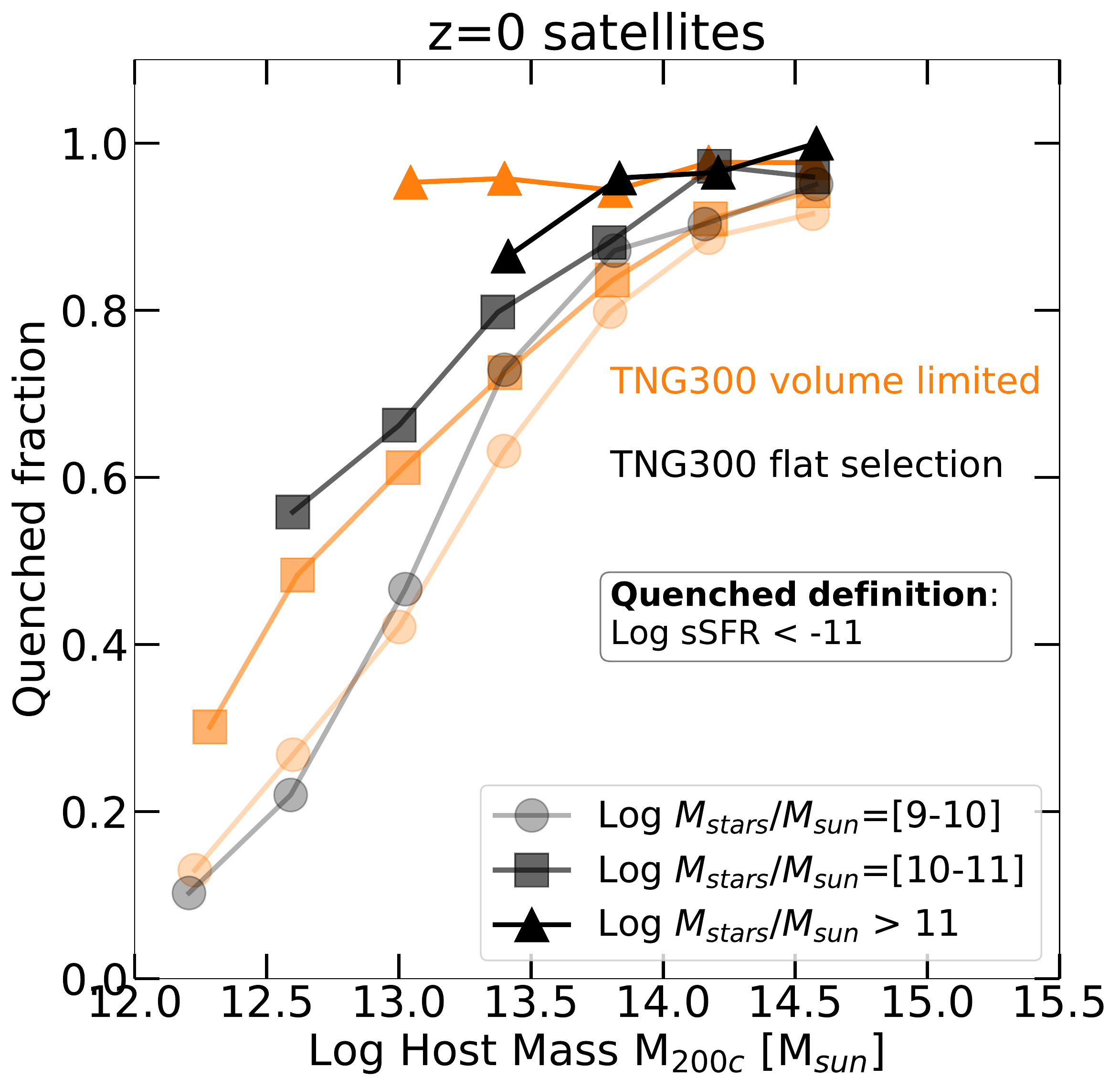}
\includegraphics[width=0.46\textwidth]{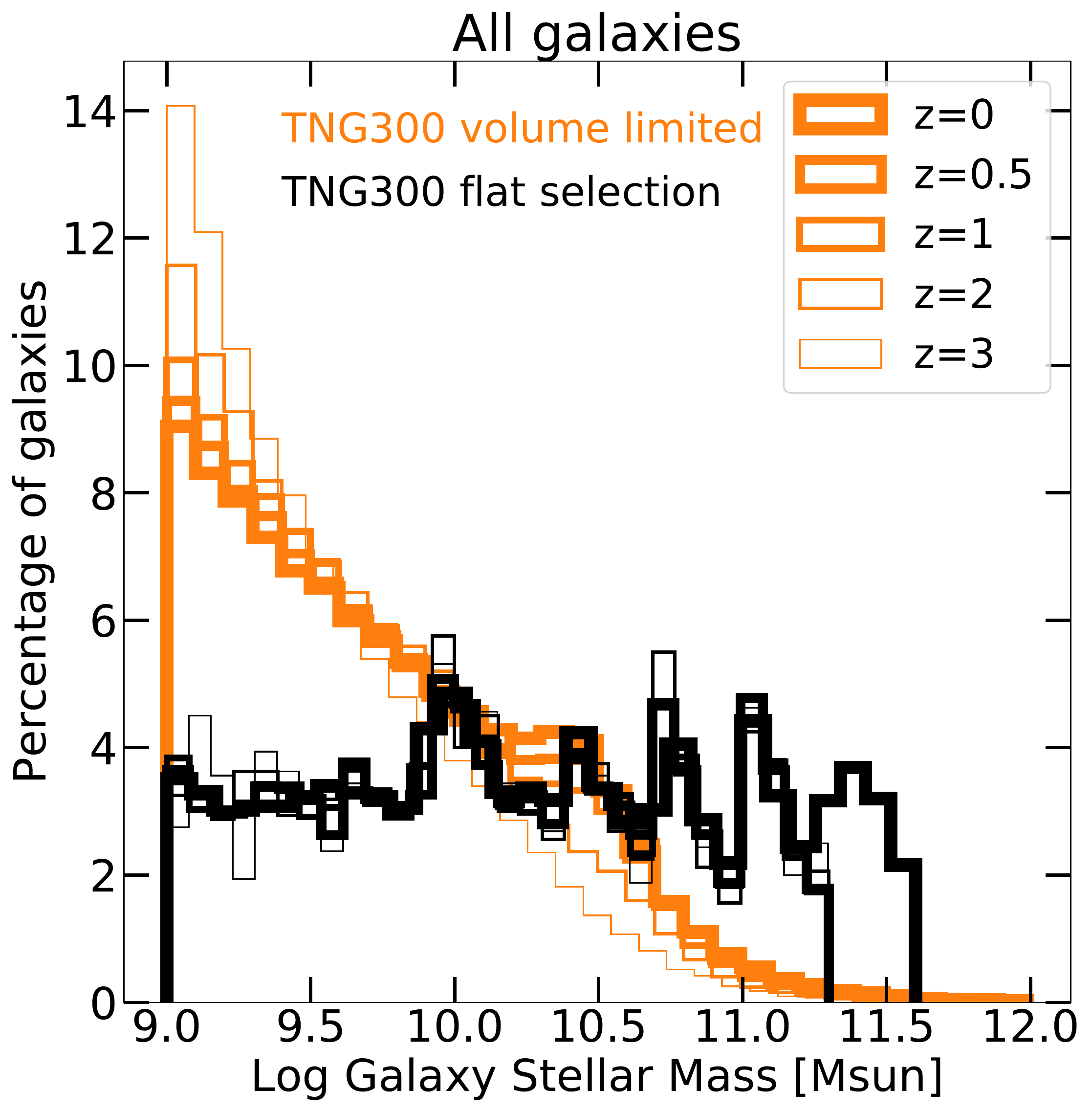}
\includegraphics[width=0.48\textwidth]{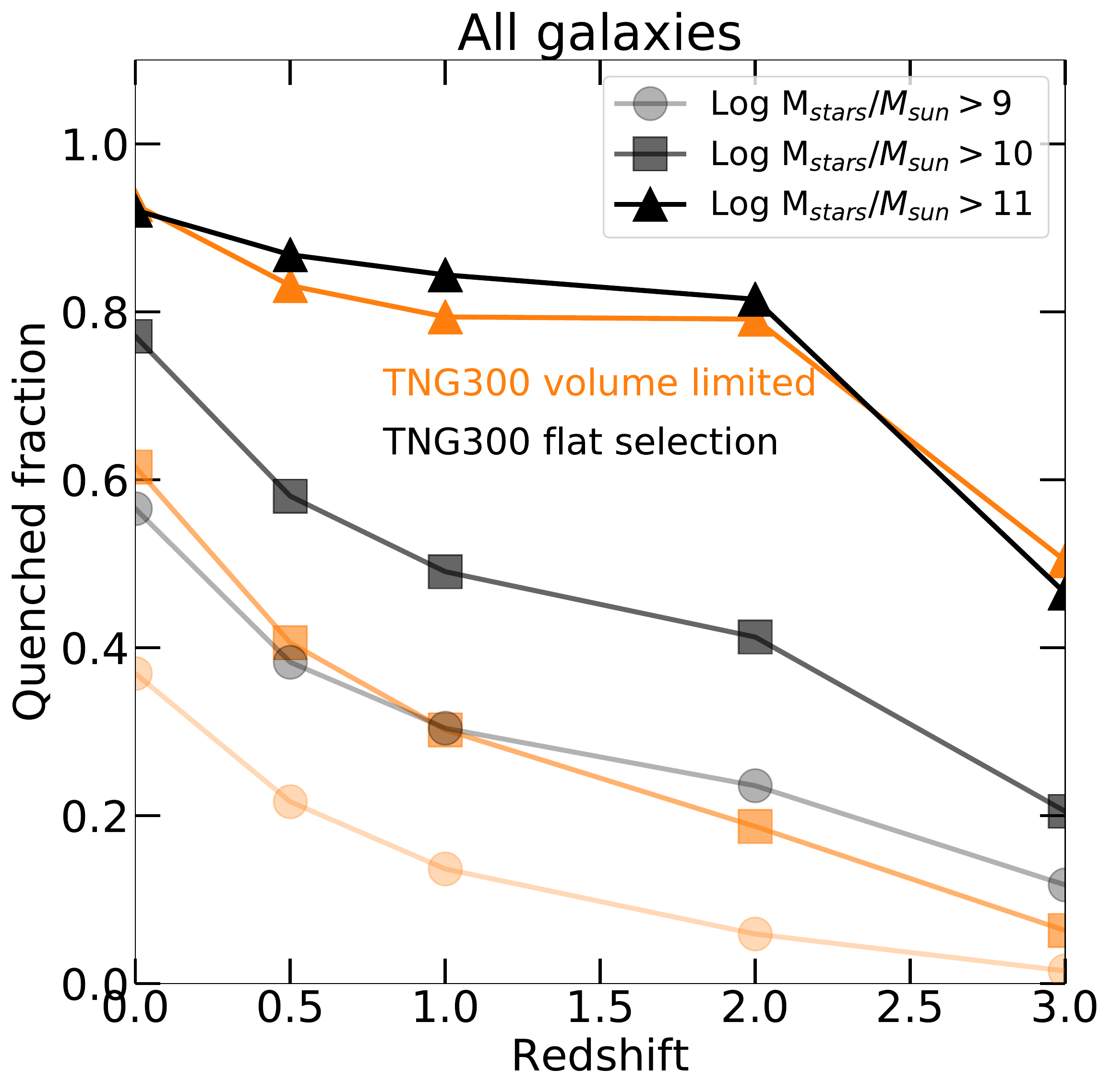}
\caption{\label{fig:MstarsDistr} {\bf Different sample mass distribution}. We construct two fundamentally different galaxy distributions, both out of the galaxies simulated within the TNG300 volume: a volume-limited sample (orange), and a flat distribution (black). First, we consider satellite galaxies residing in all host halos above $10^{12}\,\Ms$ (top row), showing the satellite mass distributions themselves (top left) and the resulting quenched fractions as a function of host mass (top right), for three different satellite galaxy stellar mass bins. In this case, the impact on inferred quenched fractions is small: the flat distribution produces up to $\sim\,$10\,percent higher quenched fractions, particularly for intermediate mass hosts. Second, we consider all galaxies with $M_\star > 10^9\,\Ms$ together with their redshift evolution, showing the resulting mass distributions (lower left) and quenched fractions as a function of redshift (lower right) for three different minimum stellar mass thresholds. Here the two synthesized galaxy samples have substantially different quenched fractions. The flat sample, as is more common observationally, produces quenched fractions $\sim\,20$\,percent higher than the volume-limited sample, as would be intrinsically drawn from a cosmological simulation volume, for all most massive galaxy mass bins.
}
\end{figure*}

\subsubsection{Misclassification of centrals and satellites}
\label{sec:misclassification}

Because group finding in observational galaxy surveys is inexact, regardless of using spectroscopic or photometric redshifts, some level of misclassification is always present. Namely, some central galaxies are inevitably misidentified as satellites of higher mass haloes, while at the same time some fraction of satellites are misidentified as isolated centrals \citep[see e.g.][]{2015Campbell,2020Tinker}. This leads to a contamination effect for any measured quenched fraction, and it has only recently become appreciated that the resulting bias can be important for comparisons \citep[in the GAMA survey:][]{2020Bravo}.

Whereas we do not attempt to replicate the typical observational procedures for galaxy group finding in the projected sky by adopting the simulation data, we quantify for this effect by randomly reassigning labels from `central' to `satellite', and vice versa, with given mixture fractions. That is, we take some percent of centrals $f_{\rm cen}$, at random, and relabel them as satellites. Simultaneously, we take some percent $f_{\rm sat}$ of satellites, at random, and relabel them as centrals. We denote this transformation by [cen,sat] = [$f_{\rm cen}$,$f_{\rm sat}$]. For a given observational group finder, these fractions can be estimated by careful mocks on synthetic catalogs.

Our fiducial choice is based on our comparisons below to SDSS catalog data as described in \cite{2012Wetzel,2013Wetzel}. Namely, averaging across all halo masses, the purity and completeness of satellites are both $\sim\,$80\,percent, while the purity and completeness of centrals is slightly $\gtrsim 80\,$percent \citep[][Appendix Figure C1]{2011Tinker}. This implies that the average fraction of galaxies that are classified as centrals when they are in fact satellites, and vice-versa, is about one in ten. For our purposes, this translates into [cen,sat] = [10,20], such that we relabel 10 percent of centrals to satellites, and 20 percent of satellites to centrals. We neglect the additional complexity of the host halo mass dependence of these contamination fractions.
A similar assessment of the satellite/central misclassification is also shown in terms of averaged SFRs by \textcolor{blue}{Stevens et al. (2020)}.

Even if we do not show it here, we have quantified the impact of this reassignment. At the high-mass end ($\MS>10^{10.5}\Ms$) of both centrals and satellites, the misclassification does not imply any significant difference in terms of quenched fractions. Conversely, at the low-mass end ($\MS<10^{10}\Ms$), we find a higher (lower) quenched fractions for centrals (satellites) of about 10 percentage points.
We show the impact of this reassignment in concert with other analysis choices below in Section \ref{sec:mock} when we quantitatively compare to SDSS data.

\subsubsection{Matching the stellar mass distribution}

The fraction of quenched galaxies is a strong function of mass. As a result, subtleties in the mass distributions of both host (halo mass) and galaxy (stellar mass) can influence comparisons. Here, instead of matching to a given observational mass distribution, we perform a more illustrative experiment. Specifically, we generate two very differently shaped mass distributions from TNG300: a volume-limited sample, as naturally emerges from any cosmological volume simulation with no additional cuts, and a `flat' sample, with uniform statistics in bins of logarithmic mass. This can occur in practice in a survey where a magnitude limit is convolved with increasing volume towards higher redshift, such that progressively brighter galaxies are included more frequently and in turn make the sample mass distribution more top heavy than in reality.

Figure \ref{fig:MstarsDistr} shows these two mass distributions, constructed for $z=0$ satellites in TNG300: the nominal volume-limited case (orange), where the number of galaxies declines rapidly towards larger masses, and an artificially flat distribution (black). For the latter, we randomly sub-select satellites residing in hosts with $M_{\rm halo} > 10^{12} \Ms$ to obtain an equal number within fixed 0.2 dex bins.

The corresponding quenched fractions as a function of host mass $M_{200c}$ are also shown in Figure \ref{fig:MstarsDistr} (upper right panel), where we focus on $z=0$. Satellites are stacked in three stellar mass bins bins: $\MS=10^{9-10} \Ms$ (circles), $\MS=10^{10-11} \Ms$ (squares) and $\MS>10^{11} \Ms$ (triangles). We define quenched systems using the log(sSFR) $< -11\,\rm{yr}^{-1}$ definition. Although the two galaxy mass distributions are quite different, we see that the resulting quenched fractions differ by at most $\sim\,$10\,percent at any host mass, where the flat distribution returns higher values.

A much stronger impact is found when we consider all galaxies, both centrals and satellites, as a function of redshift. In Figure \ref{fig:MstarsDistr} (lower right panel) we show the median quenched fractions for three lower limits of galaxy stellar mass: $\MS > 10^{\{9,10,11\}} \,\Ms$ (circles, squares, and triangles). For the two lower stellar mass thresholds, at all redshifts, the flat galaxy mass distribution results in quenched fractions that are $\sim\,20$\,percentage points higher. Such open-ended mass selections are often considered observationally, but these act very differently in practice between the two samples. Bins in the volume-limited selection are dominated by galaxies near the lower edge of the mass cut, while for the flat selection bins are dominated by much higher mass galaxies, depending on the volume.

Comparisons of quenched fractions between simulations and observational surveys (and in general among any galaxy sample) must account for the impact of the differing shape of the sampled mass distributions.


\begin{table*}
\renewcommand{\arraystretch}{1.15}
\centering
\begin{tabular}{c|c|c|c|c|c|c|c}
\hline
Observations/ & Reference & Survey/ & Redshift & Log $\MS$ &  Log $M_{\rm host}$  & SFR tracer & Quenched \\
Models & & simulation & & range ($\Ms$)& range ($\Ms$)& & criteria \\
\hline
\hline
Observations & \cite{2011Mcgee} & SDSS  & 0.08 & 9.4-11.2 & - & SED (UV) & Log sSFR = -11 \\
 ($z\simeq 0.1$) & \cite{2019Schaefer} & SAMI  & $<0.11$ & 8.2-11.5 & $>$12.5 &  H$\alpha$ & Log sSFR = -12 \\
& \cite{2019Davies} & GAMA  & 0.1-0.2 & 9-11.4 & 11.4-15&  H$\alpha$ & Log sSFR = -10.5 \\
 & \cite{2013Wetzel} & SDSS & 0.1 & 9.5-11.5 & 12-15 &  H$\alpha$/D$_n$4000 & Log sSFR = -11 \\
\hline
Observations  & \cite{2014Lin} & Pan-STARRS1 & 0.2-0.5 & 9.25-11.35 & 12.5-14 & SED &Log sSFR = -11\\
 ($z\simeq 0.35$)  & \cite{2018Jian} & HSC  & 0.2-0.5 & 9.1-11.9 & $>$13.6 &  SED (UV-opt) & Log sSFR = -10.1  \\
 & \cite{2017Wagner} & CLASH & 0.15-0.4 & 10.1-11.4 & 14.8-15.7  &   SED & D$_n$(4000) \\
\hline
Observations   & \cite{2017Fossati} & 3D HST/CANDELS  & 0.5-0.8 & 9.6-11.4 & $>$13 &  - & UVJ  \\
($z\simeq 0.65$)   & \cite{2018Jian} & HSC  & 0.5-0.8 &9.1-11.9 & $>$13.6 & SED (UV-opt) & Log sSFR = -10.1 \\
& \cite{2014Lin} & Pan-STARRS1 & 0.5-0.8 &9.25-11.35 & 12.5-14 & SED &Log sSFR = -11 \\
 & \cite{2017Wagner} & CLASH & 0.4-0.8 & 10.1-11.4 & 14.8-15.7 &  SED & D$_n$(4000)\\
\hline
\hline
Models & \cite{2015Schaye} & EAGLE & 0 & 9-12 & 14-14.5  & - & Log sSFR = -11 \\
       & \cite{2017Bahe} & Hydrangea & 0-2 & 9.25-11.35 &  14-14.5 & - & Log sSFR = -11 \\
       & \cite{2019Tremmel} & RomulusC/25 & 0 & 8-11.2 & $>$14 & -  & Log $\Delta$MS $<$ 1\,dex \\
       & \cite{2020Ayromlou} & L-Galaxies (TNG) & 0-2 & 9-12 & 14-14.5 & - & Log sSFR = -11 \\
 \hline
\end{tabular}
\caption{\label{tab:observations} Sumary of the observational data used for comparisons in Figures \ref{fig:Q_frac_obs} and \ref{fig:models}. Column 1: Type of data, observations or models. Column 2: References. Column 3: Survey. Column 4: Redshift range. Column 5: Stellar mass range. Column 6: host mass range. Column 7: SFR tracers. Column 8: threshold used to divide star-forming from quenched galaxies. In all results considered here, a Chabrier IMF \protect\citet{2003Chabrier} is adopted but for \protect\cite{2014Lin} which use a Salpeter IMF \protect\citet{1955Salpeter}.}
\end{table*}

\section{Comparison to observations and other theoretical models}
\label{observations}

\subsection{TNG quenched fractions versus observations}

In this section we compare the TNG model to several observational datasets which measure the quenched fractions of central and satellite galaxies separately. These observations are based upon a diversity of data with different characteristics which we enumerate in Table \ref{tab:observations} -- redshift range, stellar and host mass range, star-formation rate tracer, criteria to separate star-forming versus quenched galaxies, and adopted initial stellar mass function. Comparison requires adopting appropriate analysis choices particular to each dataset.


\begin{figure*}
\centering
\includegraphics[width=0.47\textwidth]{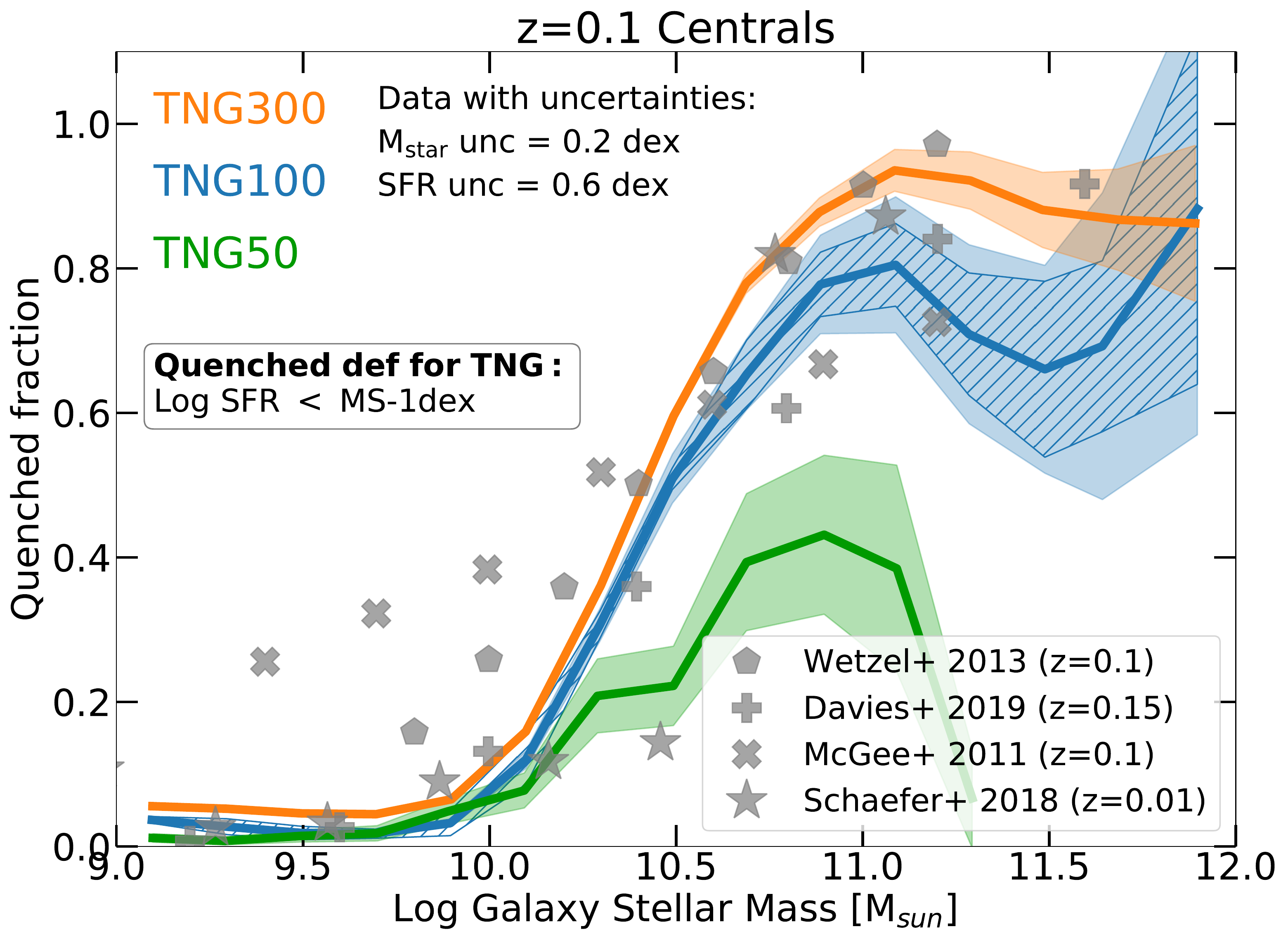}
\includegraphics[width=0.47\textwidth]{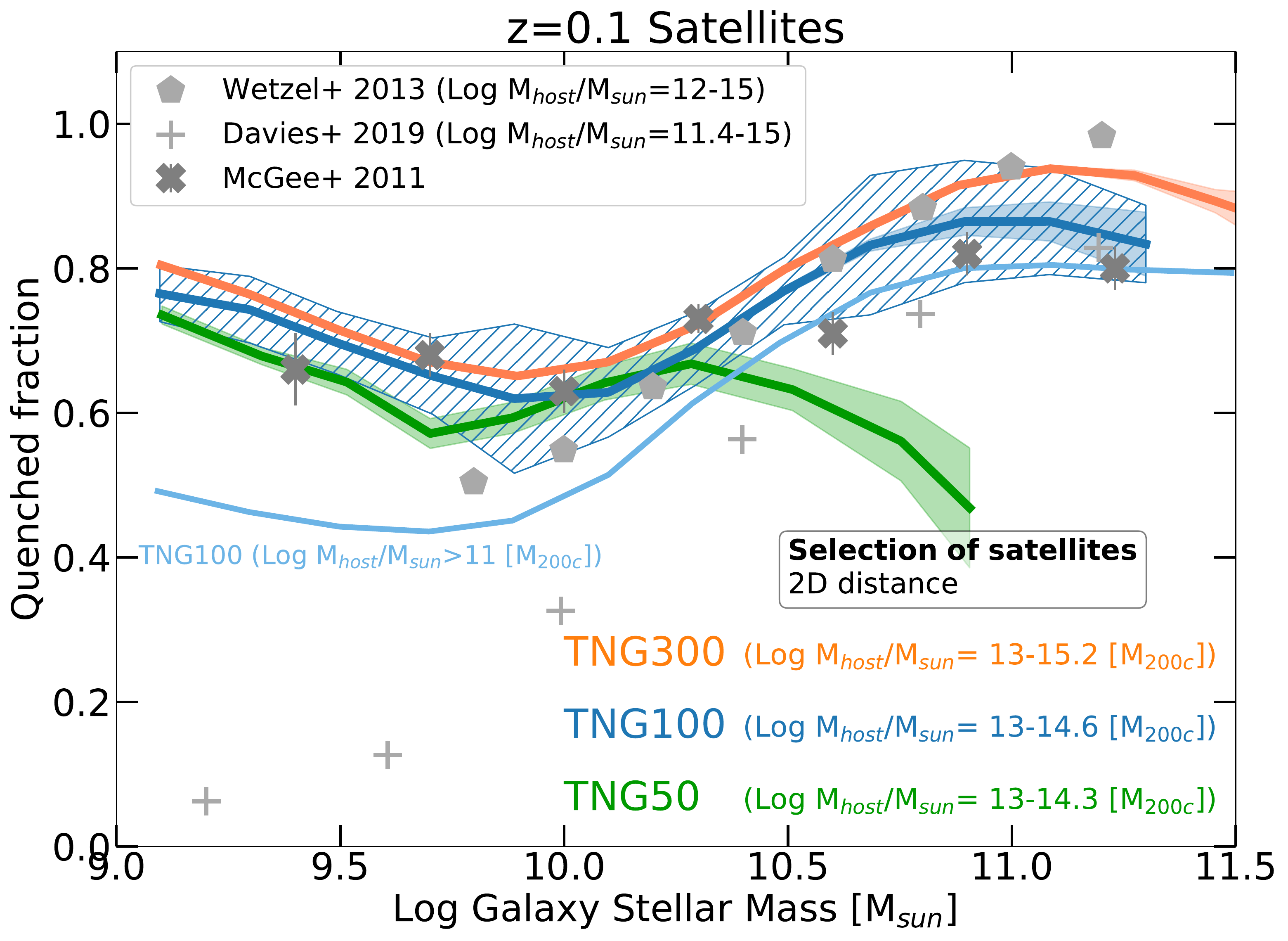}
\includegraphics[width=0.47\textwidth]{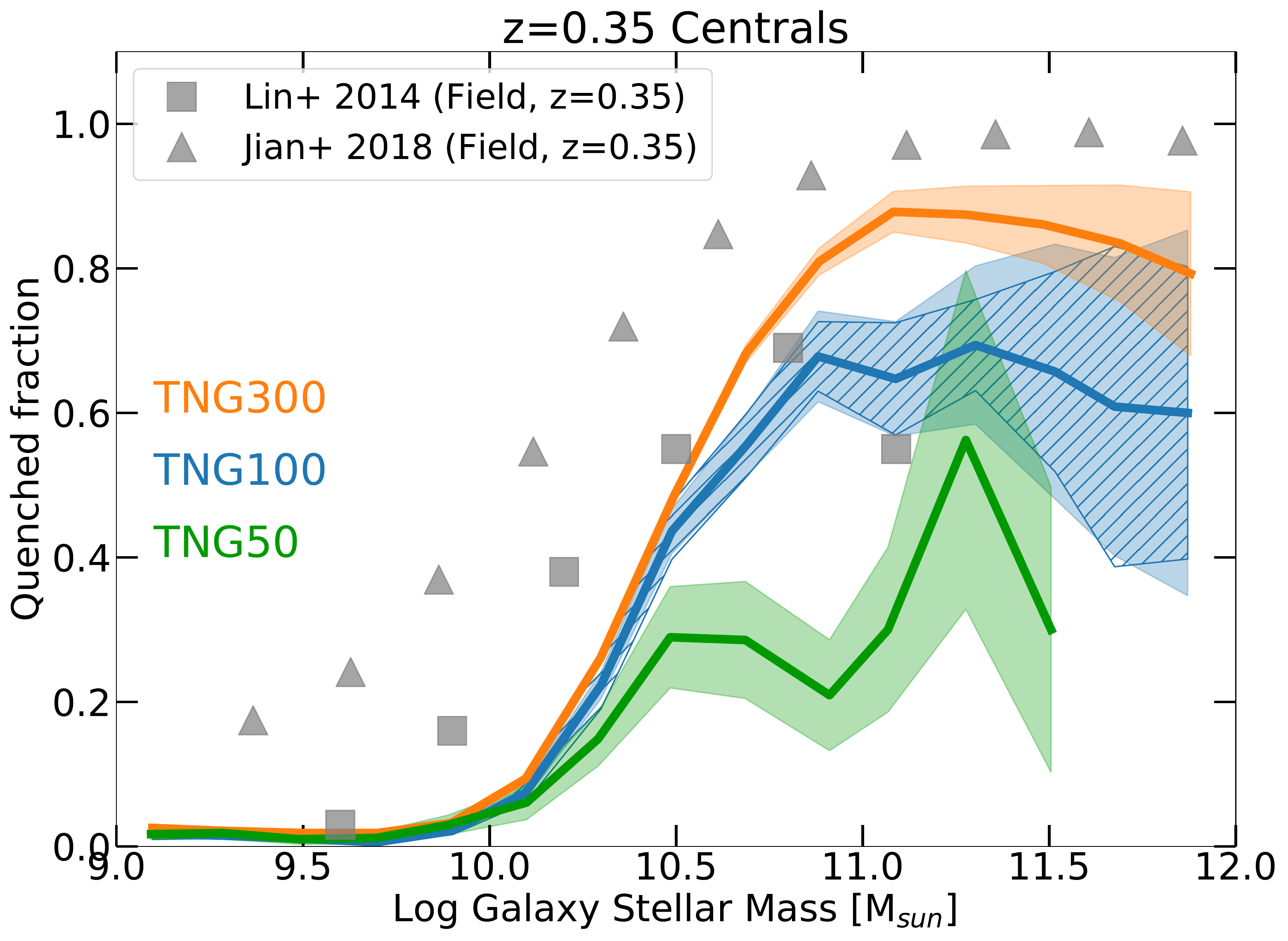}
\includegraphics[width=0.47\textwidth]{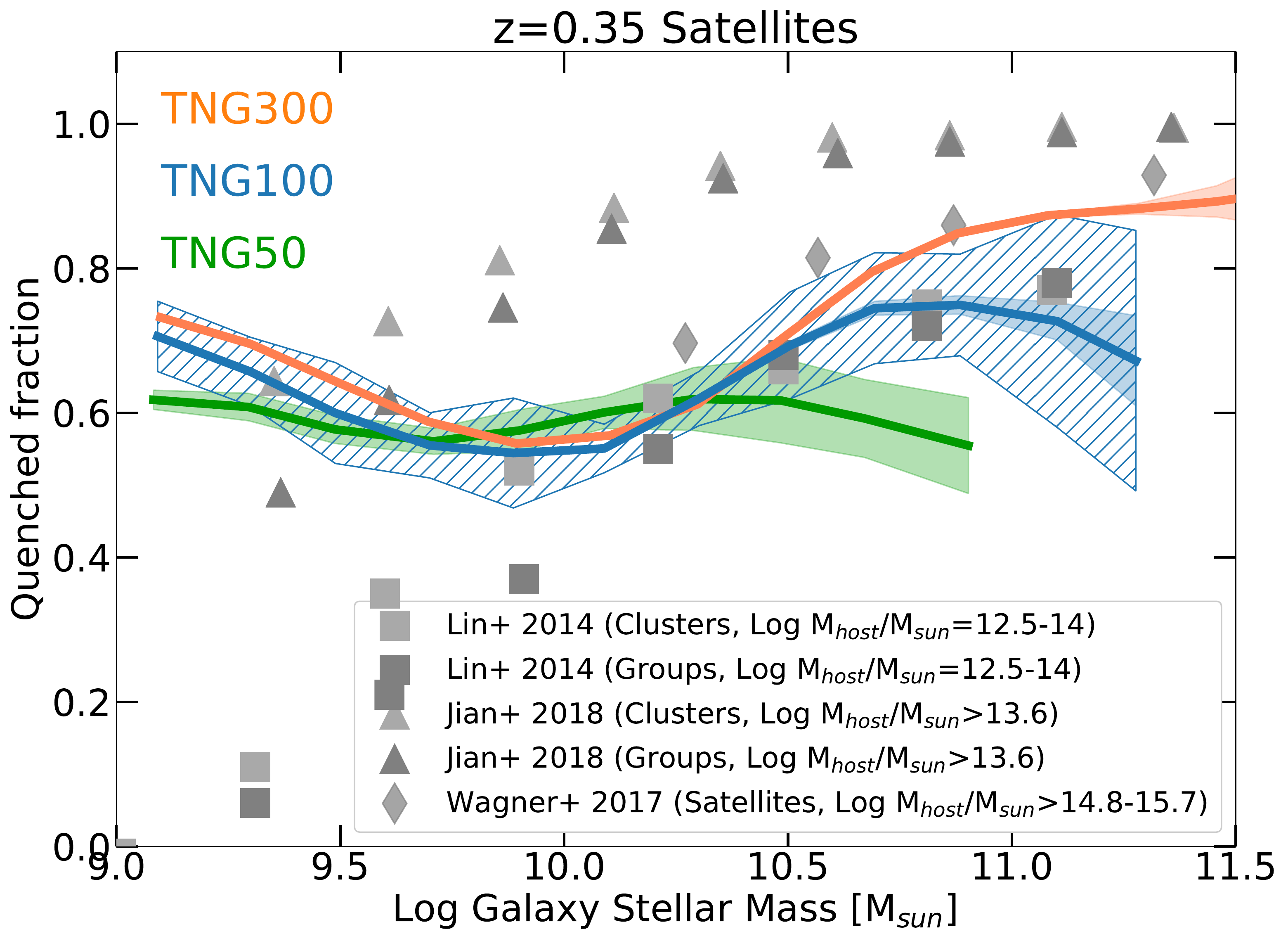}
\includegraphics[width=0.47\textwidth]{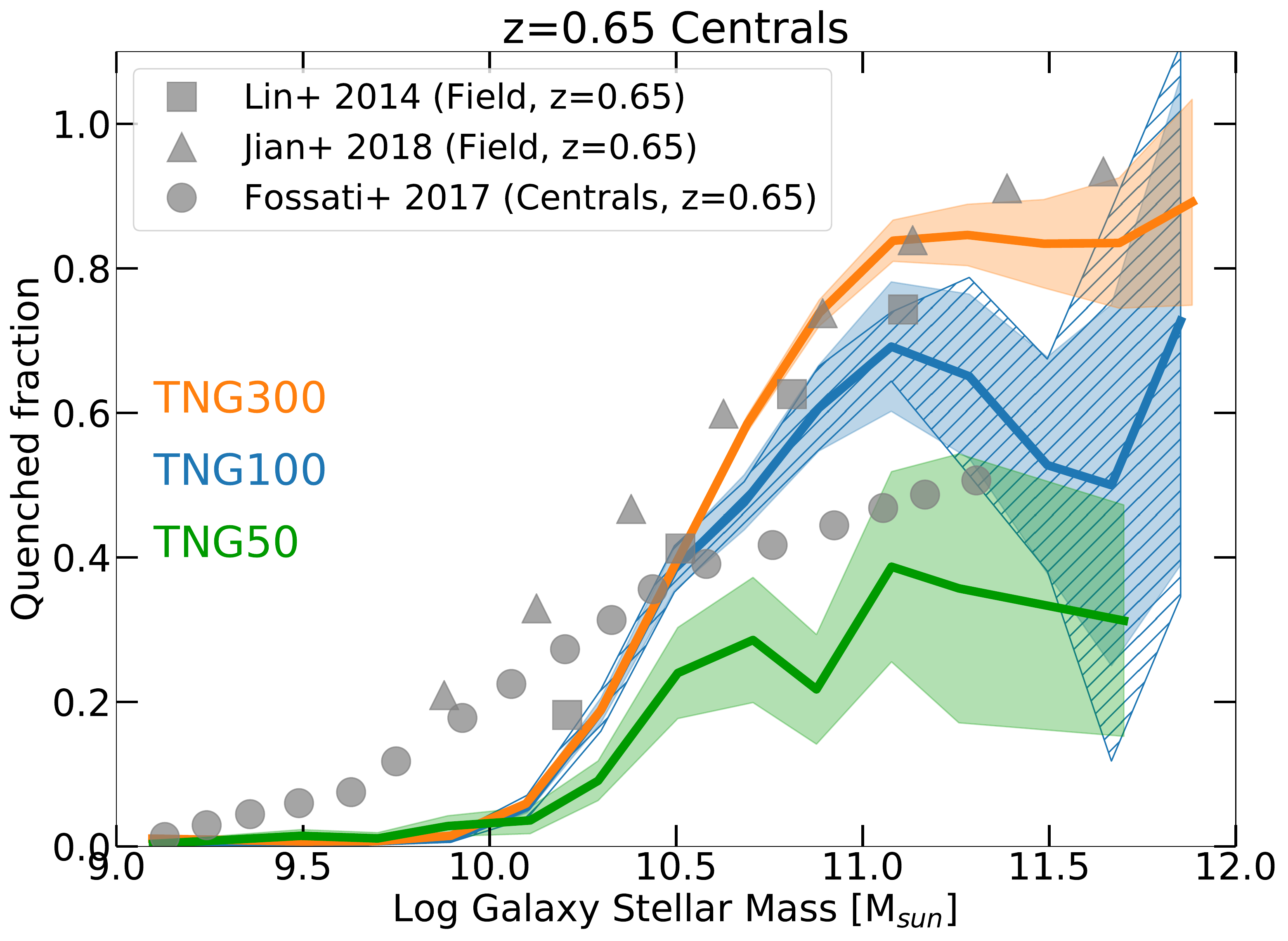}
\includegraphics[width=0.47\textwidth]{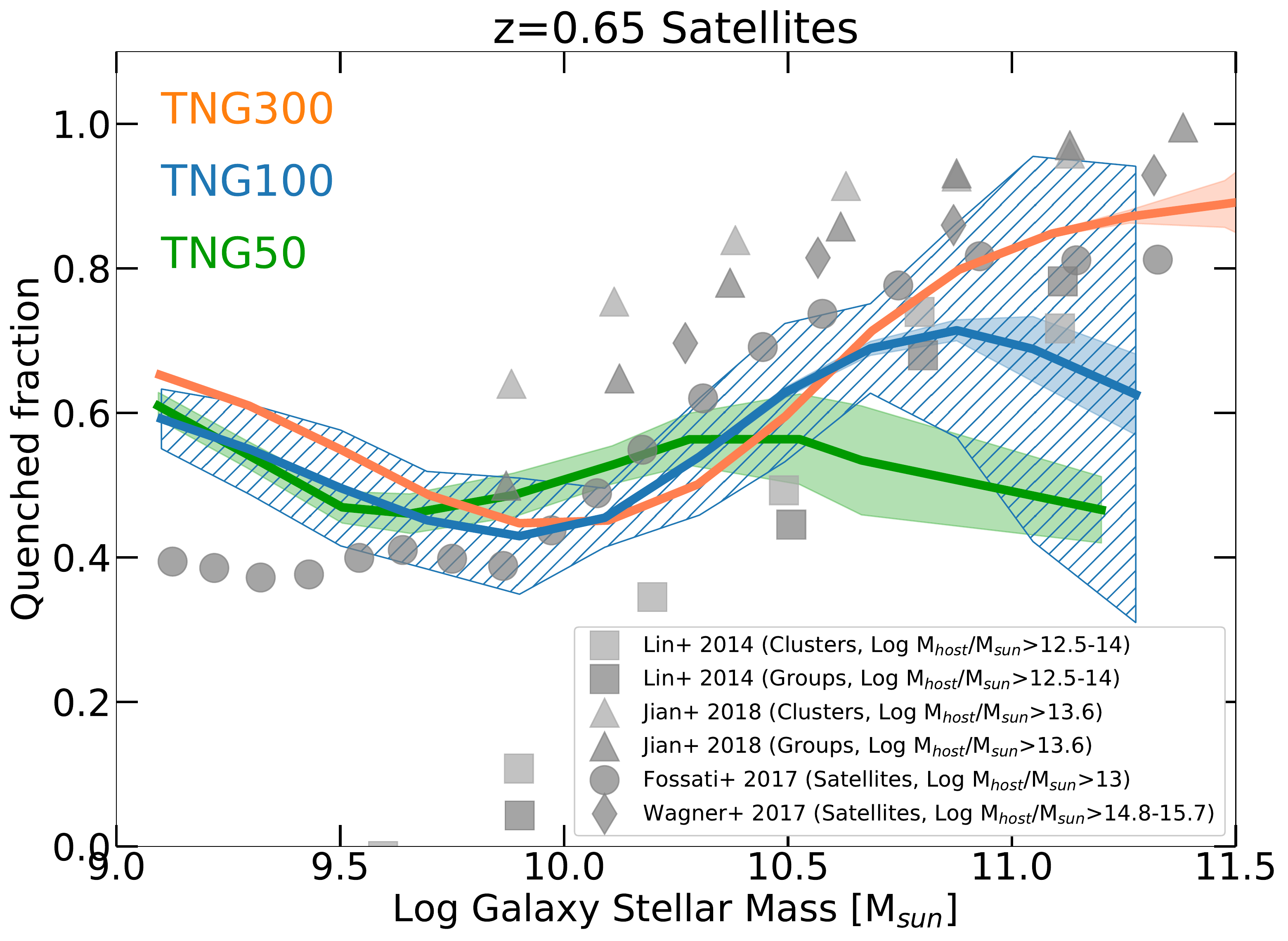}
\caption{\label{fig:Q_frac_obs} {\bf Nominal comparison to observations at low redshift}. Quenched fractions in TNG100 (blue), TNG300 (orange) and TNG50 (green) for central galaxies (left column) and satellites (right column) at $z=0.1$ (top), $z=0.35$ (middle) and $z=0.65$ (bottom). Shaded areas represent the Poissonian errors while the striped areas around TNG100 represent the cosmic variance uncertainties.
We roughly mock the satellite selections of the diverse datasets (see text) which are shown with different grey symbols and indicated in the legend: \protect\cite{2013Wetzel} (SDSS, pentagons) and \protect\cite{2011Mcgee} (SDSS/GEEC, crosses), \protect\cite{2019Davies} (GAMA, plus), \protect\cite{2019Schaefer} (SAMI, stars), \protect\cite{2014Lin} (Pan-STARRs, squares), \protect\cite{2018Jian} (Subaru HSC, triangles), \protect\cite{2017Fossati} (3HST/CANDELS, circles), \protect\cite{2017Wagner} (CLASH, diamonds). Overall, the TNG quenched fractions are qualitatively (and often, quantitatively) in good agreement with observations as a whole, although discrepancies are visible in certain regimes.}
\end{figure*}

We first show a number of ``at-face-value'' comparisons, adopting analysis choices common to all the selected datasets. Figure \ref{fig:Q_frac_obs} presents the quenched fractions of central galaxies (left column) and satellites (right column), for three redshifts: $z=0.1$ (top row), $z=0.35$ (middle row) and $z=0.65$ (bottom row). The observational surveys are denoted with different grey symbols as indicated in the legends: \cite{2013Wetzel} (SDSS, pentagons) and \cite{2011Mcgee} (SDSS/GEEC, crosses), \cite{2019Davies} (GAMA, plus), \cite{2019Schaefer} (SAMI, stars), \cite{2014Lin} (Pan-STARRs, squares), \cite{2018Jian} (Subaru HSC, triangles), \cite{2017Fossati} (3HST/CANDELS, circles), \cite{2017Wagner} (CLASH, diamonds). For clarity, we exclude observational error bars.

We adjust the observational data points only to transform them from the Salpeter (s) to Chabrier (c) IMF where required, adopting \mbox{log($M_\star^{\rm c}$) = log($M_\star^{\rm s}$) - 0.24} and \mbox{log($\rm{SFR}^{\rm c}$) = log($\rm{SFR}^{\rm s}$) - 0.15}. We do not attempt to mock or reproduce the operational definitions of SFRs and stellar mass estimates, nor the apertures within which they are measured: all these differ across the observational analyses.

For comparison, we select satellites from the TNG100 (blue curves), TNG300 (orange curves) and TNG50 (green curves) simulations residing within hosts more massive than $10^{13} \, \Ms$ at each redshift. Only for TNG100 satellites at $z=0.1$ we additionally show the quenched fractions taking the minimum host mass to be $10^{11} \Ms$ (thin blue curve). For TNG, SFRs and stellar masses are taken within twice the stellar half mass radius, and the SFRs are instantaneous (see Section~\ref{sec:SFRs}).
In all panels and for all the simulations, shaded areas represent the Poissonian error. Furthermore, to account for the limited simulation volumes, we give estimates of the cosmic-variance uncertainties by measuring the quenched fractions in eight sub-boxes of $\sim$50 Mpc on a side for TNG100 and by computing the \textit{jackknife} error represented by the striped blue areas in all panels. We note that the smallest volume simulation, namely TNG50, must be affected by at least such an amount of uncertainties due to sample/cosmic variance.

For group membership we require satellites to fall within a cylinder of radius $R_{200}$ and depth encompassing the entire FoF along the z-direction (random) of the simulation. We account for systematic uncertainties by adding random Gaussian errors to the simulated data ($\sigma = 0.2$ dex for $M_\star$, and $\sigma = 0.6$ dex for SFR, as before). We define quenched galaxies as having log(SFR) one dex or greater below the MS, which is not identical to all the observational definitions employed. However, as demonstrated in Figure~\ref{fig:definitions}, at low redshift the definition based on the distance from the recursive MS gives very similar results to that with a fixed SFR cut, which is the choice adopted by most observational works. We also note that the adopted host mass range does not exactly match all the observations (see Table \ref{tab:observations}).

Overall, the TNG model returns qualitatively and, in some cases, quantitatively good agreement with the observations as a whole, at both low and high stellar masses, for centrals and satellites, and at all redshifts available.
Particularly, for massive centrals at any redshift,  TNG50 returns a lower quenched fractions with respect to TNG100 and TNG300 and to observations, likely due to a combination of a low statistics and mass resolution (Appendix~\ref{appendix_B}).

In more detail, the best agreement is found at low redshift $z=0.1$ and for low-mass centrals ($\lesssim 10^{10} \, \Ms$) where the TNG model is remarkably consistent with the SAMI and GAMA data \citep{2019Davies,2019Schaefer}. Both simulations and observations shown quenched fractions of less than 5 percent in this regime. Similar agreement is also present at the high mass end ($\MS \gtrsim 10^{11} \, \Ms$), where the quenched fractions of TNG galaxies are consistent with both SDSS \citep{2013Wetzel} and SAMI surveys. The simulations exhibit reasonable agreement with \cite{2019Davies} and \cite{2011Mcgee}, which have a larger scatter than the other samples. For massive satellites ($\gtrsim 10^{10} \, \Ms$, right panel) we see good quantitative agreement with \cite{2013Wetzel} and \cite{2011Mcgee}. We note that the mismatch for low-mass satellites is, at least in part, apparent and due to different host mass selections (see findings of Figure~\ref{fig:hostmass}), as we suggest with the thinner and lighter TNG100 curve in the top right panel that include hosts down to $10^{11} \, \Ms$. However, in TNG we do not recover the very low quenched fractions of low-mass satellites as inferred from GAMA observations \citep{2019Davies}. We explore this possible issue in more detail below.

At intermediate redshift $z=0.35$ TNG exhibits a fraction of quenched centrals systematically lower than what is found in observations, across stellar mass. This is particularly true in comparison to \cite{2018Jian} where the discrepancy is up to $\sim\,$40\,percentage points. This also appears to be the case for satellites, where TNG generally has quenched fractions lower than \cite{2018Jian} by $\sim\,20$\,percentage points, although we note the significant discrepancy between this dataset and all other observations, by roughly the same amount. On the other hand, the quenched fractions of high-mass TNG satellites broadly agree with \cite{2014Lin} and \cite{2017Wagner}, to better than $\sim\,10$\,percent. We speculate that disagreement for low-mass centrals may also be driven by differing galaxy classifications. For example in both \cite{2014Lin} and \cite{2018Jian}, and differently from TNG and \cite{2017Wagner}, this classification is based on the richness of groups (N$_{\rm rich}$): `field' galaxies are those not associated with any groups with N$_{\rm rich}>$ 2, `groups' span 10 $<$ N$_{\rm rich}<$ 25 and `clusters' are defined as having N$_{\rm rich}>$ 25. Because of this and possibly other reasons, we note that some observational estimates appear mutually inconsistent, depending on mass regime.

At $z=0.65$ TNG massive centrals ($\gtrsim 10^{10.5} \, \Ms$) show quenched fractions in striking agreement with both \cite{2014Lin} and \cite{2018Jian}. On the other hand, the comparison to 3DHST/CANDELS galaxies \citep{2017Fossati} shows an offset, with the simulations having a quenched fraction lower by $\sim\,20$\,percentage points at the low-mass end, while also being higher by $\sim\,$25\, percentage points at the high-mass end. We reiterate that observed fractions of quenched massive satellites at this redshift are mutually inconsistent, and differ by up to 40\,percent. The degree to which the simulations are consistent with data therefore depends on the particular dataset, as well as the stellar and host mass regimes. More targeted comparisons between TNG and other observational datasets at these intermediate redshifts, but with no distinction between central and satellite galaxies, can be found in Figure 3 of \citealt{2019Donnari}.

In conclusion, the salient features of the observed quenched fractions of both centrals and satellites are broadly recovered by the TNG model. However, due to a combination of different systematics, sample selections, host mass ranges, membership definitions, and the issue of central/satellite misclassification, it is challenging to draw a satisfactory comparison. This is true not only for model versus observation comparisons, but also among observations themselves \citep{2019Donnari,2014Speagle}. Particularly at the low mass end, models and observations rarely show good agreement \citep[see also][]{2017Bahe,2019Tremmel}.

\subsection{A more robust comparison: mocking the SDSS measurement choices and sample selection}
\label{sec:mock}

\begin{figure*}
\centering
\includegraphics[width=0.45\textwidth]{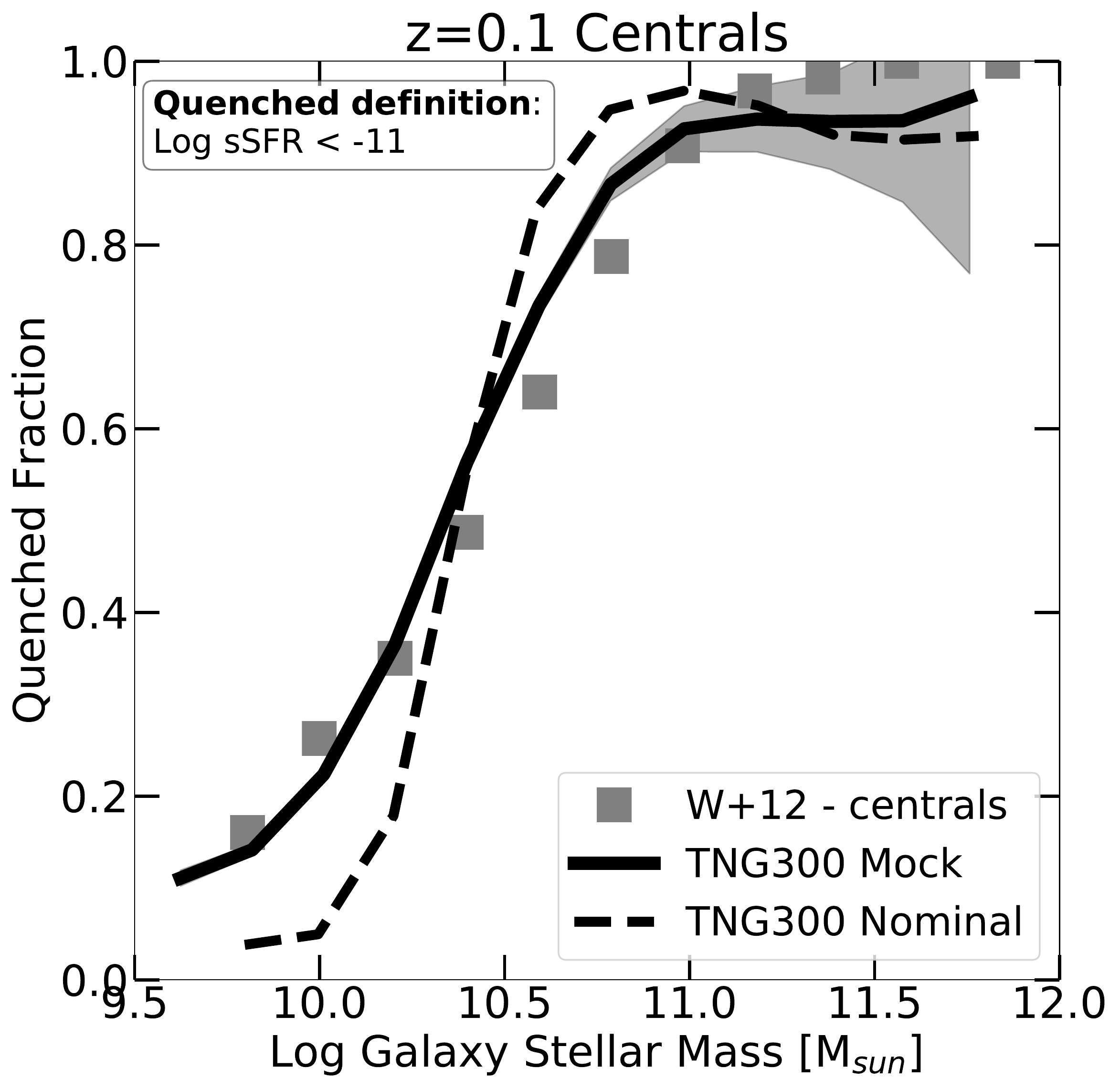}
\includegraphics[width=0.45\textwidth]{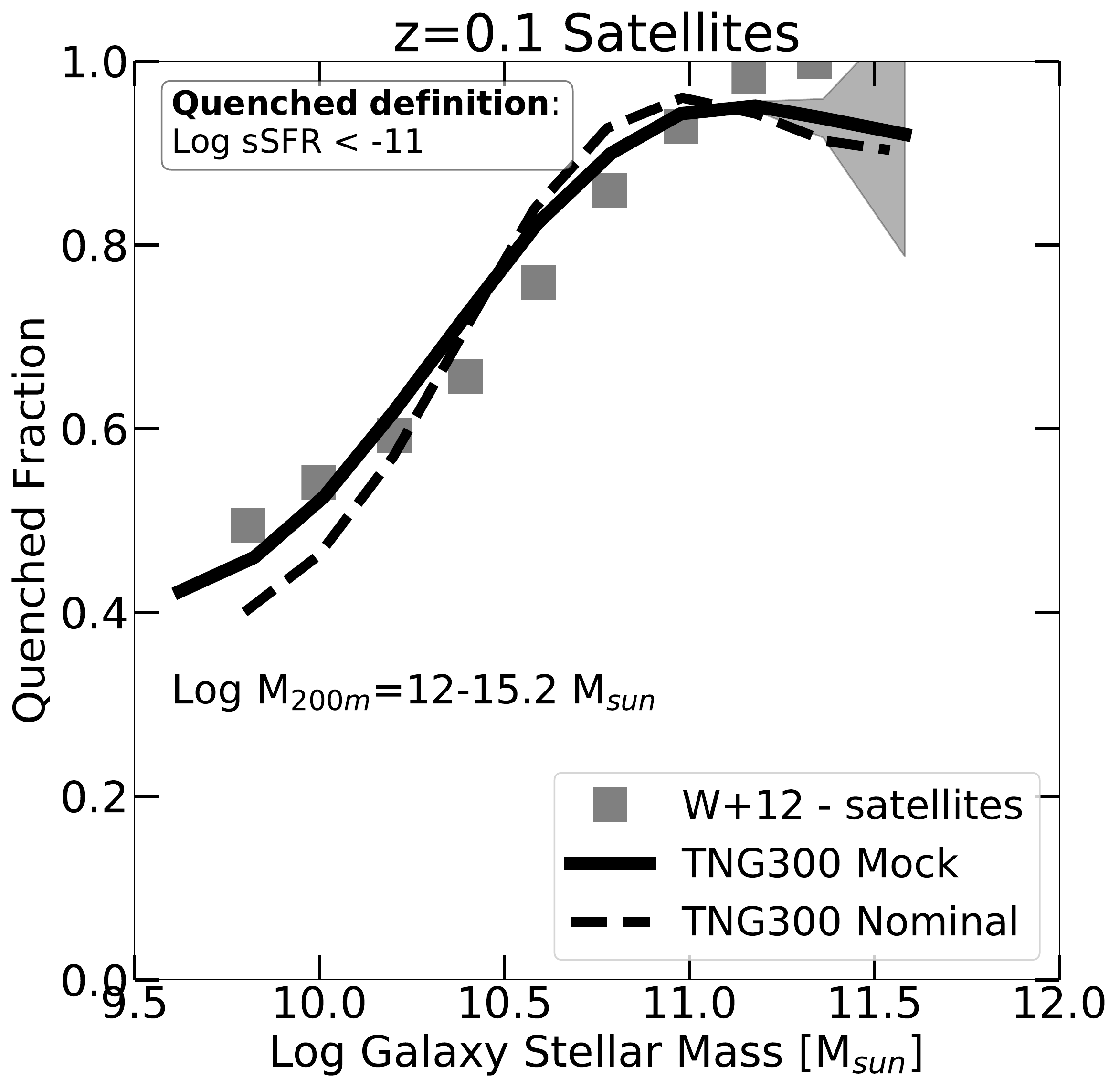}
\includegraphics[width=0.44\textwidth]{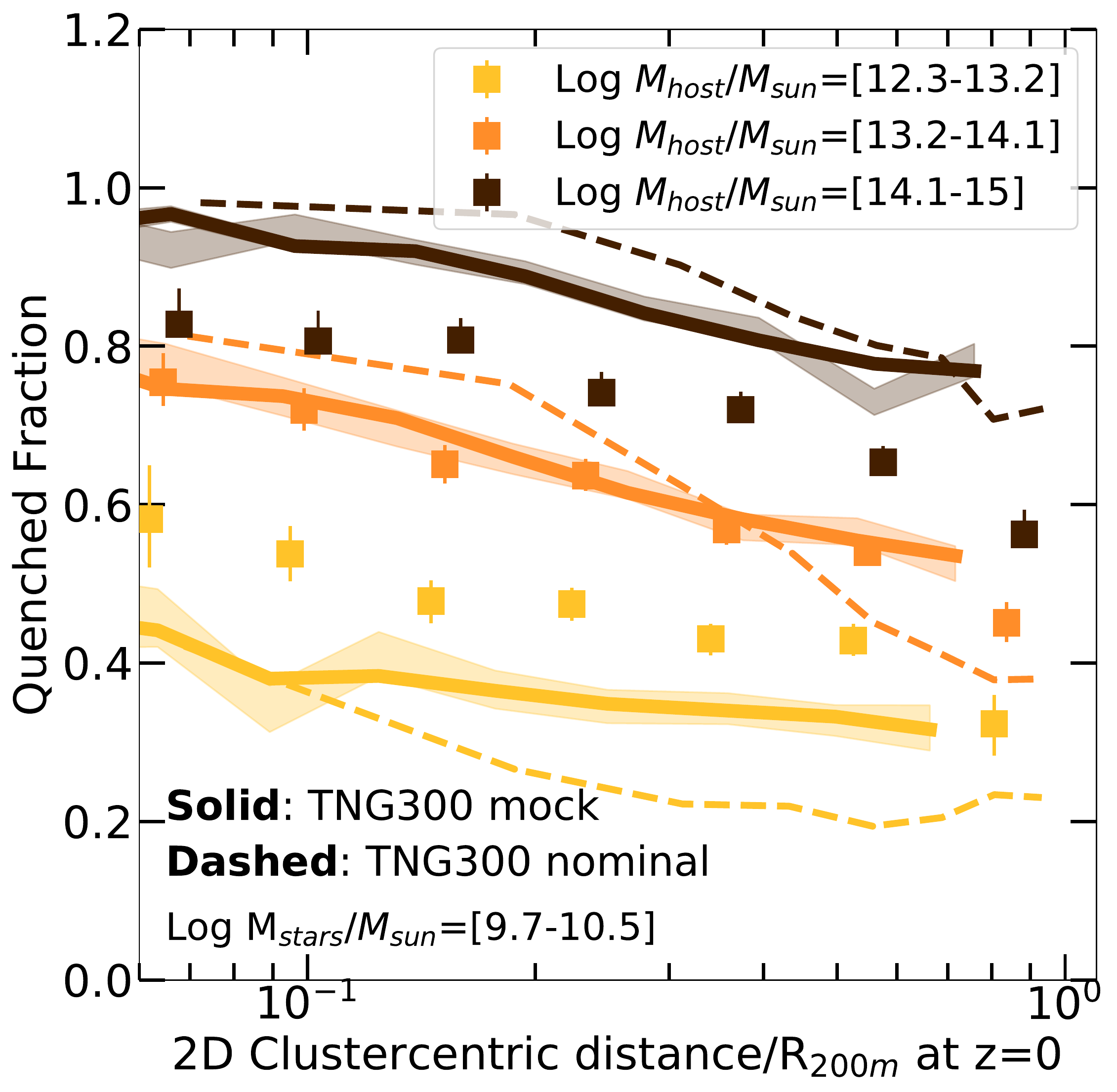}
\includegraphics[width=0.44\textwidth]{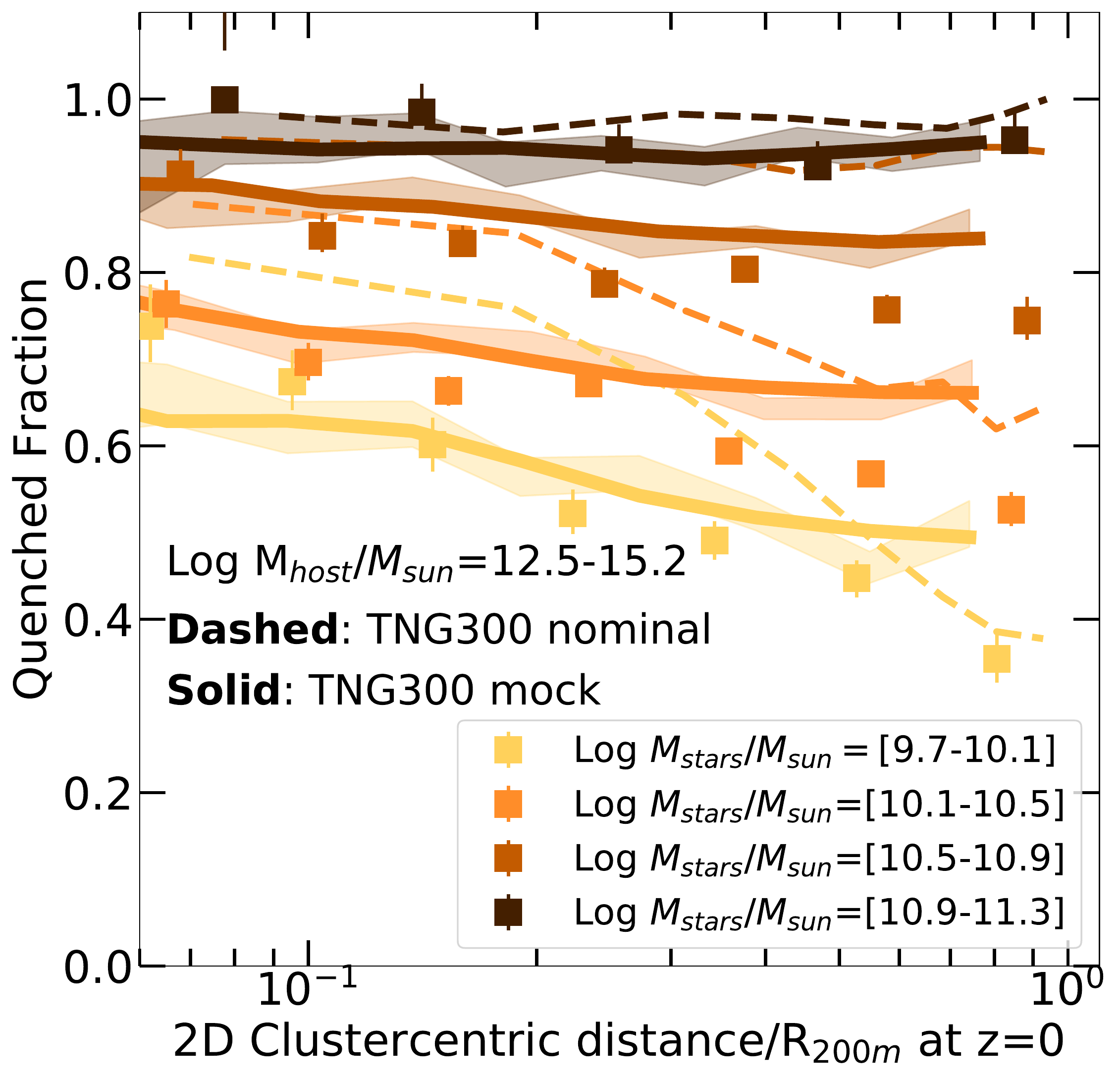}
\includegraphics[width=0.45\textwidth]{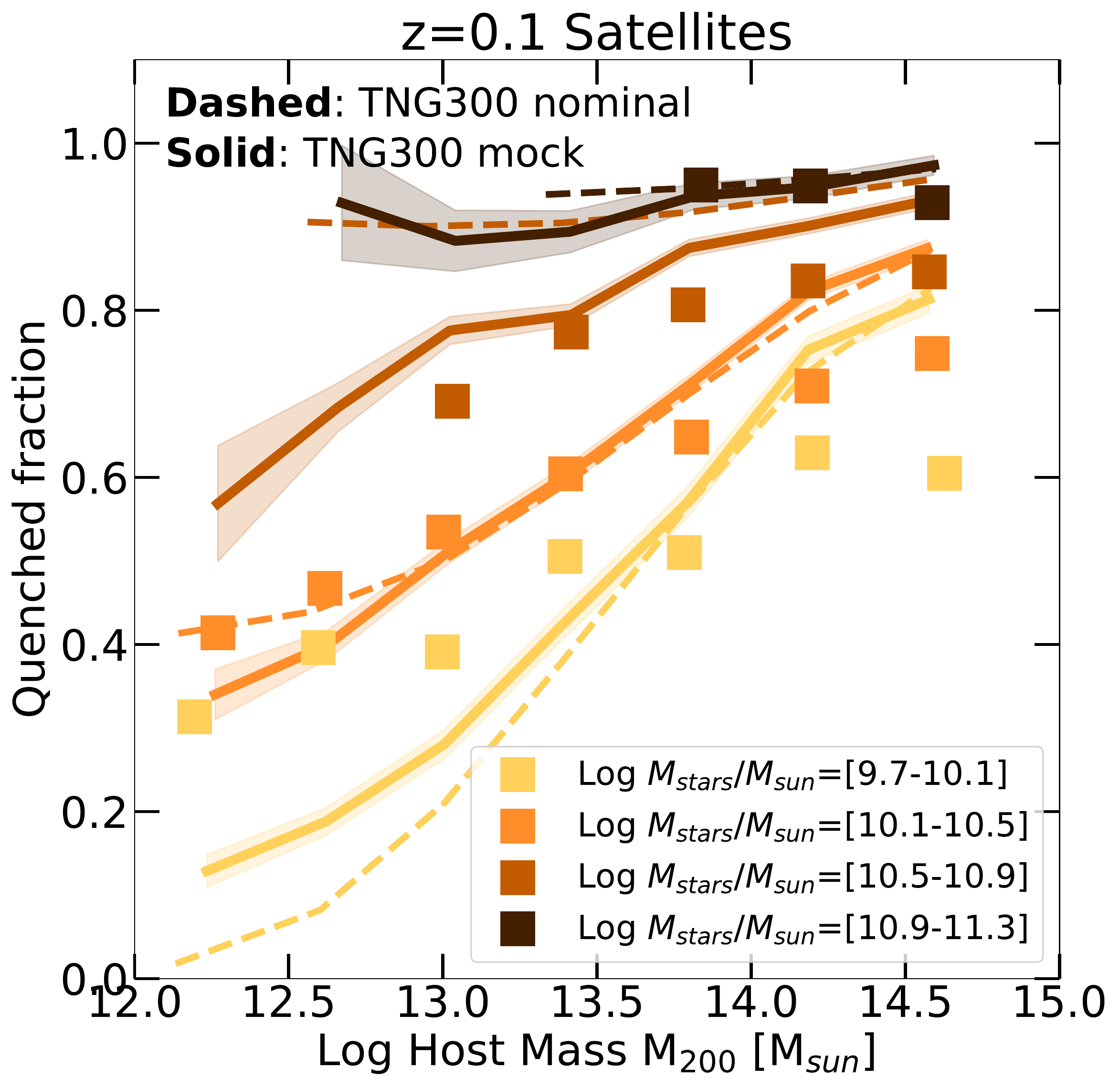}
\includegraphics[width=0.45\textwidth]{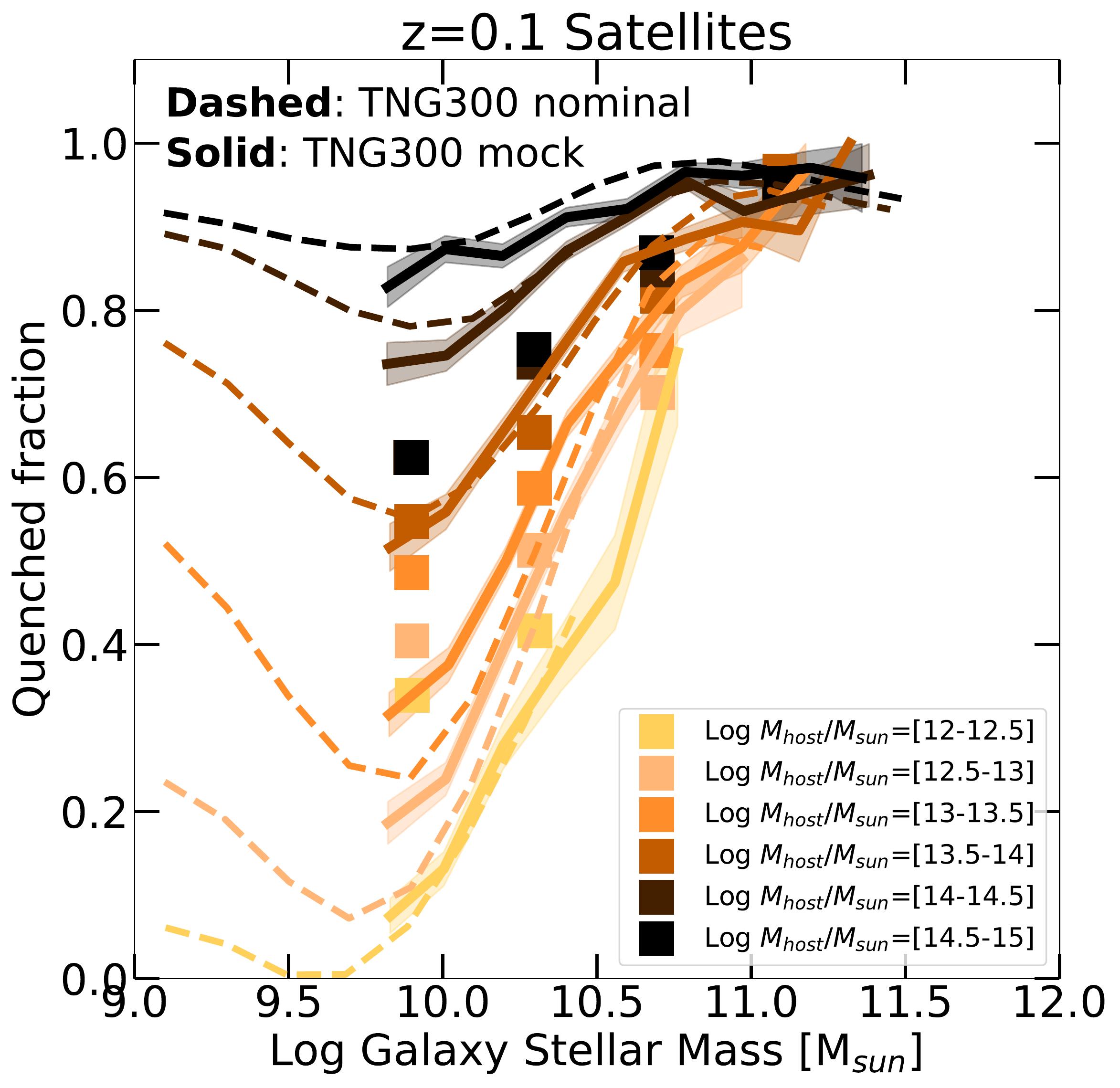}
\caption {\label{fig:mock} {\bf TNG300 nominal versus TNG300 mock versus SDSS results.} Quenched fractions as a function of stellar mass (top), halocentric distance (middle), and host mass and galaxy stellar mass (bottom), at $z=0.1$. In all panels, we show results of the nominal i.e. direct TNG300 results (dashed) and the more-realistically mocked TNG300 results (solid, see text for more details).
Shaded areas represent the Poissonian errors. The global quantitative agreement between TNG and SDSS is excellent, as visible in the top row, while specific trends as a function of halocentric depend on the galaxy stellar and halo mass regime under consideration, generally being best for intermediate group-mass halos.}
\end{figure*}

We proceed with a more robust and well-posed comparison, by focusing on data from the Sloan Digital Sky Survey (SDSS) as compiled and analyzed by \cite{2012Wetzel} and \cite{2013Wetzel}, by using the same SDSS catalogs used in their papers. To do so we adopt the most important choices and definitions of those works, their systematic impacts on TNG quenched fractions having been demonstrated in Section \ref{choices}. In particular, we

\begin{itemize} 
    \item classify satellite galaxies based on a two-dimensional, projected, cylindrical selection geometry (see Figure \ref{fig:2D3D});
    \item adopt the same virial radius $R_{\rm 200m}$ and corresponding host mass definitions, based on a sphere whose mean density is 200 times the \textit{mean} density of the Universe;
    \item define satellites as quenched using the observationally employed criterion of log(sSFR) $ < -11\,\rm{yr}^{-1}$ (see Figure \ref{fig:definitions});
    \item account for the effects of misclassification between centrals and satellites with fractions [cen,sat] = [10,20] (see Section \ref{sec:misclassification}). 
    \item include appropriate systematic uncertainties with 0.2 dex random errors for stellar masses, 0.3 dex for halo masses, and 0.6 dex for SFRs (see Figure \ref{fig:errors}),
    \item match the galaxy stellar mass and host mass distributions (see Figure \ref{fig:MstarsDistr} and further details in Appendix \ref{appendix}).
\end{itemize}

Both SFR and stellar masses in TNG are based on the simulations output, with no mocking strategy, and are evaluated within twice the stellar half mass radius, as per Section~\ref{sec:SFRs}.
However, even if we do not mock the tracers used by SDSS to infer galaxies SFRs, we expect H$\alpha$ to return SFRs on very short time scales -- roughly 5-10 Myr -- and therefore closer to the instantaneous SFR values that we adopt.

The results are shown in Figure \ref{fig:mock}: global quenched fraction as a function of galaxy stellar mass (top panels), halocentric distance (middle panels) and host mass (bottom panels). We select galaxies, centrals and/or satellites where appropriate, from the TNG300 snapshot at $z=0.1$, the median observed redshift.

We denote with dashed curves the direct output of the simulation (`TNG300 nominal'), for which we adopt previous fiducial analysis choices: $M_{200m}$ and $R_{200m}$, log(sSFR) $ < -11\,\rm{yr}^{-1}$, satellite selection based on spherical 3D geometry, unmatched mass distributions, and with no added systematic errors. In contrast, we show with solid curves the results obtained when applying the previously enumerated and more correct measurement choices sample selections (`TNG300 mock'). In all panels of Figure \ref{fig:mock} square symbols indicate observational data from SDSS \citep{2012Wetzel,2013Wetzel}, grey in the top panel and color-coded in the middle and bottom panels for different mass bins, which are matched to TNG.

The importance of the complete mock procedure is apparent in both top panels. As our ultimate objective is to make a statically robust and systematically correct comparison with data, the difference between the dashed (nominal) and solid (correctly mocked) lines is significant. Specifically, the full mock brings the quenched fractions of TNG galaxies, both centrals and satellites, ensemble averaged across the mass-matched galaxy population, into quantitative agreement with the SDSS results. Here a key effect is the inclusion of the Gaussian errors, which make the trends with mass shallower and boost the low-mass normalization. These differences are small, but noticeable.

When comparing quenched fractions versus halocentric distance (middle panels), the un-mocked simulation results (dashed lines) only roughly reproduce the observed trends, and have various normalization offsets depending on distance and mass. As we have previously shown, the geometrical selection and distances based on either two-dimensional or three-dimensional methods are critical here (see Figure \ref{fig:2D3D}). Once we consider the mock results (solid lines), which account for the projection effects and the possible effects of satellite/central misclassification, the agreement with SDSS data at intermediate host ($10^{13-14} \, \Ms$) and stellar masses  ($10^{10-11} \, \Ms$) is significantly better. While agreement at the group scale is excellent (see curve for host mass $10^{13.2-14.1} \, \Ms$), the TNG quenched fractions for satellites in lower (higher) mass hosts is roughly lower (higher) by up to $\sim\,$20 percentage points than observed (middle left panel).

The dependence on host halo mass is therefore stronger in the TNG simulations than in the data. Here we are fundamentally probing environmental processes, that are not related to SMBH-driven quenching and that are fully emerging features of the underlying physical model: therefore this excessively strong dependence on host mass points towards an issue in the interplay between stripping processes and the background halo medium, i.e. either in the restoring gravity of satellite galaxies and/or the background density of the intragroup medium. It is, however, unclear what aspect of the TNG model could be modified to impact this particular observable trend. On the one hand, the low-mass end satellite quenched fractions in TNG are consistent within 10-15 percentage points across variations in baryonic mass resolution of a factor of 1024 (see Appendix~\ref{appendix_B} and Appendix A of \citealt{2021Donnari}). On the other hand, as the quenching of low-mass satellites in high-mass hosts is due in similar proportions to pre-processing in smaller hosts at earlier times and environmental effects of the hosts in which they are found today (see \citealt{2021Donnari}), it is difficult to pin point in what mass and redshift regime the main culprit of the disagreement may lie. Finally, at the low host mass end, the yellow curves and data points include satellite galaxies in hosts whose centrals have comparable stellar masses, i.e. small groups with possibly two dominant galaxies: these are systems for which larger systematic issues may arise also observationally in terms of e.g. host mass labeling. 

Considering the dependence on satellite stellar mass (middle right panel), we find generally good agreement to better than $\sim\,10$\, percent. The only noticeable tension is that the data favors a quenched fraction which declines more rapidly with halocentric distance than in the simulations. This is particularly clear for the lowest mass satellites, below $\MS < 10^{10} \Ms$, where the data decline by nearly 40\,percent from the host center to the virial radius, while the decline in TNG is less than 20\,percent. As above, this could be related to the radial profiles of background gas density in the hosts being too shallow, which would cause ram-pressure stripping to increase more weakly with decreasing radius. For instance, feedback from both supernovae and SMBHs are known to significantly distribute baryons in the CGM and IGM of galaxies in the TNG simulations \citep{2018Pillepich_model,2019Nelson_50,2020Terrazas}, and this baryonic redistribution could be indirectly probed by environmentally induced quiescence.

Finally, we stack satellites at all distances to provide an in-depth view of the results in the top right panel of Figure~\ref{fig:mock}. We show the dependence on host halo mass for different satellite mass bins (Figure~\ref{fig:mock}, lower left panel), as well as the dependence on satellite mass for different host mass bins (Figure~\ref{fig:mock}, lower right panel). As before, we see excellent agreement for intermediate mass hosts, but the trend with host mass is somewhat too large in TNG, which results in quenched fractions below the SDSS data for the lowest mass hosts, and above the SDSS data for the most massive hosts. This is only true for intermediate and low mass satellites with $\MS \lesssim 10^{10.5} \Ms$ -- more massive satellites are in good agreement, although here the quenched fractions are set by SMBH-driven quenching in the field (see \citealt{2021Donnari}). For the most massive clusters available in the sample, $M_{\rm host} \simeq 10^{14.5} \Ms$, the quenched fractions of intermediate (low) mass TNG satellites are too high by $\sim$\,15 ($\sim$\,20) per cent, as shown by the orange (yellow) lines.

Overall, we see that the impact of the mock procedure (solid versus dashed lines) is moderate for most of the relevant comparisons of these two bottom panels, rarely leading to changes of more than $\sim$\,10\,percent in the regimes where data exists. We note that the good \textit{global} agreement between TNG300 satellites and \cite{2012Wetzel} (top right panel) is driven by the broad host mass bin of $M_{200m} = 10^{12-15.2} \Ms$. After we separate satellites into different host mass bins (bottom right panel), the level of agreement clearly depends on host mass, being best for hosts which are common though not the most sampled in SDSS (see Appendix \ref{appendix}).

Finally, we recall that we have adopted instantaneous star-formation rates based on the gas in the TNG simulations. On the other hand, the SFRs in \cite{2012Wetzel} are derived from a combination of different tracers: H$\alpha$ emission (for sSFR $> 10^{-11}\,\rm{yr}^{-1}$), multiple emission lines (for $10^{-12}\,\rm{yr}^{-1} < \rm{sSFR} < 10^{-11}\,\rm{yr}^{-1}$), or the amplitude of the 4000$\AA$ Balmer break D$_n 4000$ (for sSFR $< 10^{-12}\,\rm{yr}^{-1}$), coupled to AGN contamination and fiber aperture corrections. Even more detailed future mocks will be required to assess the impact of such specific tracers.


\begin{figure*}
\centering
\includegraphics[width=0.48\textwidth]{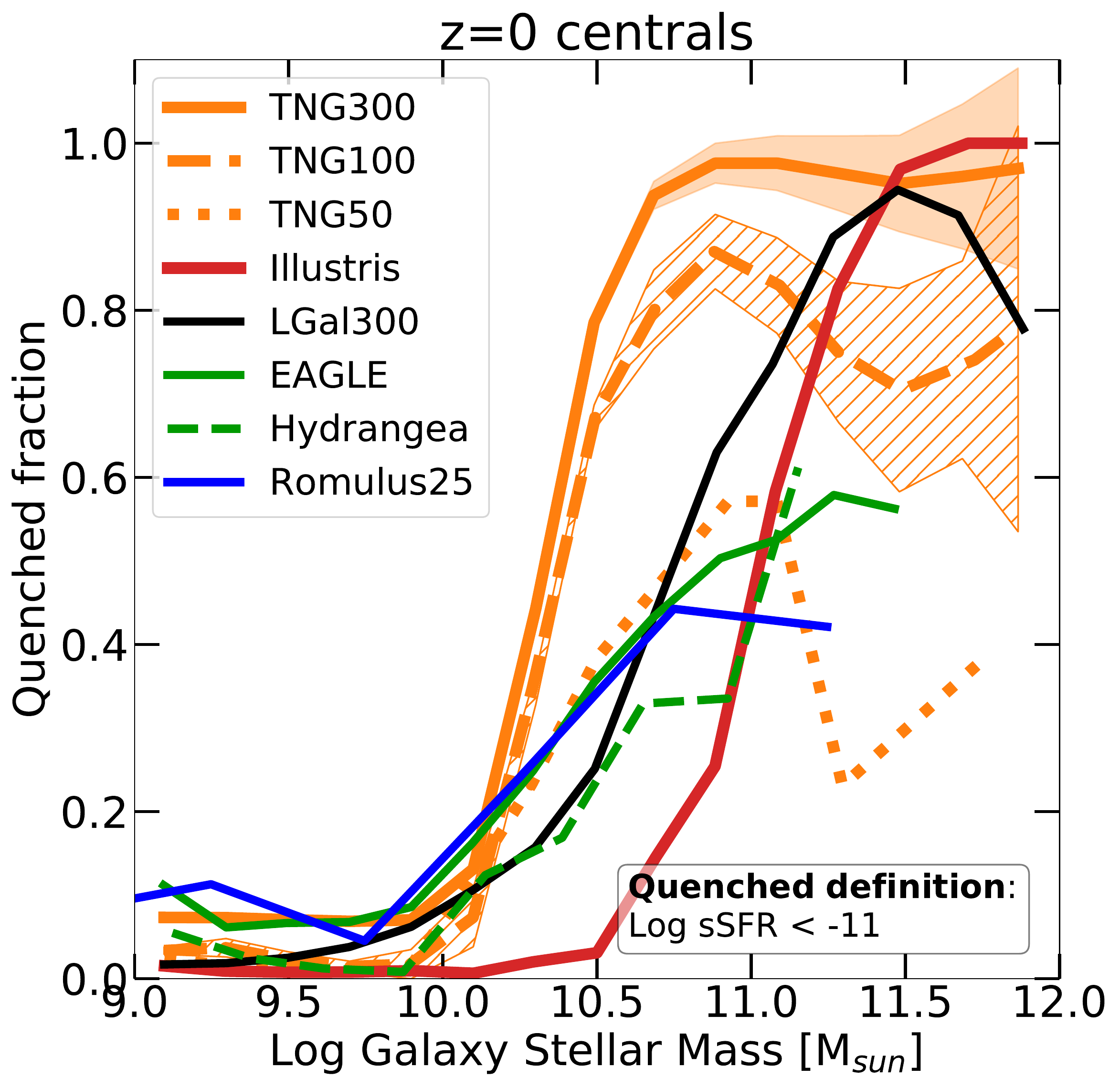}
\includegraphics[width=0.48\textwidth]{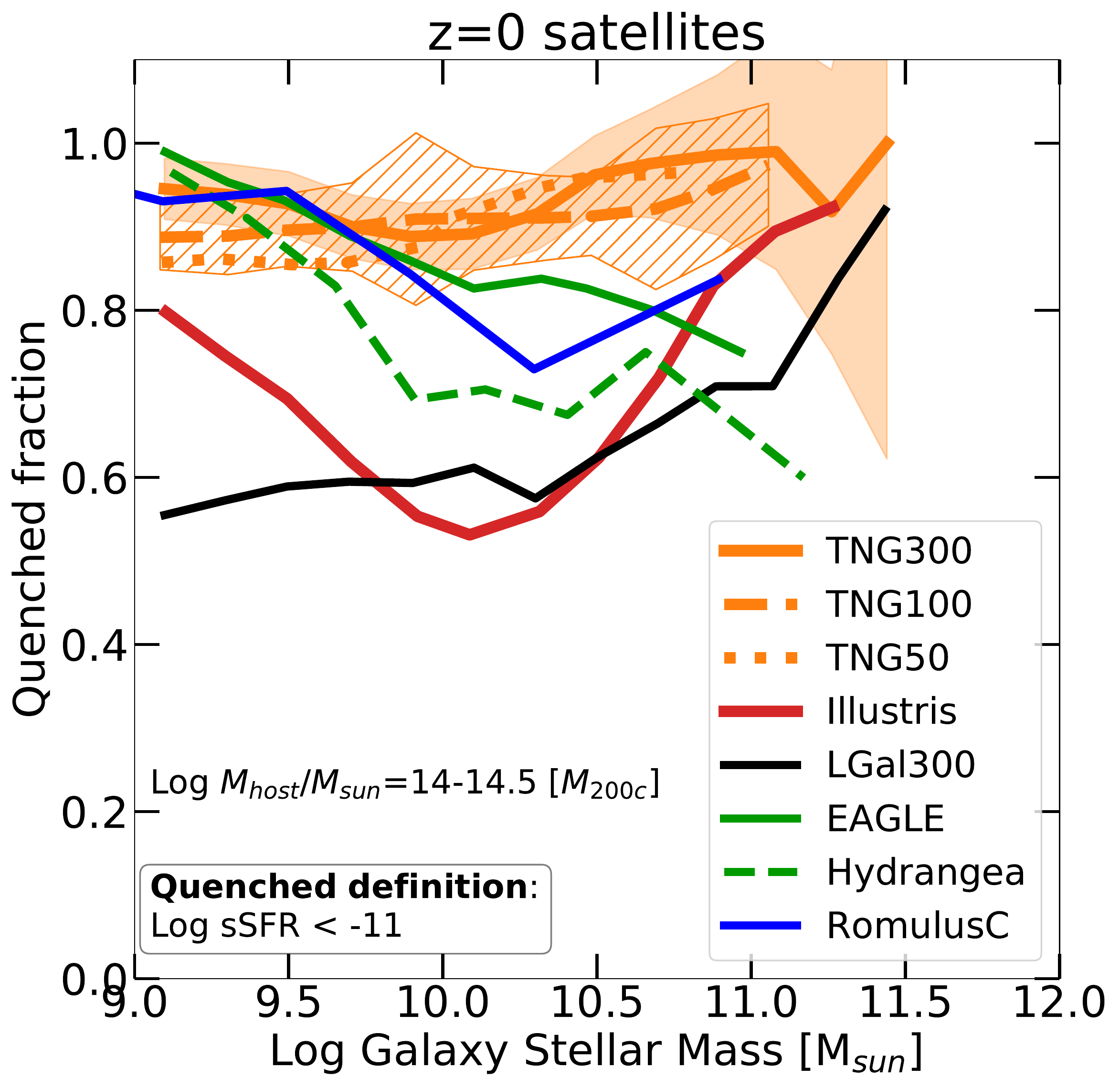}
\includegraphics[width=0.48\textwidth]{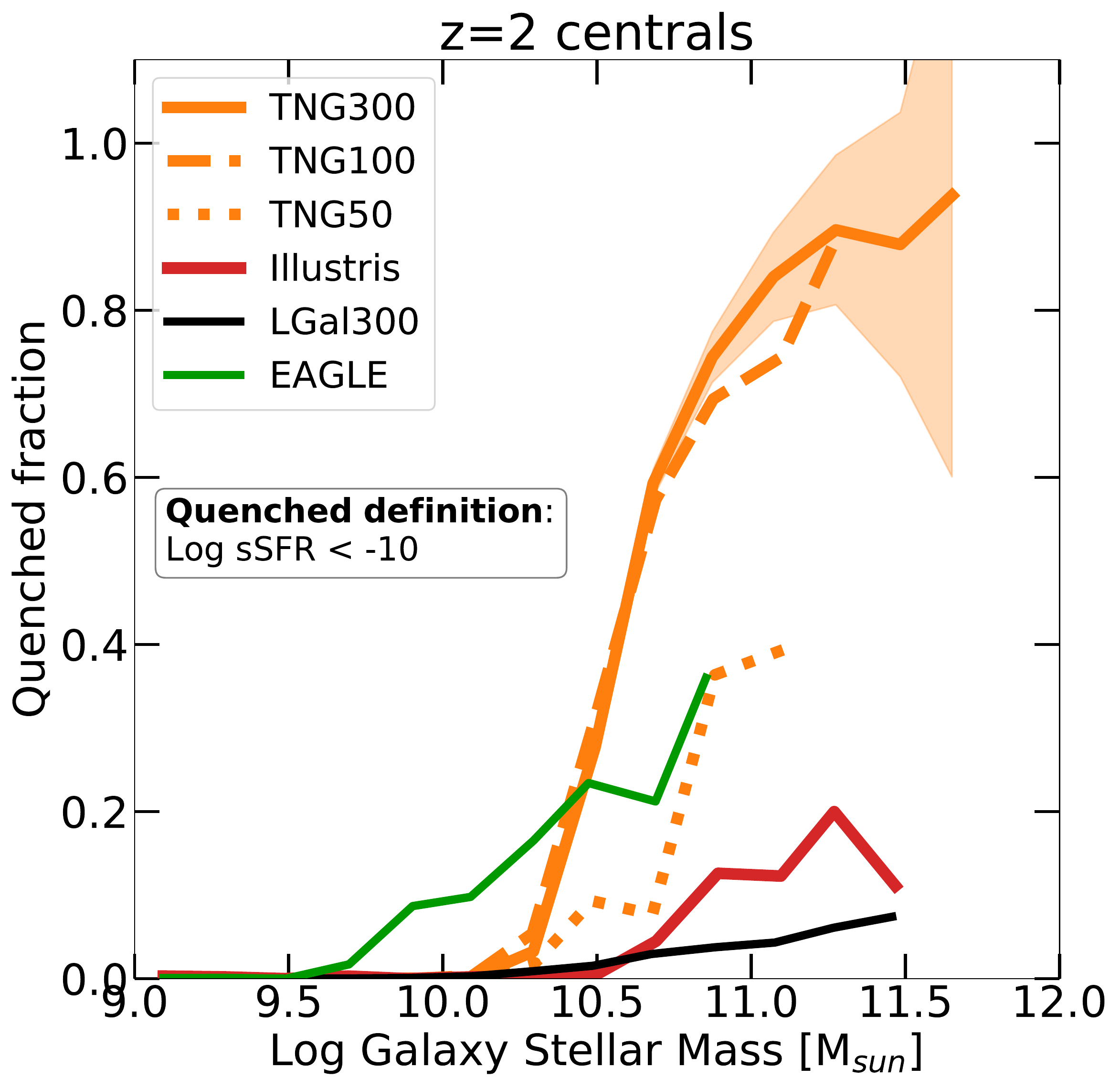}
\includegraphics[width=0.48\textwidth]{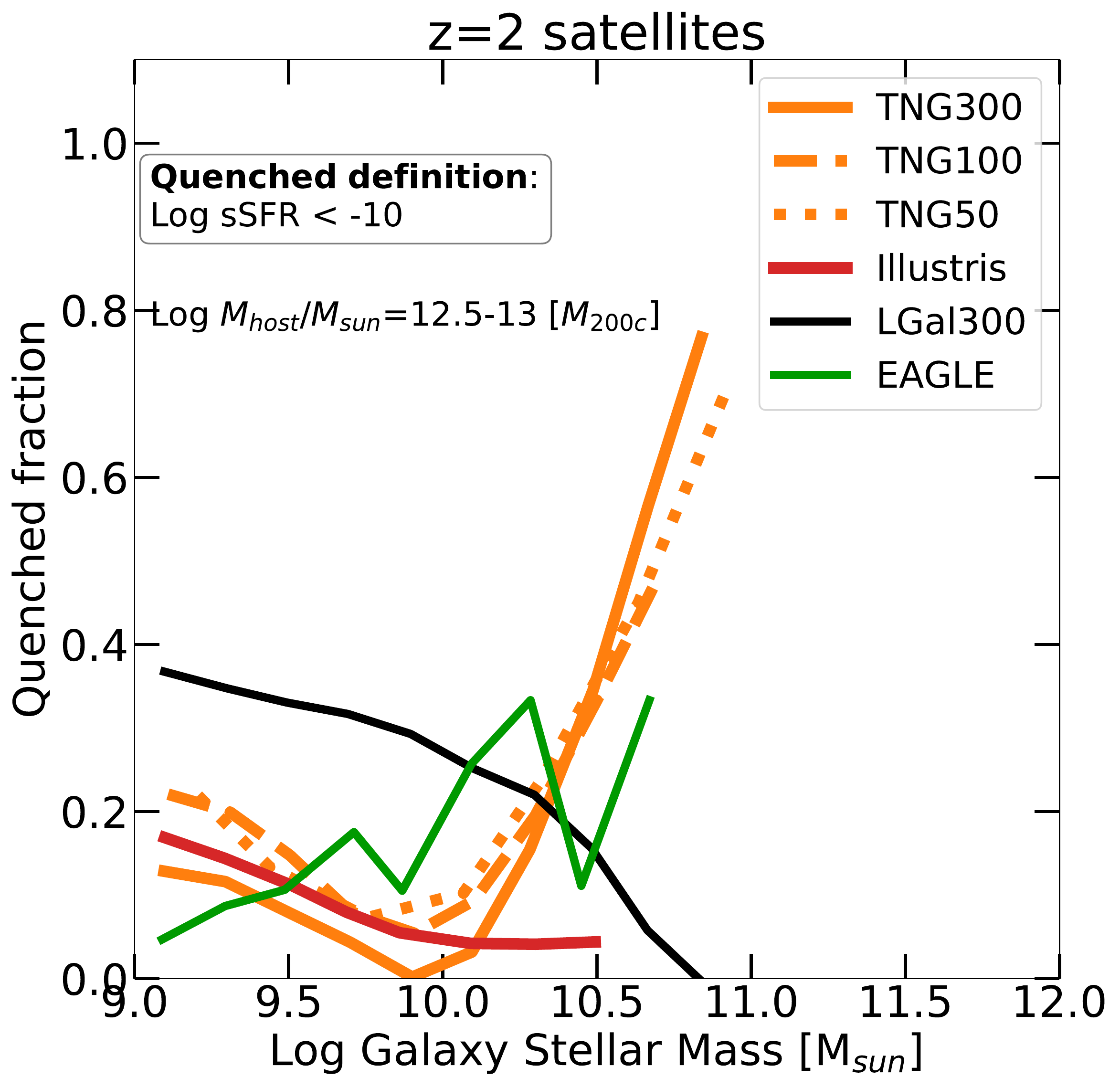}

\caption{\label{fig:models} {\bf Comparison to other models}. Quenched fractions of central galaxies (left) and satellite galaxies (right) at $z=0$ (top) and $z=2$ (bottom) in TNG300 (orange solid), TNG100 (orange dashed), TNG50 (orange dotted), Illustris (red), EAGLE and Hydrangea (green solid and dashed, respectively), RomulusC and Romulus25 (blue), and L-Galaxies (black). For TNG, Illustris, EAGLE and Hydrangea curves, SFRs and stellar masses are measured within twice the stellar half-mass radius. For the satellite selections, we uniformly take all galaxies above a stellar mass of $10^9 \Ms$ within hosts of mass $10^{14-14.5} \Ms$ at $z=0$ and $10^{12.5-13}$ at $z=2$. Orange shaded areas denote estimates of the Poisson errors, shown for TNG300 for reference, whereas the striped areas around TNG100 give estimates of the cosmic variance uncertainties. While the trends of quenched fraction with mass for $z=0$ centrals show the same qualitative trends between models, they differ in the details, particularly in the rapidity of the transition towards the quenched population. Satellite galaxy quenched fractions are even more disparate, with different models having a diversity of trends with mass. High redshift ($z=2$) offers one of the most interesting regimes for comparison, as some models generate substantative populations of quenched galaxies at these early epochs, while others do not.}
\end{figure*}

\subsection{TNG quenched fractions versus other models}

We conclude with a comparison of several theoretical models, contrasting TNG versus the original Illustris simulation \citep{2014MNRASVogel,2014vogel, 2014Genel,2014Torrey,2015Sijacki,2015Nelson}, as well as EAGLE \citep{2015Crain,2015Schaye,2015Furlong}, RomulusC and Romulus25 \citep{2019Tremmel}, Hydrangea \citep[a set of zoom-in clusters simulated with the EAGLE model,][]{2017Bahe}, and the semi-analytical model \textsc{L-Galaxies}, which has been run directly on the TNG dark matter only simulations \citep[][see Table \ref{tab:observations} for more details]{2020Ayromlou}. Similarly, the results we present for EAGLE are based on a re-analysis with the TNG version of \textsc{Subfind}, enabling an apples-to-apples comparison by insuring that galaxies and their properties are identified and measured in exactly the same way.

For the remaining simulations (Romulus and Hydrangea) we note that results for quenched fractions are extracted from the literature, and thus may invoke different definitions and methodologies at a number of levels. Specifically, the TNG, Illustris, and EAGLE curves all use our fiducial definitions exactly, whereby e.g. the galaxy stellar masses and SFRs are evaluated within twice the stellar half mass radius; however, for the other simulation results we do not have any information about the aperture within which the mass and SFR are evaluated. The L-Galaxies and Hydrangea results adopt the same quenched galaxy selection as that used for the Illustris, TNG, and EAGLE curves, while in Romulus the definition of quenched is based on a fit of the median values of the SFR within 0.1 dex bins of stellar mass in the range $10^{8-10}~\Ms$. We have shown, however (Section \ref{sec:choices}), that at low redshift, and particularly at $z=0$, the quenched fractions do not change significantly between these two definitions.

Figure \ref{fig:models} shows the quenched fractions of centrals (left panel) and satellites (right panel) at $z=0$ (top) and $z=2$ (bottom) as a function of galaxy stellar mass. For the TNG simulations we show the Poissonian errors for TNG300 (denoted with shaded areas) and the cosmic variance uncertainties for TNG100 at $z=0$ (denoted with striped areas). In principle, the jackknife error computed for TNG100 gives a lower estimate for the sample/cosmic variance error in the boxes smaller than TNG100, namely TNG50, Romulus and Hydrangea. We compare all simulation results, where available, with different line colors, grouped by underlying galaxy formation model. For all models we include satellites residing in hosts with $z=0$ masses of $10^{14-14.5} \Ms$, which enables a comparison between the different models which otherwise have significantly different volumes and statistical sampling of massive hosts.
As we have explored in our companion paper \citep{2021Donnari}, low-mass satellites are overall more quenched than centrals at fixed stellar mass. This holds true regardless of redshift, highlighting the role of the environment in the quenching process. Conversely, high mass galaxies are quenched regardless of whether they are centrals or satellites, mainly due to SMBH feedback, at least in the TNG model. These two features are broadly common to all the simulations we explore here. 

Specifically, at $z=0$ and for all the models, the quenched fraction of centrals (upper left panel) is a monotonic function of the stellar mass, less than 10\,percent below $\sim 10^{10} ~\Ms$, increasing to $\sim$50 percent in EAGLE, Hydrangea, and RomulusC, and up to $80-100$\,percent in TNG, Illustris and L-Galaxies, particularly for galaxies more massive than $\sim 10^{11} ~\Ms$. Illustris, TNG and L-Galaxies all show a relatively distinct transition between star-forming and quenched centrals, albeit at a different transitional mass scale\footnote{We refer the reader to \citet{2018Weinberger,2019Donnari,2021Donnari}, for the main differences in quenching properties between TNG100 and TNG300, and between TNG and Illustris. The differences at high mass end in TNG50 centrals in comparison to TNG100 and TNG300 is due to a combination of resolution effects and sample variance, as the volume of TNG50 is considerably smaller than the other TNG runs -- see Appendix \ref{appendix_B} for more details.}\footnote{Even if not shown, we have checked that the difference between TNG and L-Galaxies is not due to the different adopted stellar mass definition, whereby stellar masses are measured within twice the stellar half mass radius for the TNG model and in practice accounting for all gravitationally-bound stellar mass for L-Galaxies). See Appendix~\ref{appendix_A} for a quantification of such a choice on TNG300}. The origin of the different `quenching mass scales` between TNG and L-Galaxies, for instance, ultimately lies in the differing treatments of supermassive black hole feedback \citep{2020Ayromlou}, and current low-redshift data can discriminate on both the locus and steepness of the transition (see top left panel of Figure~\ref{fig:mock}). On the other hand, EAGLE, RomulusC and Hydrangea show a weaker trend with galaxy stellar mass, with quenched fraction rising more slowly.

The quenched fractions of satellites versus mass at $z=0$ (upper left panel) are more diverse. In \cite{2021Donnari} we have shown that this is a non-monotonic trend with stellar mass in TNG, which occurs due to the combination of environmental and mass quenching processes. Here we see that TNG100, TNG300 and TNG50 are all in good agreement, notwithstanding the different numerical resolution (see also Appendix\ref{appendix_B}), with satellite quenched fractions of almost $\sim\,$90\,percent, largely independent of stellar mass at this host mass scale. This is $\sim\,$30\,percentage points higher than in Illustris. The three other hydrodynamical simulations also rise towards low-mass: L-Galaxies is the only exception, with a monotonically increasing quenched fraction from $\sim\,$60 to $\sim\,$90 percent across this mass range. RomulusC, Hydrangea and EAGLE all have the opposite trend: lower quenched fractions for higher stellar masses.

At the high mass end, when satellite galaxies exceed $\gtrsim 10^{10.5} \Ms$, differences start to become attributable to relative (in)efficiencies of SMBH feedback between the models. On the other hand, mismatches below this mass traces different efficiencies of environmental effects, which could be due to differences in background gas properties as discussed earlier, or to more subtle effects for instance in the numerical treatment of the hydrodynamics and resulting changes in gas stripping \citep{2007Agertz, 2012Sijacki}.

Turning to higher redshifts, the models become much more varied in their predictions. Redshift $z=2$ is a particularly illuminating regime, as current models often fail to reproduce the observed quenched fractions of massive galaxies, or similarly the observe space density of massive quenched galaxies \citep[e.g.][]{2020Valentino}. For example, the original Illustris simulation produces essentially no quenched centrals already by $z=2$ or $z=3$, whereas the TNG model does \citep{2019Donnari}. 

The bottom panels of Figure \ref{fig:models} show the fractions of quenched centrals and satellites at $z=2$ for the available models. To take the time evolution of the star-formation main sequence normalization with redshift, we change the definition of quenched galaxies to log(sSFR/yr) $< -10$. We also change the host halo mass selection for satellites to a lower bin of $10^{12.5-13}\Ms$ to accommodate the available statistics. The differences in central quenched fractions at the massive end, above $10^{11}\Ms$, are clearly highly in the TNG model, and the same is also true for satellites. This suggests that both secular quenching through supermassive black hole feedback, as well as environmentally-driven quenching in high density regions, are processes which are both already in place in the early $z \sim 2$ Universe, as also inferred in recent observations \citep[e.g.][]{2017Glazebrook,2020Forrest}.


\section{Summary and conclusions}
\label{summary}

Together with our companion paper \citep{2021Donnari} we have used the IllustrisTNG simulations to investigate the diverse physical pathways of galaxy quenching and their observational signatures. In that work we demonstrate how the TNG model produces a picture of both mass-driven and environmentally-driven quenching. Here we focus on the large-volume TNG300 cosmological hydrodynamical simulation to: (i) study systematic issues in inferring quenched fractions, (ii) quantitatively compare TNG quenched fractions against low-redshift observational data, and (iii) contrast TNG with other theoretical models. Our key results are:

\begin{itemize}
    \item {\bf Measurement choices}. We study the role played by different measurement choices, definitions, and sample selection issues in shaping the quenched fractions of galaxies (centrals and satellites in the $10^{9-12}\Ms$ stellar mass range) from $0 \leq z \leq 3$. Different criteria to define `quenched' translate into population-wide quenched fractions which differ by up to 70 (30) percentage points for centrals (satellites), particularly at the high-mass end of $M_\star \gtrsim 10^{10.5} \Ms$. Adopting a fixed threshold in sSFR at different redshifts fails to identify quenched galaxies due to the evolution of the star formation main sequence with time (Figure \ref{fig:definitions}). At high redshift $z=2-3$ the averaging timescale (Figure \ref{fig:timescales}) as well as the physical aperture in which SFR is measured have strong impacts on the inferred quenched fractions. For massive galaxies larger apertures produce lower quenched fractions, by up to $\sim\,$50 percent (Figure \ref{fig:apertures}). We show that adding systematic uncertainties to theoretical values for stellar mass and SFR is important when comparing to observational data, as it changes not only the values of quenched fractions for low-mass galaxies, but also the trend and/or slope with mass (Figure \ref{fig:errors}).
    \\
    \item {\bf Sample selection}. Adopting a satellite selection that accounts for the two-dimensional projection effects and geometry inherent in observations, rather than theory-space choices more common in simulations, reduces the quenched fractions as a function of halocentric distance by $\sim$ 10-30 percentage points, especially for low mass satellites ($\lesssim10^{10.5}\Ms$) and hosts ($\lesssim10^{14}\Ms$): Figure \ref{fig:2D3D}. Quenched fractions as a function of redshift evaluated for galaxies above a minimum stellar mass threshold depend sensitively on the shape of the sample mass distributions. For example, a flat in log mass, rather than volume-limited, sample leads to inferred quenched fractions higher by $\sim$\,20\, percentage points, at all redshifts, for all but the most massive galaxy samples (Figure \ref{fig:MstarsDistr}).
    \\
    \item {\bf Comparison to observations}. We make a quantitative comparison to observational results at low redshift, with a particular focus on the SDSS catalog, for which we have created detailed mocks by matching the key analysis choices of \cite{2012Wetzel,2013Wetzel}. We find that the TNG quenched fractions compare well with a selection of observational estimates, for both centrals and satellites (Figure \ref{fig:Q_frac_obs}), being in striking quantitative agreement for global central and satellite quenched fractions at $z=0.1$. For satellites, as a function of halocentric distance, the level of (dis)agreement depends on the host halo and galaxy stellar mass regime under consideration, being best for intermediate, group-mass scale hosts ($\lesssim10^{13-14}\Ms$), and worst for the lowest mass satellites in either the lowest ($\sim10^{12}\Ms$) or highest mass hosts ($\sim10^{14.5-15}\Ms$, Figure \ref{fig:mock}). 
    \\
    \item {\bf Comparison to other models}. We contrast the fraction of quenched TNG galaxies versus other theoretical models: Illustris, Eagle, Hydrangea, Romulus25/C, and the semi-analytical model L-Galaxies. Although the trends of quenched fraction with galaxy mass for $z=0$ centrals have qualitatively similar behavior, the models differ in how rapidly the population transitions to predominantly quenched. Satellite quenched fractions are even more diverse -- although all models produce a large fraction of quenched satellites in hosts with masses of $10^{14-14.5} \Ms$, the normalization and even trend with galaxy mass can differ. At $z=2$ some models produce a significant population of quenched galaxies, while others do not. This fraction, for massive centrals, is higher in TNG than other models, reaching $\sim\,$80\,percent already at this early epoch. Together with satellite quenching at $z=2$, this shows how environment and mass quenching in the early Universe are a model dependent, and thus informatively constraining, phenomenon (Figure \ref{fig:models}).
\end{itemize}

Our study highlights the importance of carefully considering the many observational effects and choices that can play a role in any comparison between theory and data, particularly for a quantity as complex as the quenched fractions of galaxies. Our results suggest that the only way to make meaningful comparisons across galaxy samples, in fact of any type and origin, is by matching the operational definitions and choices for quenched galaxies, galaxy apertures, SFR tracers, sample mass distributions, etc. Overall we have also validated the outcome of the TNG model, demonstrating its broad agreement with observation data, although here we have focused on the local Universe at low redshift. The signatures of different feedback models and the ways in which different simulations generate the dichotomy between the star-forming and quenched galaxy populations are encoded into the details of galaxy quiescence, which offers a powerful probe on theoretical models for galaxy evolution. Future, careful, and targeted comparisons to other observational datasets will be invaluable, particularly at intermediate and high redshift.

\section*{Data Availability}

Data directly related to this publication and its figures is available on request from the corresponding author. The IllustrisTNG simulations themselves including the more recent TNG50 data, are publicly available and accessible at \url{www.tng-project.org/data}.

\section*{Acknowledgements}

We thank Elad Zinger, Mohammadreza Ayromlou, Bruno Henriques, Sydney Sherman, Shardha Jogee, and Danilo Marchesini for useful discussions, and Andrew Wetzel for providing data files.
MD acknowledges support from the Deutsche Forschungsgemeinschaft (DFG, German Research Foundation) -- Project-ID 138713538 -- SFB 881 (``The Milky Way System'', subproject A01). FM acknowledges support through the Program "Rita Levi Montalcini" of the Italian MIUR.
MV acknowledges support through an MIT RSC award, a Kavli Research Investment Fund, NASA ATP grant NNX17AG29G, and NSF grants AST-1814053, AST-1814259 and AST-1909831.
The IllustrisTNG simulations were run on the Hazel Hen Cray XC40-system at the High Performance Computing Center Stuttgart (HLRS) as part of projects GCS-ILLU and GCS-DWAR of the Gauss Centre for Supercomputing (GCS). Additional calculations were run at the Max Planck Computing and Data Facility (MPCDF).

\bibliographystyle{mnras}
\bibliography{biblio}


\appendix

\section{Effects of aperture on  the galaxy stellar mass}
\label{appendix_A}
\begin{figure}
\centering
\includegraphics[width=0.48\textwidth]{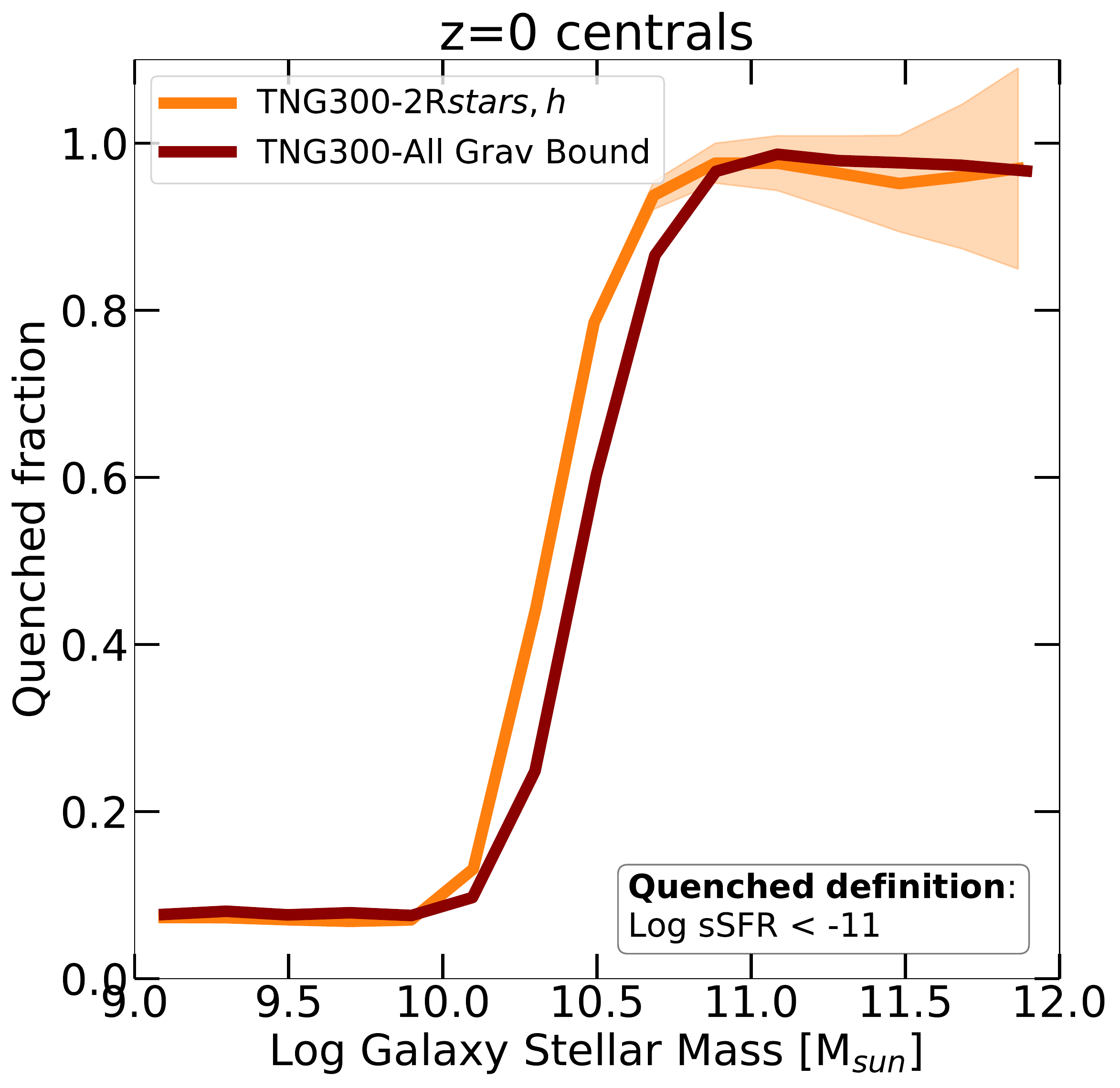}
\caption{\label{fig:MstarsDefinition} {\bf Effect of aperture on the galaxy stellar mass}. Quenched fractions of TNG300 centrals at $z=0$ for two different stellar mass definitions: within twice the stellar half mass radius (orange curve) and accountinf for all gravitationally bound particles (dark red curve). Different stellar mass definitions imply a slightly shifted quenched fraction in the x-axis direction.}.
\end{figure}

Whereas throughout the paper we pay great attention to the way the quenched fractions are measured, also differences in the way galaxy stellar masses are estimated may a priori have an impact on how the quenched fractions from different sample compares. In Figure \ref{fig:MstarsDefinition}, we quantify this possibility by showing the quenched fractions from TNG300, at fixed definition, vs. galaxy stellar masses: the latter are obtained within twice the stellar half mass radius (orange curve, as in the main body of the paper) and accounting for all the stellar particles that are gravitationally bound to a galaxy's subhalo (red curve). A different stellar mass choice implies a shift in the x-axis direction, but this is noticeable only in the steep portion of the trend, where it can be mistaken for a difference in quenched fraction of up to 30 percentage points at fixed galaxy mass.

\section{Host and stellar mass distributions: TNG versus SDSS sample}
\label{appendix}

\begin{figure*}
\centering
\includegraphics[width=0.30\textwidth]{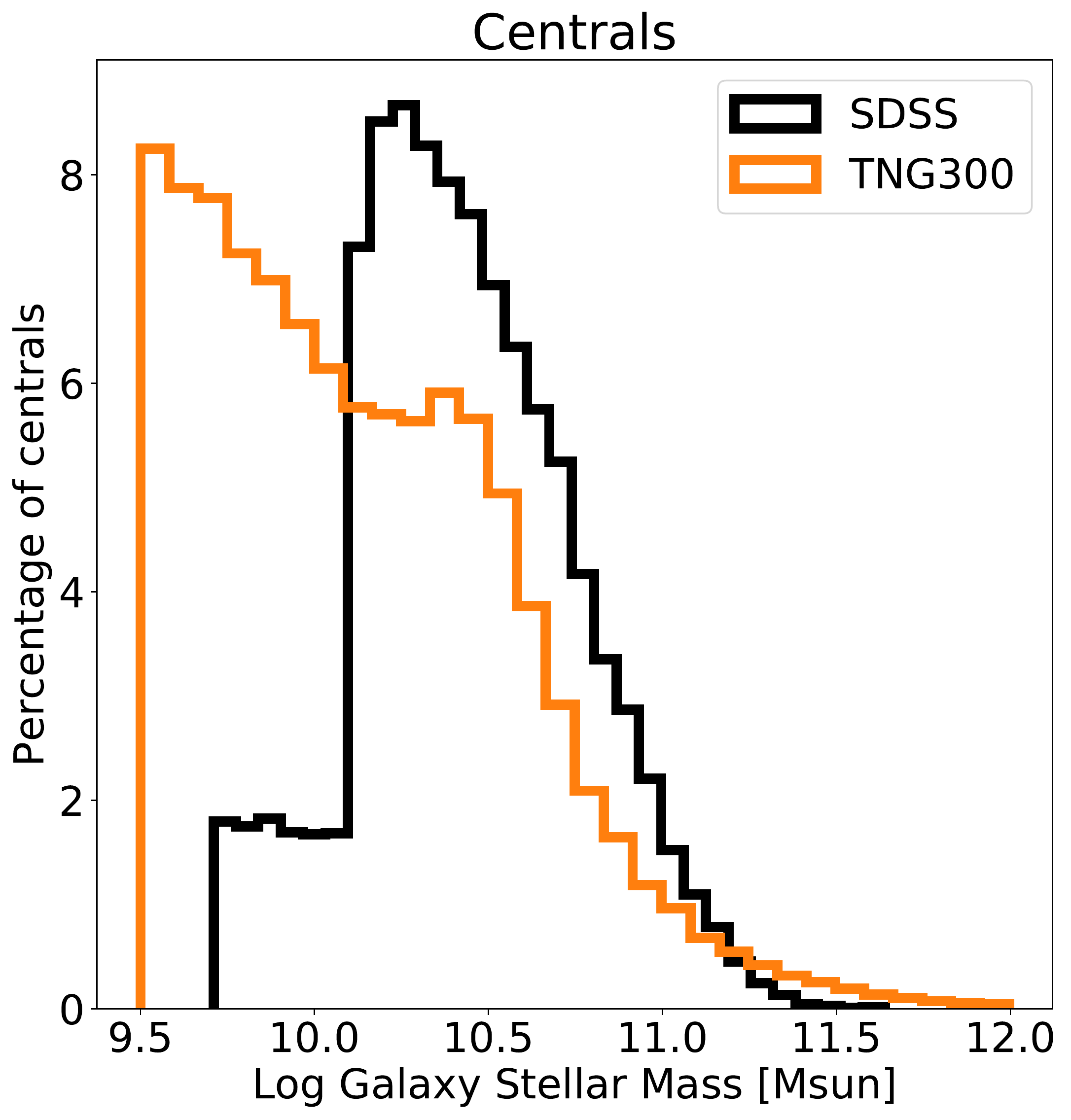}
\includegraphics[width=0.30\textwidth]{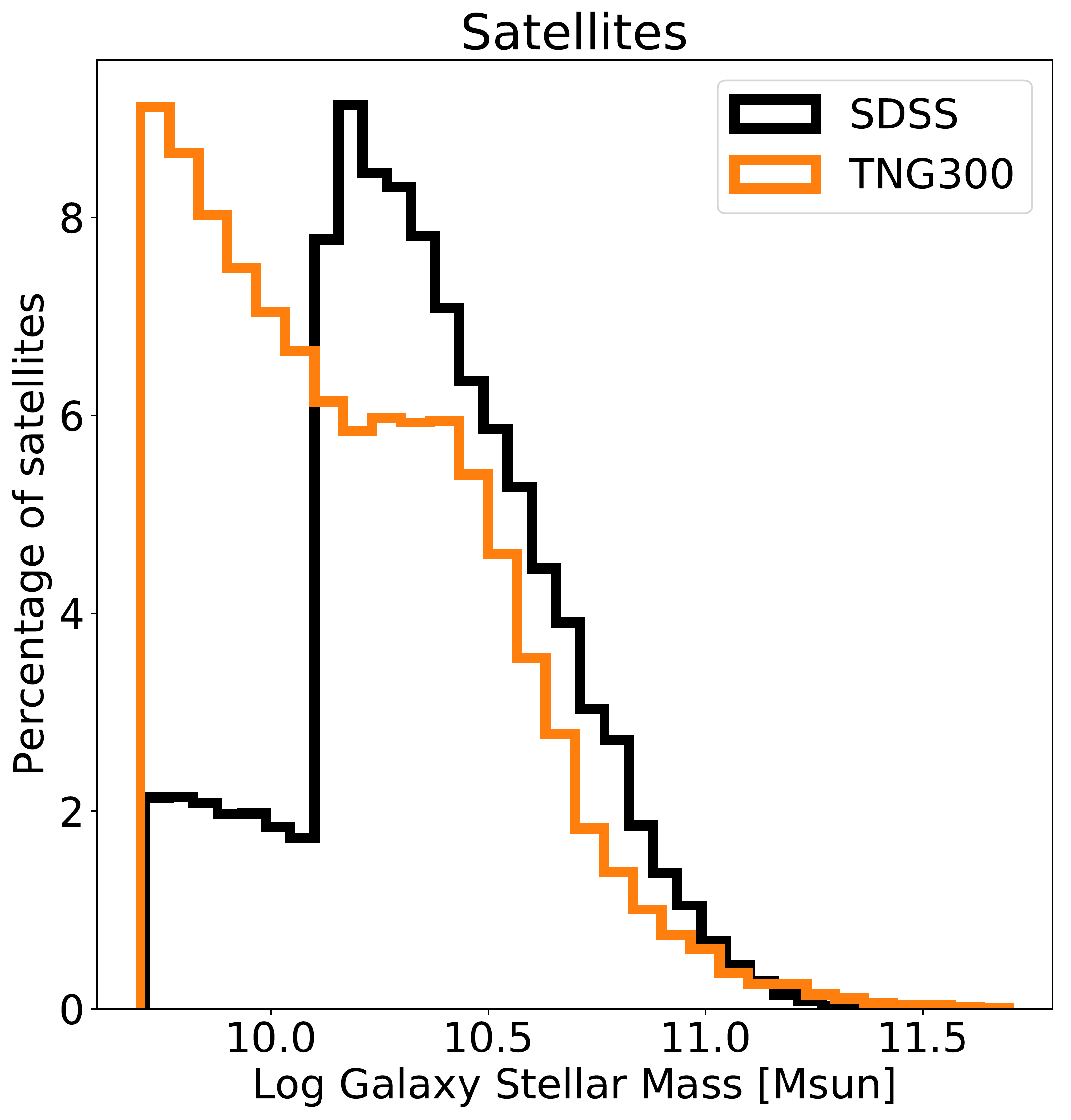}
\includegraphics[width=0.30\textwidth]{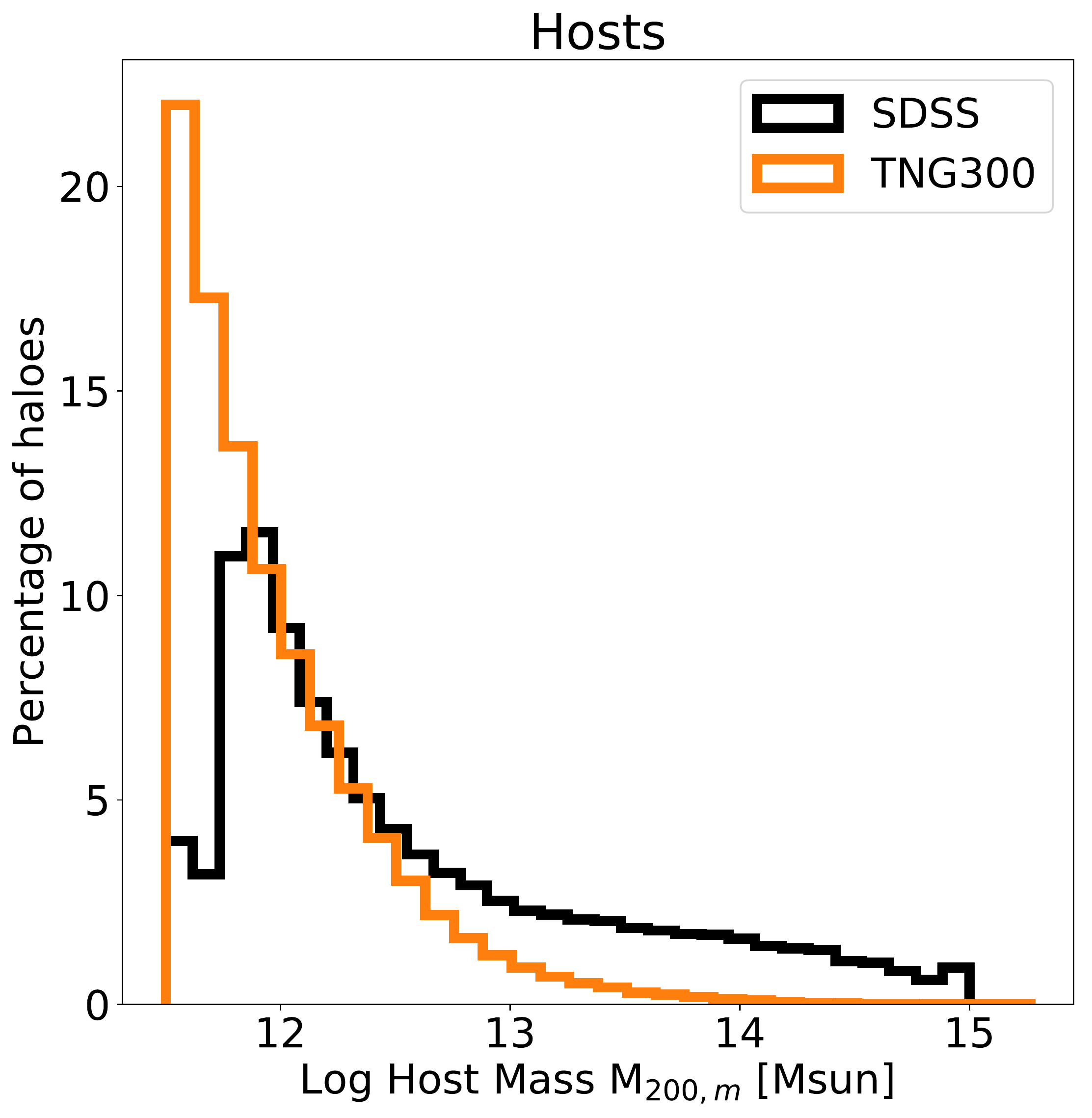}
\includegraphics[width=0.30\textwidth]{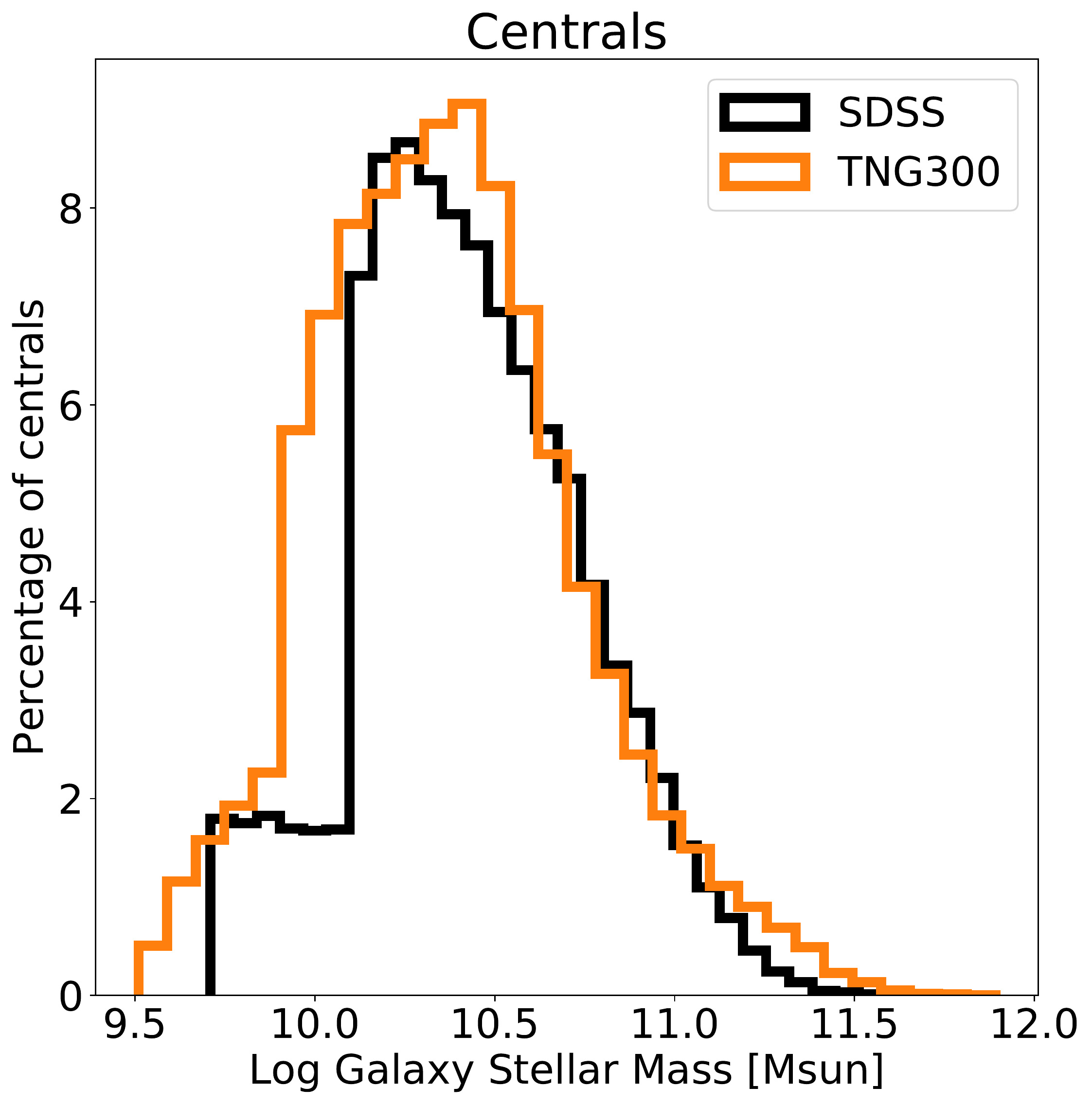}
\includegraphics[width=0.30\textwidth]{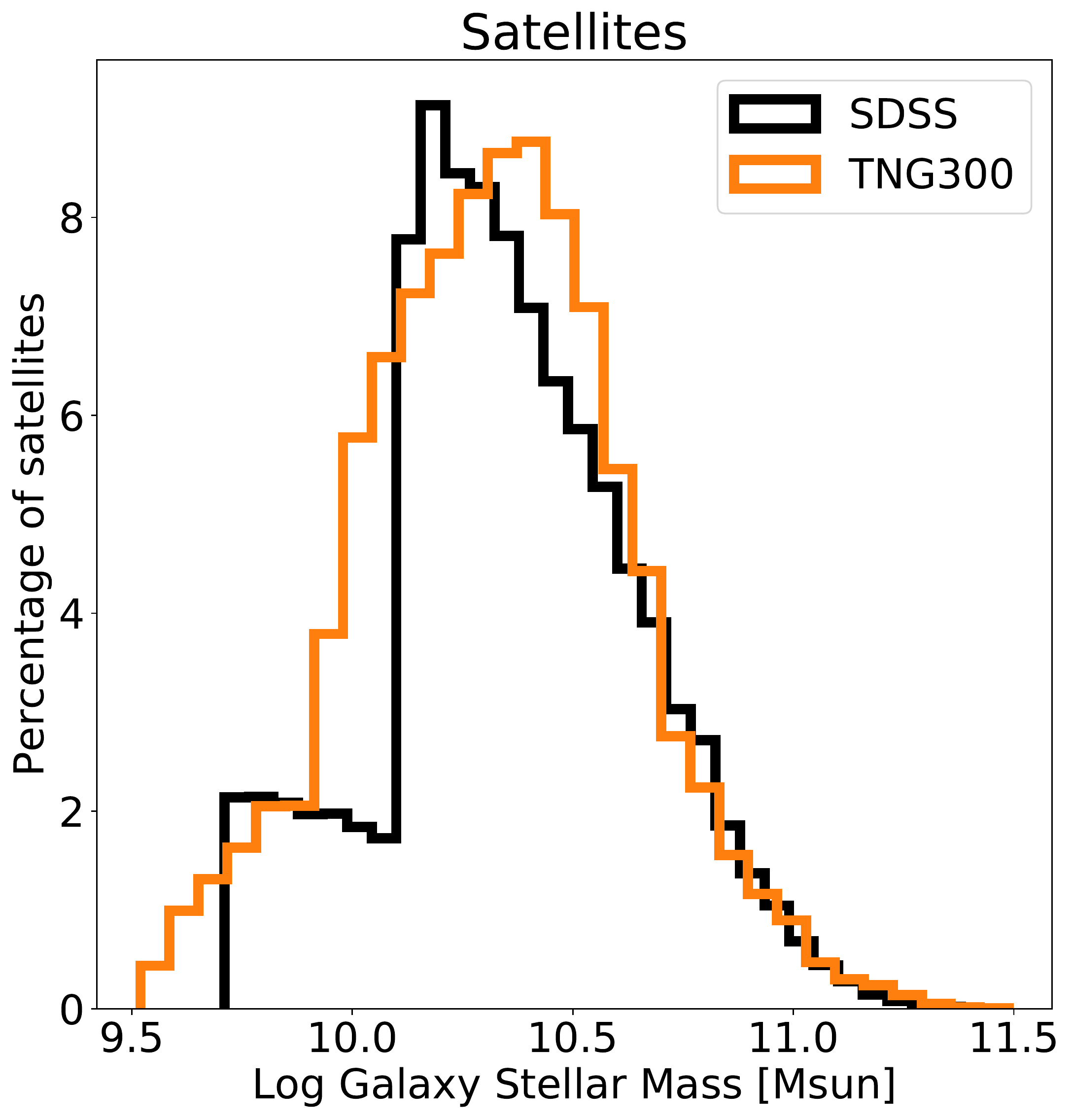}
\includegraphics[width=0.30\textwidth]{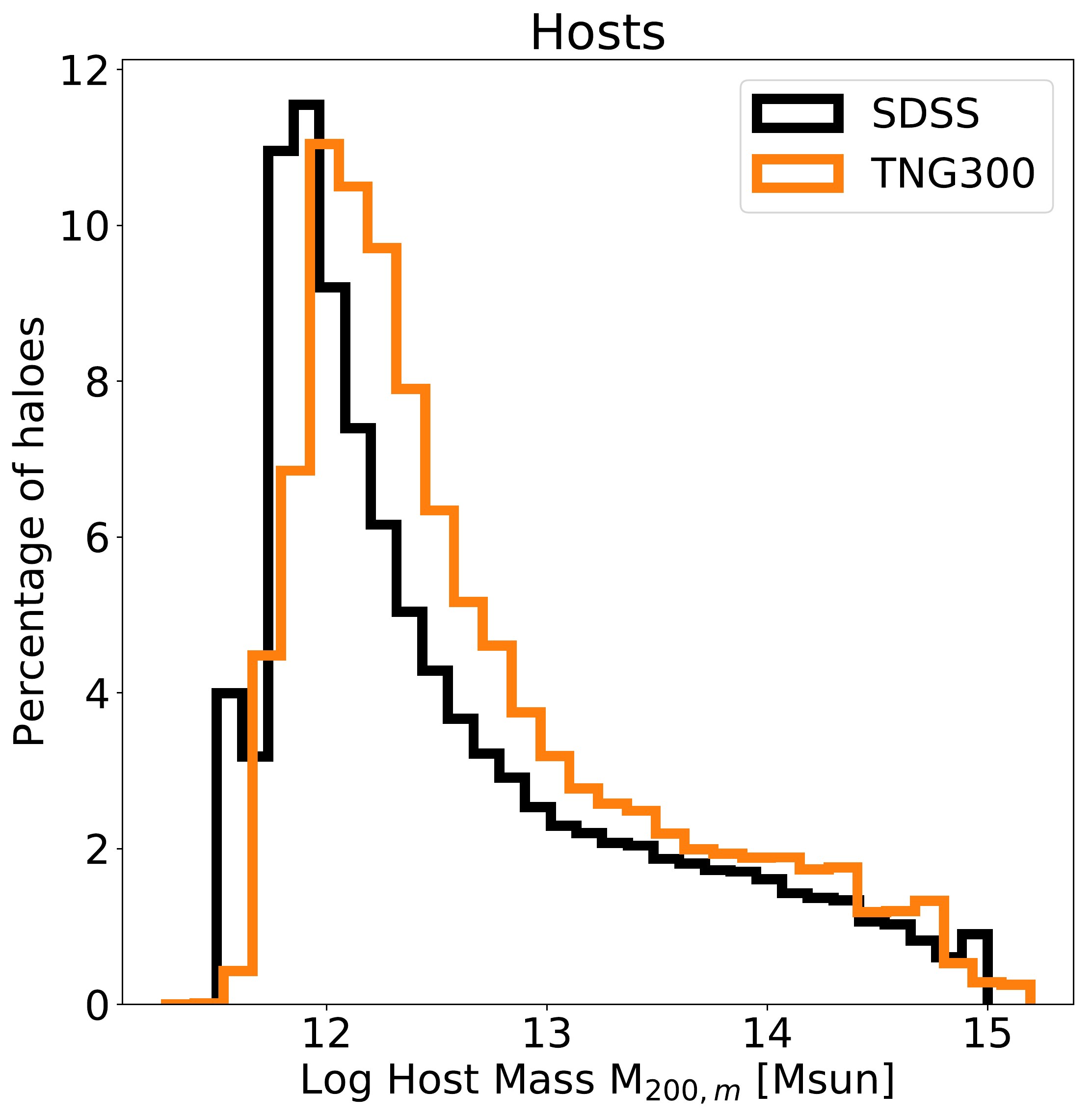}
\caption {\label{fig:distributions} {\bf Host and stellar mass distribution of TNG300 and SDSS}. The mass distribution of centrals (left), satellites (middle) and hosts (right) in TNG300 (orange) and SDSS (black). The distribution of SDSS galaxies is the same in the top and bottom row, while for TNG300 we show the distribution of the raw data, i.e. the output of the simulation (top), and the distribution of the mock catalog (bottom).}
\end{figure*}

As discussed in Section \ref{sec:mock}, in order to properly compare the outcome of TNG300 with SDSS data we create a mock catalog of TNG300 satellites and centrals, by matching measurement choices and sample selections as adopted in \cite{2012Wetzel} and \cite{2013Wetzel} (see Figure \ref{fig:mock}). Importantly, since the SDSS catalogs used to compare with TNG include \textit {only} galaxies with stellar mass in the range $\MS = 10^{9.7-11.6} \Ms$ and haloes in the range $M_{200m} = 10^{11.4-14.9} \Ms$, we match the stellar mass distribution of both centrals and satellites, separately, as well as the host mass distribution and halocentric distance \citep[see][for more details]{2011Tinker}. Practically, for each SDSS galaxy with given properties, we randomly select a TNG300 galaxy satisfying: a stellar mass within 0.2 dex, a host mass $M_{200m}$ within 0.1 dex, and a 2D projected distance/R$_{200m}$ below 10 percent from the corresponding observed values. We note that the host mass distribution and distances from the cluster centre have been matched only for satellite galaxies.

In Figure \ref{fig:distributions} we show the resulting mass distributions of centrals (left panels), satellites (middle panels) and host masses (right panels), for TNG300 (orange curves) and SDSS data (black curves). The top panels shows the intrinsic distribution in the TNG300 simulation. The bottom panels show the mocked distributions, which are used for the observational comparisons.

\section{Numerical resolution effect on quenched fractions}

\label{appendix_B}
We show here the resolution effect for the different resolution realizations of TNG50, namely TNG50-2 and TNG50-3 characterized by a baryonic mass of $6.8\times 10^5 \Ms$ and $5.4\times 10^6 \Ms$, respectively, in comparison to $8.5\times 10^4 \Ms$ for the flagship run.

\begin{figure*}
\centering
\includegraphics[width=0.46\textwidth]{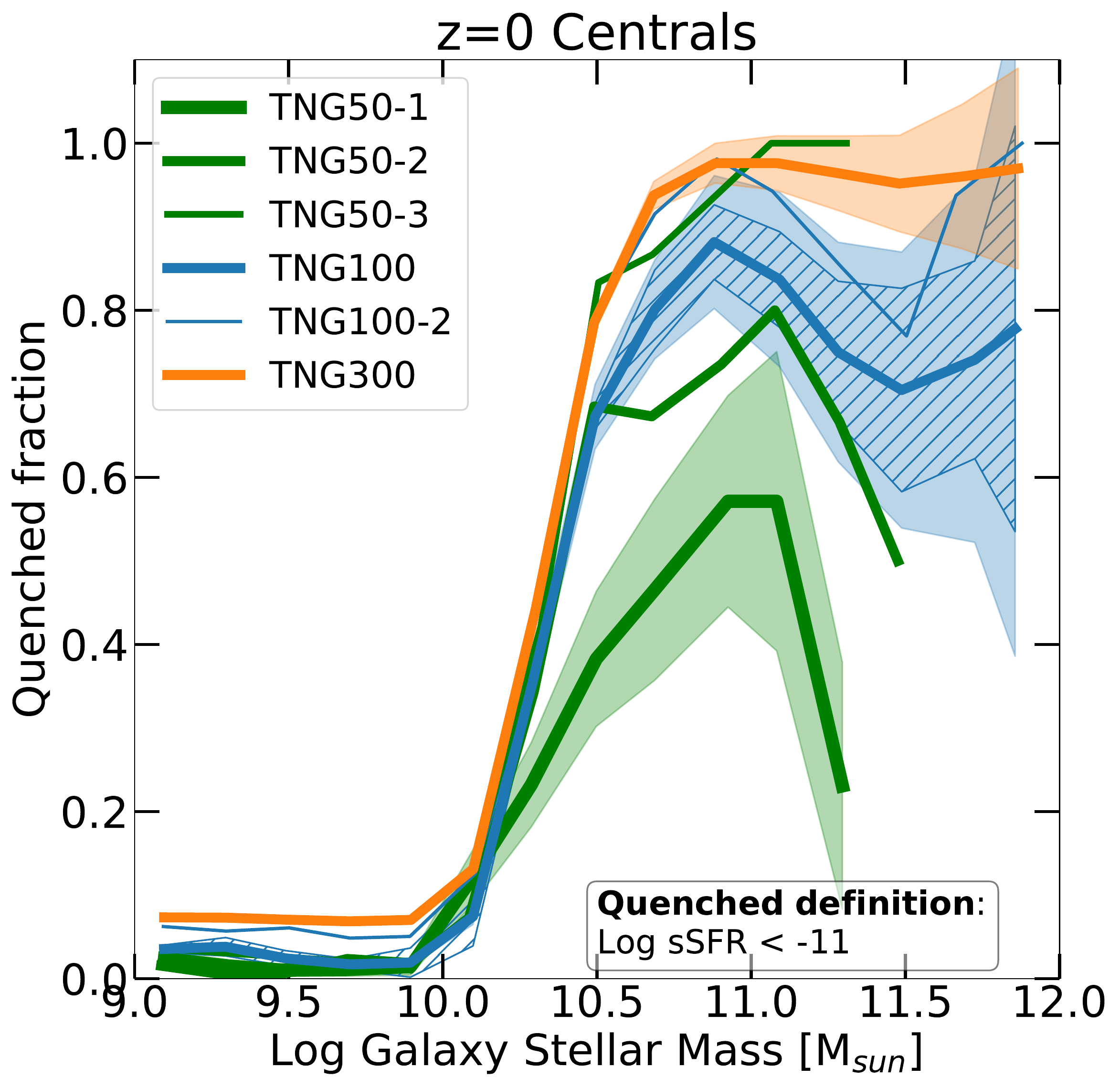}
\includegraphics[width=0.46\textwidth]{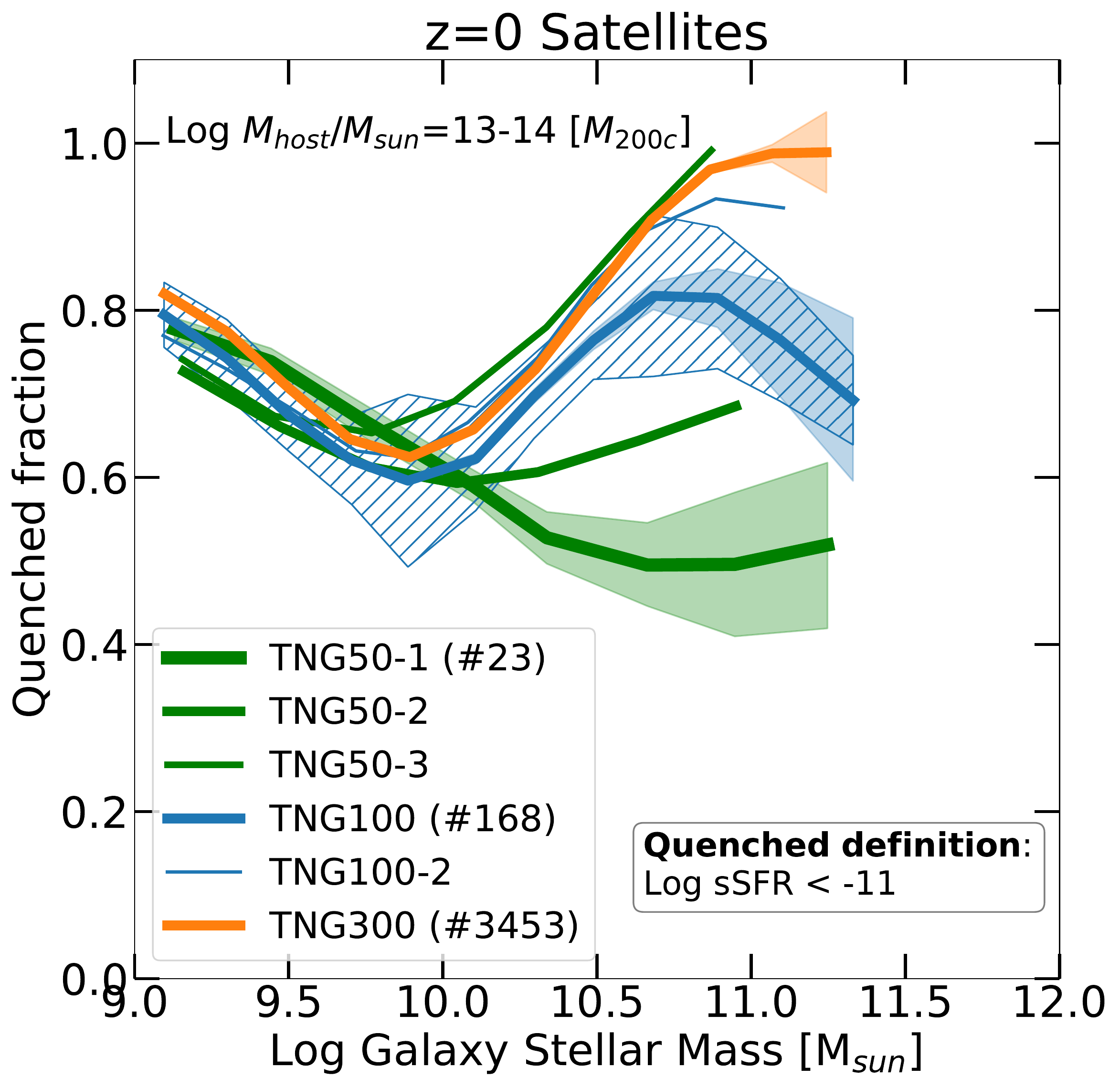}
\caption {\label{fig:resolution} {\bf Resolution study of the quenched fractions as a function of stellar mass}. Fraction of quenched galaxies as a function of stellar mass for $z=0$ centrals (left) and satellites (right) according to the three different resolution levels of TNG50 (green), two resolution levels of TNG100 (blue), and TNG300 (orange). On the right panel, in parenthesis we show the number of hosts with masses $10^{13-14} \Ms$.}
\end{figure*}

In Figure \ref{fig:resolution} we show the quenched fractions of centrals (left) and satellites (right) at $z=0$ according to the three resolution levels of TNG50 (green), represented with different curve thickness. In all simulations, we select satellites residing in hosts with masses $10^{13-14} \Ms$.
To separate quenched and star-forming galaxies we use Log (sSFR/yr)$<-11$.
For comparison, we also show results of TNG100, TNG100-2 (blue) and TNG300 (orange). We remind here that TNG100 has a mass resolution in between TNG50-2 and TNG50-3, while TNG100-2 and TNG300 have a mass resolution somewhat similar to TNG50-3.  

For centrals, as already shown in \cite{2021Donnari}, a lower resolution returns a higher quenched fraction at fixed galaxy stellar mass, by which amount depending on mass scale. For satellites, we see here that, while the quenched fractions agree very well (within 10 percent) across {\it all} the TNG resolution levels and across all satellite masses for more massive hosts (see top right panel of Figure~\ref{fig:models}), for group-mass hosts ($10^{13-14} \Ms$) the TNG simulations agree to a similar degree only at the satellite low-mass end, while exhibiting a resolution-dependent trend for massive galaxies, $\MS \gtrsim 10^{10.5}\Ms$. This would seem to suggest that, at least in TNG, the effects of environmental quenching processes are less susceptible to numerical resolution than the effects of SMBH feedback. However, besides the resolution, the different statistical sampling across the different TNG simulated volumes plays a role in determining the differences we see at the high mass end of both centrals and satellites, since the volume of TNG50 is considerably smaller than TNG100 and TNG300. At $z=0$ in TNG50 there are 23 group-mass hosts, compared to e.g. 168 in TNG100. We have verified that when we randomly downsample the TNG100 host population to 23 objects as is the case for TNG50, in about 30-40 percent of the cases also TNG100 returns quenched fractions similar or even lower than the average ones of TNG50 at the $\MS \sim 10^{11}\Ms$ mass scale. This suggests that the differences between TNG50 and TNG100/TNG300 that we see in Figure \ref{fig:resolution} are due to a combination of resolution effects and sample variance, the former probably to a larger degree.

\begin{figure*}
\centering
\includegraphics[width=0.46\textwidth]{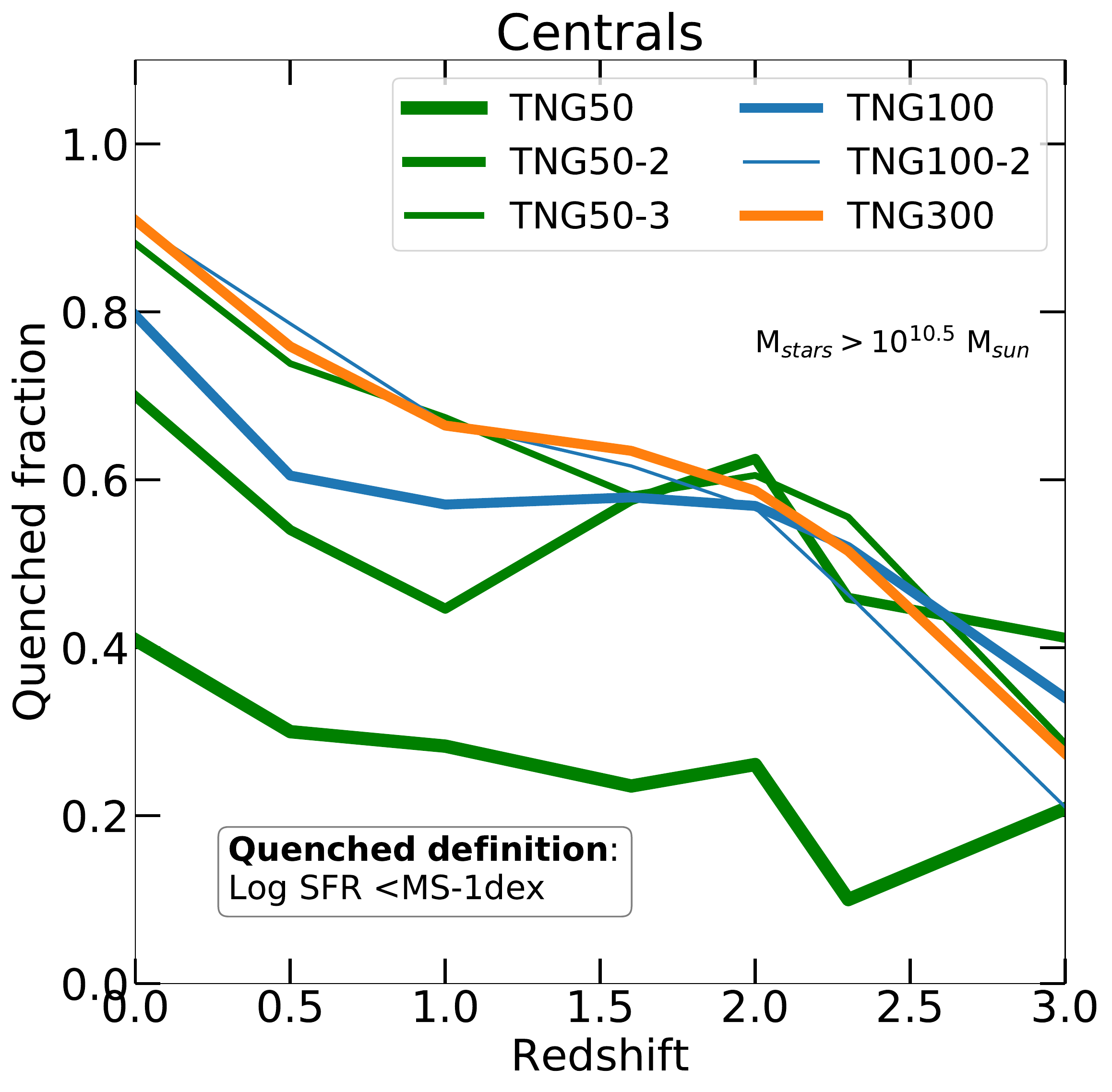}
\includegraphics[width=0.46\textwidth]{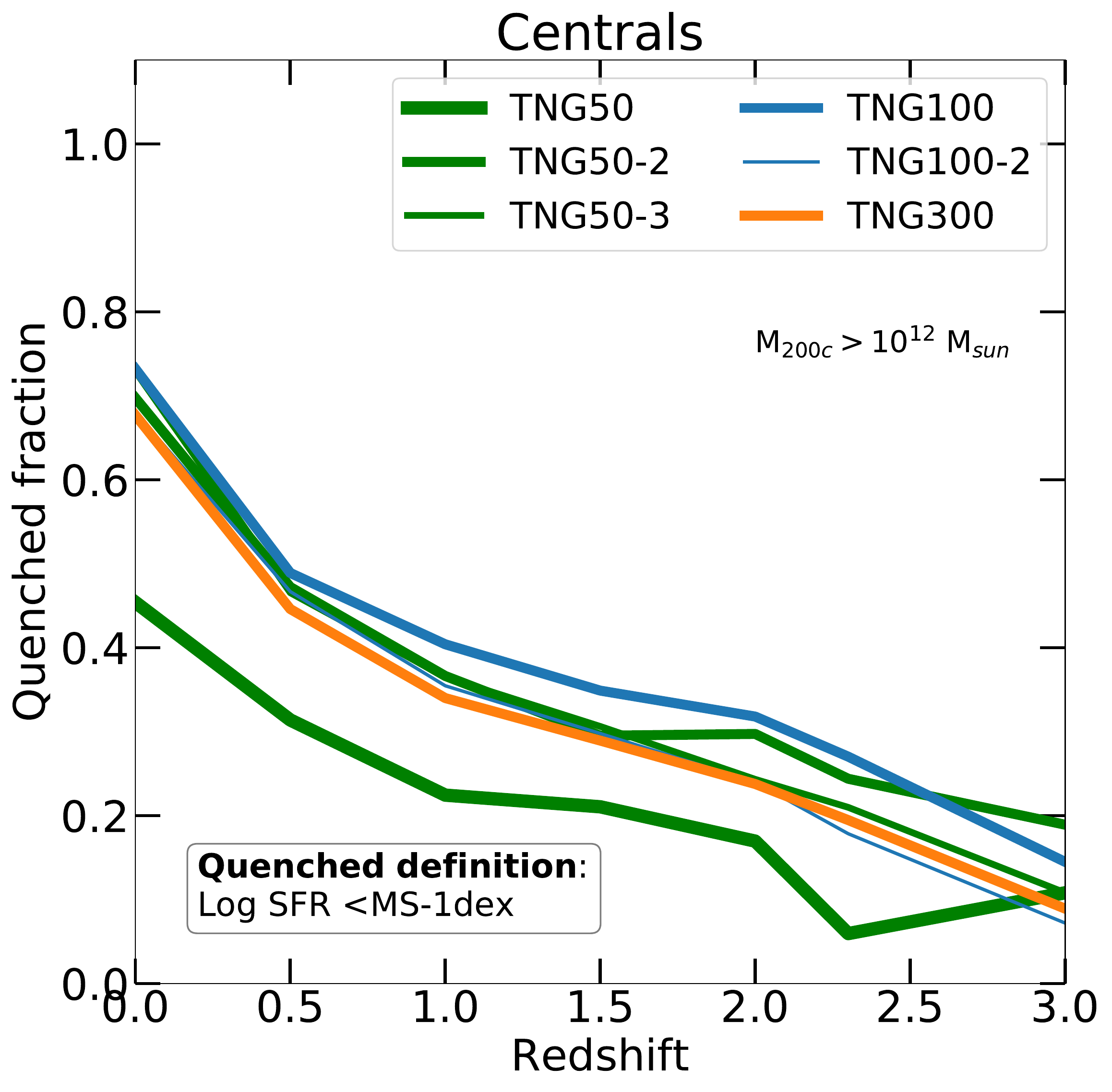}
\caption {\label{fig:QfracVSz} {\bf Resolution study of the quenched fractions as a function of redshift}. Fraction of quenched centrals as a function of redshift for all central galaxies with $\MS>10^{10.5} \Ms$ (left) and for all central galaxies residing in haloes with $M_{200c}>10^{12} \Ms$ (right), according to the three different resolution levels of TNG50 (green), the two resolution levels of TNG100 (blue), and the highest resolution level of TNG300 (orange).}
\end{figure*}

Figure \ref{fig:QfracVSz} shows the quenched fractions as a function of redshift for all centrals with stellar mass larger than $10^{10.5}\Ms$ (left panel) and for all central galaxies in haloes with $M_{200c}>10^{12} \Ms$ (right panel). Similarly to Figure \ref{fig:resolution} we show the three resolution runs of TNG50 (green), two runs of TNG100 (blue) and TNG300 (orange) spanning a range of particle mass resolutions across a factor larger than 100. 

Overall, when galaxies are selected according to their stellar mass (left panel), the lower resolution runs return higher quenched fractions, with the actual amount depending on the redshift and simulation. Indeed, while TNG50 shows quenched fractions lower than all the other runs regardless of redshift ($\sim$ 10 percentage points at $z=3$ and up to 40-50 percentage points at $z=0$), at $z<3$ TNG300, TNG100-2, and TNG50-3 agree well with TNG100, showing a quenched fraction of about 40 percent on average at e.g. $z=2-3$. Interestingly, while all runs return a quenched fraction that is generally monotonically decreasing with redshift, individual runs may exhibit fluctuations at specific times, possibly due to sample variance.

On the right panel of Figure \ref{fig:QfracVSz}, central galaxies are taken with a cut in their host halo mass instead of galaxy stellar mass. Here, the resolution trends seen so far (higher fractions for lower resolution runs) are much less pronounced. TNG100, TNG300, TNG50-2 and TNG50-3 return almost the same quenched fractions at $z\lesssim 3$ within 10 percentage points. As in the left panel, TNG50 shows lower fractions than all the other runs at all redshifts, but, in this case, the difference is smaller, capped at 20 percentage points of differences at most.

The results presented in Figures~\ref{fig:resolution} and \ref{fig:QfracVSz} suggest that, within the IllustrisTNG model, numerical resolution impacts the results for the quenched fraction vs. galaxy stellar mass or vs. redshift also because it impacts the stellar mass of galaxies that form within haloes (see Discussions in \citealt{2018Pillepich, 2020Engler}). Furthermore, an improved implementation of the subgrid scheme for SMBH feedback that is less susceptible to numerical resolution, in fact, may lessen the differences. We remind here that, besides the softenings, also the BH kernel-weighted neighbour number \citep{2017Weinberger} that defines the BH accretion and feedback region is changed with resolution, with a scaling based on the baryonic target mass \citep{2018Pillepich_model}. A different scaling might help enhance the convergence of the quenched fractions of high-mass galaxies in the TNG model across the whole range of resolution levels probed by the TNG runs.

\label{lastpage}
\end{document}